\pdfoutput=1
\documentclass[a4paper,fleqn,usenatbib]{mnras}

\usepackage{amsmath}
\usepackage{amssymb}
\usepackage{bm}
\usepackage{graphicx}
\usepackage{multirow,colortbl}
\usepackage{verbatim}
\pdfmapfile{+newtx.map}
\usepackage{newtxtext,newtxmath}
\usepackage[T1]{fontenc}
\usepackage{ae,aecompl}

\title[Equal Mass Mergers of Galaxy Clusters]{In the Wake of Dark Giants: New Signatures of Dark Matter Self Interactions in Equal Mass Mergers of Galaxy Clusters}

\author[S. Y. Kim et al.]{
Stacy Y. Kim,$^{1,2}$\thanks{E-mail: kim.4905@osu.edu (SYK)}
Annika H.~G. Peter,$^{1,2,3}$
and David Wittman$^{4}$	
\\
$^{1}$Department of Astronomy, The Ohio State University, 140 W. 18th Ave., Columbus, OH 43210, USA \\
$^{2}$Center for Cosmology and AstroParticle Physics (CCAPP), The Ohio State University, 191 W. Woodruff Ave., Columbus, OH 43210, USA \\
$^{3}$Department of Physics, The Ohio State University, 191 W. Woodruff Ave., Columbus, Ohio 43210, USA \\
$^{4}$Physics Department, University of California, Davis, CA 95616, USA}

\date{Accepted XXX. Received YYY; in original form ZZZ}

\pubyear{2016}

\begin{document}
\label{firstpage}
\pagerange{\pageref{firstpage}--\pageref{lastpage}}
\maketitle

\begin{abstract}
Merging galaxy clusters have been touted as one of the best probes for constraining self-interacting dark matter, but few simulations exist to back up this claim.  We simulate equal mass mergers of 10$^{15}$ M$_\odot$ halos, like the El Gordo and Sausage clusters, with cosmologically-motivated halo and merger parameters, and with velocity-independent dark-matter self-interactions.  Although the standard lore for merging clusters is that self-interactions lead to large separations between the galaxy and dark-matter distributions, we find that maximal galaxy-dark matter offsets of $\lesssim~20$~kpc form for a self-interaction cross section of $\sigma_\text{SI}/m_\chi$ = 1 cm$^2$/g.  This is an order of magnitude smaller than those measured in observed equal mass and near equal mass mergers, and is likely to be even smaller for lower-mass systems.  While competitive cross-section constraints are thus unlikely to emerge from offsets, we find other signatures of self-interactions which are more promising.  Intriguingly, we find that after dark matter halos coalesce, the collisionless galaxies (and especially the Brightest Cluster Galaxy [BGC]) oscillate around the center of the merger remnant on stable orbits of 100 kpc for $\sigma_\text{SI}/m_\chi = 1$~cm$^2$/g for at least several Gyr, well after the clusters have relaxed.  If BCG miscentering in relaxed clusters remains a robust prediction of SIDM under the addition of gas physics, substructure, merger mass ratios (e.g., 10:1 like the Bullet Cluster), and complex cosmological merger histories, the observed BCG offsets may constrain $\sigma_\text{SI}/m_\chi \lesssim$ 0.1 cm$^2$/g---the tightest constraint yet.
\end{abstract}



\section{Introduction}
\label{sec:introduction}

In the 1980s, a new cosmological paradigm emerged in which dark matter was cold and collisonless.  Studies of cold dark matter \citep[CDM,][]{1983ApJ...271..417F} were remarkably successful at explaining a wide range of observations, from the cosmic microwave background to the abundance and clustering of $L_*$ galaxies \citep{2006PhRvD..74l3507T, 2015PhRvD..92l3516A, 2015arXiv150201589P}.  However, it remains unclear whether CDM can explain observations on sub-galactic scales.  Cosmological simulations without gas and star formation routinely predict central halo densities significantly higher than observed \citep{2010ApJ...717L..87W, 2011ApJ...741L..29K, 2011MNRAS.415L..40B, 2012MNRAS.425.2817F, 2013ApJ...765...25N}.  Moreover, there are lingering questions as to whether the abundance of dwarf galaxies is in line with CDM predictions, with claims in both the directions that CDM produces too many small halos relative to the number of small galaxies \citep{1999ApJ...522...82K, 1999ApJ...524L..19M, 2015A&A...574A.113P}, and too few \citep{2002ApJ...572...25D, 2014MNRAS.442.2017V}.  While simulations that include gas physics have come a long way in reconciling CDM theory with observations \citep{2012ApJ...761...71Z, 2013ApJ...765...22B, 2015MNRAS.454.2092O, 2016MNRAS.tmp..504R, 2016MNRAS.457.1931S}, questions linger about the key assumptions of CDM.  Can relaxing either the cold or collisionless assumptions of dark matter's fundamental nature lead to a better fit to observations of galaxies and clusters?

The idea that dark matter may be collisional is interesting to both physicists and astronomers.  Evidence of collisionality would dethrone the dominant weakly interacting massive particle (WIMP) paradigm \citep{1985NuPhB.253..375S, 1988PhRvD..38.2357G}.  Strong self-interactions are typically predicted in models for which dark matter exists in a ``hidden sector'', in which forces between dark-matter particles are mediated by analogs to electroweak or strong forces \citep{2000PhRvL..84.3760S, 2008PhLB..662...53P, arkanihamed2009, buckley2010, feng2010, 2014PhRvD..89k5017B}.  Self-interacting dark matter (SIDM) predicts that dark-matter halos have low-density cores instead of of high-density cusps, and that these cores are especially prominent in low-baryon systems like dwarf galaxies \citep{2000PhRvL..84.3760S, 2001ApJ...547..574D, 2012MNRAS.423.3740V, 2013MNRAS.430...81R, 2016arXiv160308919D}.  Depending on the exact particle model for the scatter, SIDM can lead to halo mass functions that are either nearly identical to CDM, or significantly suppressed on small scales \citep{2014PhRvD..90d3524B, 2015arXiv151205344C, 2015arXiv151205349V}.

Merging clusters offer a unique insight into dark matter collisionality.  Clusters are the largest bound dark-matter objects in the universe, and mergers between clusters are the biggest particle colliders in the Universe.  They are thus prime systems to study dark-matter self-scattering cross sections.  Merging clusters also contain two components that bracket the extrema in collisionality---highly collisional gas and effectively collisionless galaxies---the evolution of which can be compared to that of the dark matter.

The dominant paradigm for merging clusters in SIDM is that the dark matter collisionality acts as a drag force, in which the forward momentum of the smaller cluster (``subcluster'') is reduced because of two-body interactions between particles in each respective subcluster.  After a cluster collision in this paradigm, the gas, galaxies, and dark matter separate at speeds determined by the strength of their self-interactions---those that are more collisional lag behind the less collisional components \citep{2004ApJ...606..819M}.  It has thus been proposed that measuring the offset between the dark matter and the collisionless galaxies in merging clusters could provide a means to measure the dark matter self-interaction cross section. 

Indeed, this paradigm has been used to set SIDM cross section constraints.  Galaxy-dark matter offsets have been observed in some mergers, such as the Musket Ball Cluster \citep{2013PhDT.......211D}, El Gordo \citep{2014ApJ...785...20J}, the Baby Bullet Cluster \citep{2008ApJ...687..959B}, and CIZA J2242.8+5301 \citep{2015ApJ...802...46J}, of order $100-200$ kpc (see Table \ref{tab:obs_mergers}).  Interpreted in a simple paradigm in which an offset develops if the optical depth to scattering in the interaction $\Sigma \sigma_\text{SI}/ m_\chi \sim 1$ (with $\Sigma$ being the column density of dark matter in the more massive cluster and $\sigma_\text{SI}/m_\chi$ being the SIDM elastic scattering cross section), the cross section limits are of order $\sigma_\text{SI}/m_\chi = 1 - 10\hbox{ cm}^2/\hbox{g}$.   Similar constraints come from observations of individual galaxies in clusters \citep{2015MNRAS.449.3393M, 2015MNRAS.452L..54K}.  In addition, some more complex mergers of involving more than two clusters as Abell 520 \citep[][although see also \citet{2012ApJ...758..128C}]{2012ApJ...747...96J, 2014ApJ...783...78J} and Abell 2744  \citep[Pandora's Cluster, ][]{2011MNRAS.417..333M} have revealed the presence of a ``dark core," a dark matter substructure observed to have no associated galaxies.

Complicating this picture is that not every merging cluster has an apparent offset between the galactic and dark components.  For example, the Bullet Cluster exhibits a marginal lack of offsets, implying $\sigma_\text{SI}/m_\chi \lesssim 1.25 \hbox{ cm}^2/\hbox{g}$ \citep{2008ApJ...679.1173R}.  The non-detection of offsets of individual galaxies in clusters from their subhalos, found by statistical analysis of ensembles of systems, have been interpreted as indicating that $\sigma_\text{SI}/m_\chi \lesssim 0.5 \hbox{ cm}^2/\hbox{g}$ \citep[][though see Wittman et al., in prep]{2015Sci...347.1462H}.

Partly feeding into the confusion in the interpretation of galaxy-dark matter offsets is the lack of theoretical work on their formation.  Most theoretical work has focused on the most famous merging cluster exemplar, the Bullet Cluster, which has unique kinematic properties \citep{2015MNRAS.452.3030T}.

Moreover, there has been significant confusion in the literature in mapping the microphysical properties of scatter to the macroscopic phenomenology of the merger.  Much of the modeling work on the Bullet Cluster has been undertaken assuming a microphysical model of velocity-independent contact (``hard sphere") interactions \citep{2004ApJ...606..819M,2008ApJ...679.1173R,2016arXiv160504307R}.  This was interpreted as leading to a macroscopic drag force on the subcluster, and yielding a separation between the subcluster galaxy and dark-matter distributions of $\sim 10-40$ kpc for $\sigma_\text{SI}/m_\chi \sim 1$ cm$^2$/g.   \citet{2014MNRAS.437.2865K} argue strongly against this mapping between microphysical and macrophysical models.  They demonstrate that hard-sphere interactions lead to the expulsion of particles from the subcluster (of the Bullet Cluster, the Musket Ball, and A520).  While the wake of expelled particles can exert a gravitational force on both the galaxies and remaining dark matter in the subcluster, it does not lead to a separation in the peaks of the two populations, but may lead to a difference in the centroid because dark matter dominates the wake.   \citet{2014MNRAS.437.2865K} suggest that the drag macroscopic picture only applies when the microphysical model allows relatively frequent collisions that lead to particles being retained by the halos.  Models of this sort are the ones constrained by the ``bulleticity'' method \citep{2014MNRAS.441..404H,2015Sci...347.1462H}.

However, there has not yet been a fully self-consistent set of simulations exploring the evolution of merging clusters with either set of microphysical models, sampling the types of merger configurations found in nature.  Moreover, \citet{2016arXiv160504307R} highlight the fact that the size of any type of offset depends sensitively on the specific definition of the ``center" of the distribution.  A robust quantitative mapping between observed systems and SIDM cross sections is still largely missing.

Systematically determining how SIDM imprints itself on mergers as a function of merger configuration is especially important in light of the rapidly growing sample of merging clusters and because current SIDM cross section constraints from mergers are already competitive with other methods \citep{2013MNRAS.430..105P, 2016PhRvL.116d1302K}.  Merging clusters are being identified at a nearly exponential rate due to millimeter and radio surveys \citep[e.g.,][]{2011JApA...32..505V, 2012ApJ...748....7M, 2015ApJ...808L...6M}.  To use these mergers to study dark-matter collisionality, and to determine which systems will lead to the best constraints, we need to know how they work.

This work is the first in our simulation program to find and understand SIDM signatures in merging clusters as a function of merger properties and microphysical SIDM models.  In this work, we have run a suite of simulations to study the formation of offsets in the simplest merger: an equal mass, binary merger.  The Baby Bullet \citep{2008ApJ...687..959B}, CIZA J2242.8+5301 \citep{2015ApJ...802...46J}, Musket Ball \citep{2012ApJ...747L..42D}, and El Gordo \citep{2014ApJ...785...20J} clusters are all believed to be equal or near-equal mass mergers.  

Adopting a simple hard-sphere, isotropic scattering model for dark matter self-interactions, we find merger phenomenology unlike those predicted in any previous study.  In mergers of 10$^{15}$ M$_\odot$ clusters we find that small offsets ($<50$ kpc) form around pericenter passage between the \emph{peaks} of the dark-matter and galaxy distributions in addition to those between centroids, but are short-lived and from only under a narrow range of cross sections $\gtrsim 1\hbox{ cm}^2/\hbox{g}$.  Offsets consistent with the 100-200 kpc offsets reported in some systems only manifest when the cross sections are so high that the subclusters dissolve, and are thus inconsistent with other merging cluster observables.

Excitingly, we find new, unique signatures of SIDM in mergers that may improve SIDM cross section constraints from mergers by an order of magnitude.  This is because we have found that self-interactions significantly alter the evolution of the cluster merger.  The most interesting of these is related to the fact that mergers trigger \emph{core sloshing} of galaxies in the dark-matter core of the cluster remnant.  Because core sloshing is slow to damp \citep{2006MNRAS.373.1451R}, SIDM generically predicts large separations between the peaks of the dark-matter and galaxy distributions (and of the brightest cluster galaxy, BCG) even in systems where the dark matter and gas appear to have reached dynamical equilibrium.  Based on the distribution of observed offsets between the light and dark components of galaxy clusters \citep{2012ApJ...757....2G, 2014ApJ...797...82L}, we show that hard-sphere SIDM cross section constraints of $\sigma_\text{SI}/m_\chi \lesssim 0.1\hbox{ cm}^2/\hbox{g}$ may be achievable.

We outline the merger and self-interaction physics responsible for the formation of offsets in equal mass mergers in Sec. \ref{sec:theory}.  We study these effects in detail with a suite of $N$-body simulations.  Details of these simulations as well as our methods to calculate offsets between the dark matter and galaxy distributions are outlined in Sec. \ref{sec:methods}.  The offsets we measured in our simulations are discussed in Sec. \ref{sec:offsets}.  These results are compared to observed offsets in Sec. \ref{sec:observability}.  We describe our most promising alternative signature of SIDM in clusters---the large separation between the BCG and peak of the galaxy distribution with respect to the center of the dark matter halo of the merger remnant---in Sec. \ref{sec:bcg_miscentering}.  Alternative methods to constrain self-interactions via mergers are described in Sec. \ref{sec:betterconstraints}.  We discuss extensions of this work to cluster mergers with unequal mass ratios and recommended next steps in Sec. \ref{sec:discussion}, then finish with concluding remarks in Sec. \ref{sec:conclusions}.


\section{THE DEVELOPMENT OF OFFSETS}
\label{sec:theory}

In this section, we present and employ a simple toy model for the macroscopic effect of hard-sphere SIDM scatters on the evolution of the merger and the development of offsets.  We connect our model to past work mapping SIDM microphysics to the development of offsets, and show that equal-mass mergers should have qualitatively different phenomenology than indicated by any prior model.  We use simulations in the remainder of this work to validate this model and to quantify the claims sketched here.

\subsection{Past work}

\begin{figure*}
\centering
\includegraphics[width=0.7\textwidth]{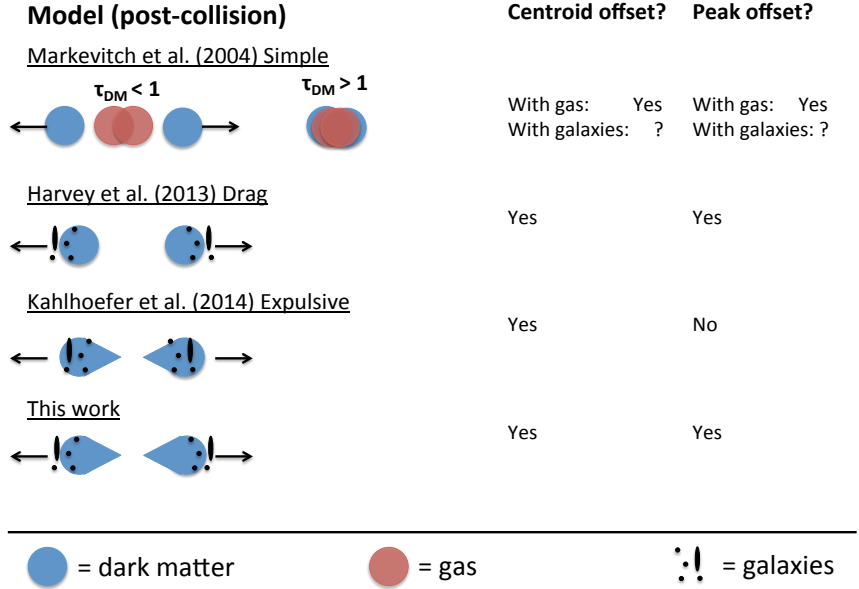}
\caption{\label{fig:cartoon_sidm} Summary of different proposed macroscopic models for SIDM interactions between merging clusters.}
\end{figure*}

Under the CDM paradigm, the evolution of an equal mass cluster merger is qualitatively simple: the merging clusters follow decaying orbits as dynamical friction transforms their bulk relative velocities into random motions, slowing the clusters until they merge.  As the dark matter and the galaxies are both effectively collisionless, their distributions evolve identically, and thus no dark matter-galaxy offsets develop.

How does the introduction of dark matter self-interactions change this picture?  The answer depends on the specific particle model for self-interactions, and on the specific kinematic properties of the merger (i.e., subcluster masses, relative speed, impact parameter).  Fig. \ref{fig:cartoon_sidm} illustrates the different macroscopic models proposed in the literature.  The dominant paradigm has been that SIDM acts like a drag force, acting on the dark matter but not the galaxies \citep{2004ApJ...606..819M}.  In its simplest incarnation, which is typically used to set conservative constraints on SIDM cross sections from clusters \citep[e.g.,][]{2012ApJ...747L..42D}, it assumes that the halos stick together like the gas if the optical depth for SIDM scatters in the merger $\tau > 1$.  Any separation of halos post-collision is interpreted as indicating that $\tau < 1$.  This model is labeled ``Markevitch et al. (2004) Simple'' in Fig. \ref{fig:cartoon_sidm}.  

While the original models assumed that the extra drag force leads to an offset between galaxies and the cluster dark-matter halo that grows with time \citep{2008ApJ...679.1173R}, more recent iterations of the drag phenomenology recognize that the extra drag force leads to a quasistatic equilibrium between the galaxy and dark-matter distributions \citep{2014MNRAS.441..404H}.\footnote{\citet{2004ApJ...606..819M} were first to note that because the galaxies are gravitationally bound to the cluster halo, offsets between the galaxies and dark matter were likely to be relatively small.}  One can make an analogy with springs.  In the absence of gravity, the equilibrium position of a mass at the end of a spring is fixed by the spring's relaxed state.  If gravity is slowly turned on, the equilibrium position shifts such that the gravitational force on the mass is balanced by the spring force.  If there is a drag-like force on the dark-matter distribution, the equilibrium position of the galaxy distribution is displaced from the dark-matter distribution in such a way that the drag force is balanced by the gravitational restoring force.  This leads to an offset that is proportional to the present drag force, which is proportional to the local density of the dark-matter cluster through which the subcluster passes \citep{2015MNRAS.452L..54K}.  One consequence of this model is that offsets should be large at the time of cluster collision, and smallest when the clusters are at apocenter.  This model is labeled ``Harvey et al. (2014) Drag" in Fig. \ref{fig:cartoon_sidm}.

However, \citet{2014MNRAS.437.2865K} argued that this macroscopic interpretation only works for specific \emph{microphysical} models for dark-matter interaction: those with frequent, small-angle scatters.  This is because when two particles from two separate clusters collide, momentum is only transferred from one halo to the other in a drag-like way if the particles are retained by the halos, and not ejected.  After the collision, the average final particle speed must be below the escape speed from the parent halo.  They argued that scatters must be frequent so that enough momentum can transfer between halos, and so that the momentum is thermalized.

Furthermore, \citet{2014MNRAS.437.2865K} claimed that there should be \emph{no} offset between the peaks of the dark-matter and galaxy distributions for particle models that dominantly have large-angle scattering, like our hard-sphere model.  They argued that most particles will be ejected from their parent halos after interactions with particles in the other halo.  Therefore, the momentum per unit mass of the halo is unchanged, and galaxies stay tied to the remaining part of the halo.  \citet{2014MNRAS.437.2865K} find that this gives rise to a difference in the \emph{centroids} of the two populations for a short amount of time, because the ejected particles are preferentially scattered opposite the halo's direction of motion in a structure that looks like a tail.  The offset disappears when that tail escapes beyond the virial radius of the host.  This implies that offsets should initially grow with time, then vanish after the tail escapes.  Interestingly, \citet{2008ApJ...679.1173R} and \citet{2016arXiv160504307R} see offsets between the dark-matter and galaxy centroids in their hard-sphere-scattering SIDM simulations of the Bullet Cluster.  They attribute the offsets to drag, although this is not actually demonstrated in their work.  \citet{2014MNRAS.437.2865K} would reinterpret these offsets as originating from the escaping tail of ejected particles.  This model is labeled ``Kahlhoefer et al. (2014) Expulsive" in Fig. \ref{fig:cartoon_sidm}.

Importantly, the overall evolution of the merger (not just the development of offsets) has not yet been explored in any model, because the focus of theory work so far has been on characterizing offsets before the first apocenter after the collision.  \citet{2004ApJ...606..819M} do comment that drag-like forces should slow down the clusters, but do not make a quantitative prediction for how it affects merger evolution.  Furthermore, neither \citet{2004ApJ...606..819M} nor \citet{2014MNRAS.437.2865K} accounted for dynamical friction, and \citet{2008ApJ...679.1173R} only simulated the Bullet Cluster up to the observed separation between the dark-matter peaks.

\subsection{New model}

\begin{figure*}[t]
\centerline{
\includegraphics[height=2.3in]{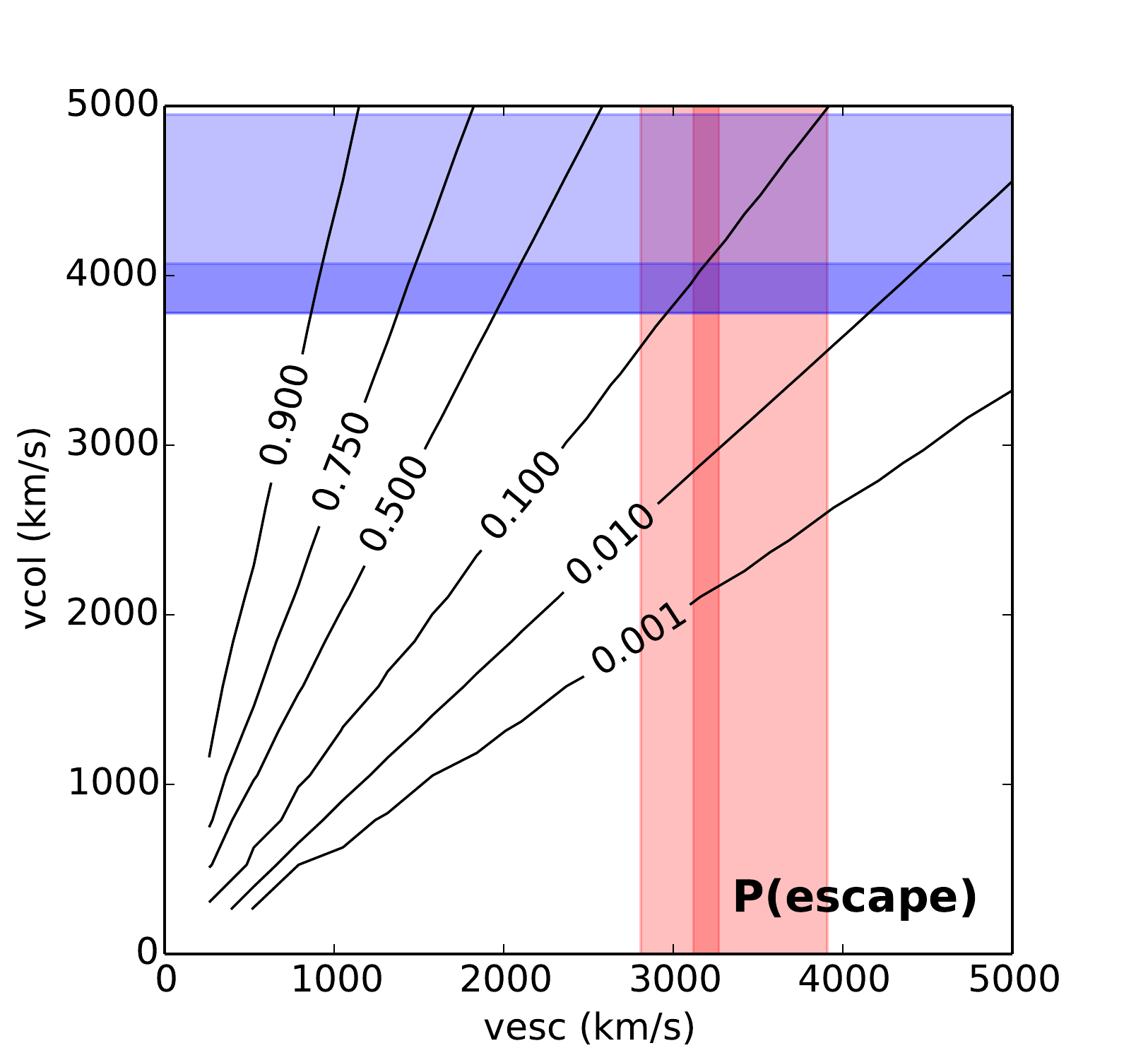}
\includegraphics[height=2.3in]{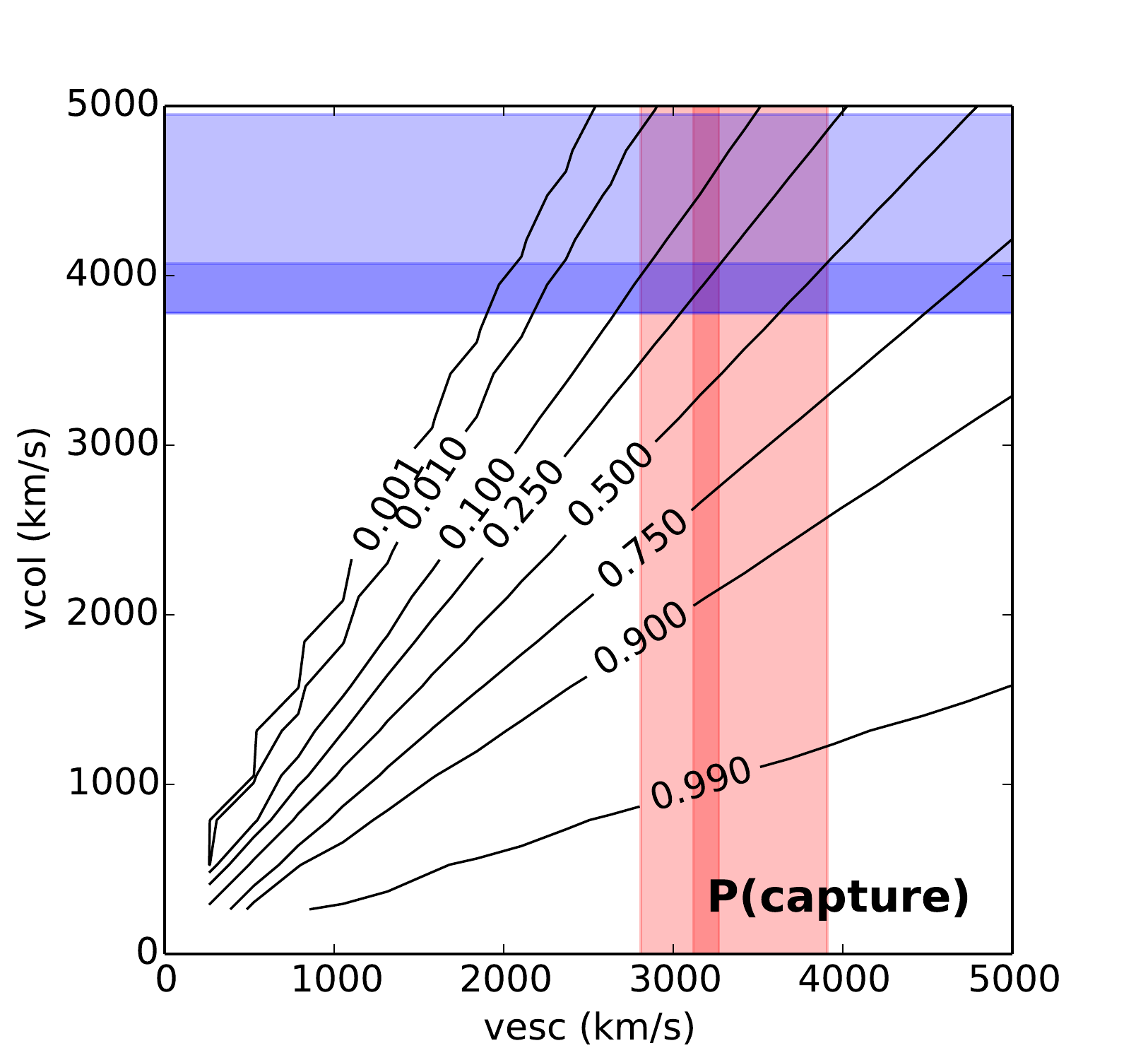}
\includegraphics[height=2.3in]{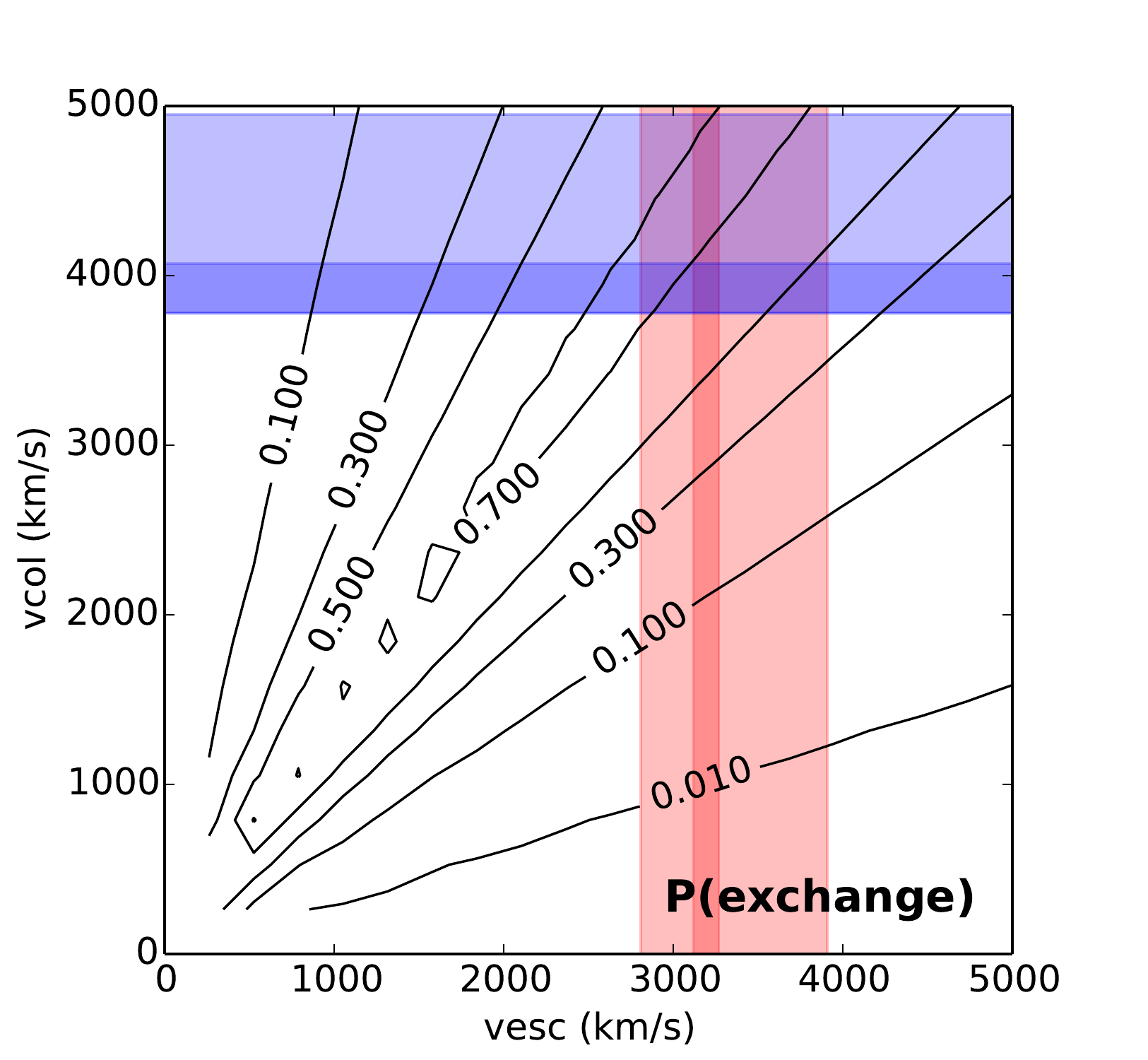}}
\caption{Probability that two particles in two halos, each with velocity dispersion $\sigma_v$, and a relative halo velocity $v_\text{col}$ at pericenter passage, would both become unbound and escape a single halo (left), that the incoming halo 2 particle would be captured (middle), or exchanges places with a halo 1 particle (right).  We set the velocity dispersion $\sigma_v \approx v_\text{esc}/4$.  The light red band highlights the escape velocities of 10$^{15}$ M$_\odot$ halos with concentrations within a factor of two given by the median mass-concentration relation of \citet{2008MNRAS.390L..64D}.  The dark red band covers the range of escape velocities of halos with the median concentration of \citet{2008MNRAS.390L..64D} but cores due to self-interaction cross sections ranging from 0 (CDM) to 10 cm$^2$/g.  The light blue band denotes the range of collision velocities in our CDM merger simulations (initialized with $v_\text{4Mpc}$ = 0 to 3000 km/s).  The dark blue band denotes the range of collision velocities in our freefall mergers with self-interaction cross sections from 0 (CDM) to 10 cm$^2$/g.}
\label{fig:poutcomes}
\end{figure*}

\begin{figure*}
\centering
\includegraphics[width=\columnwidth]{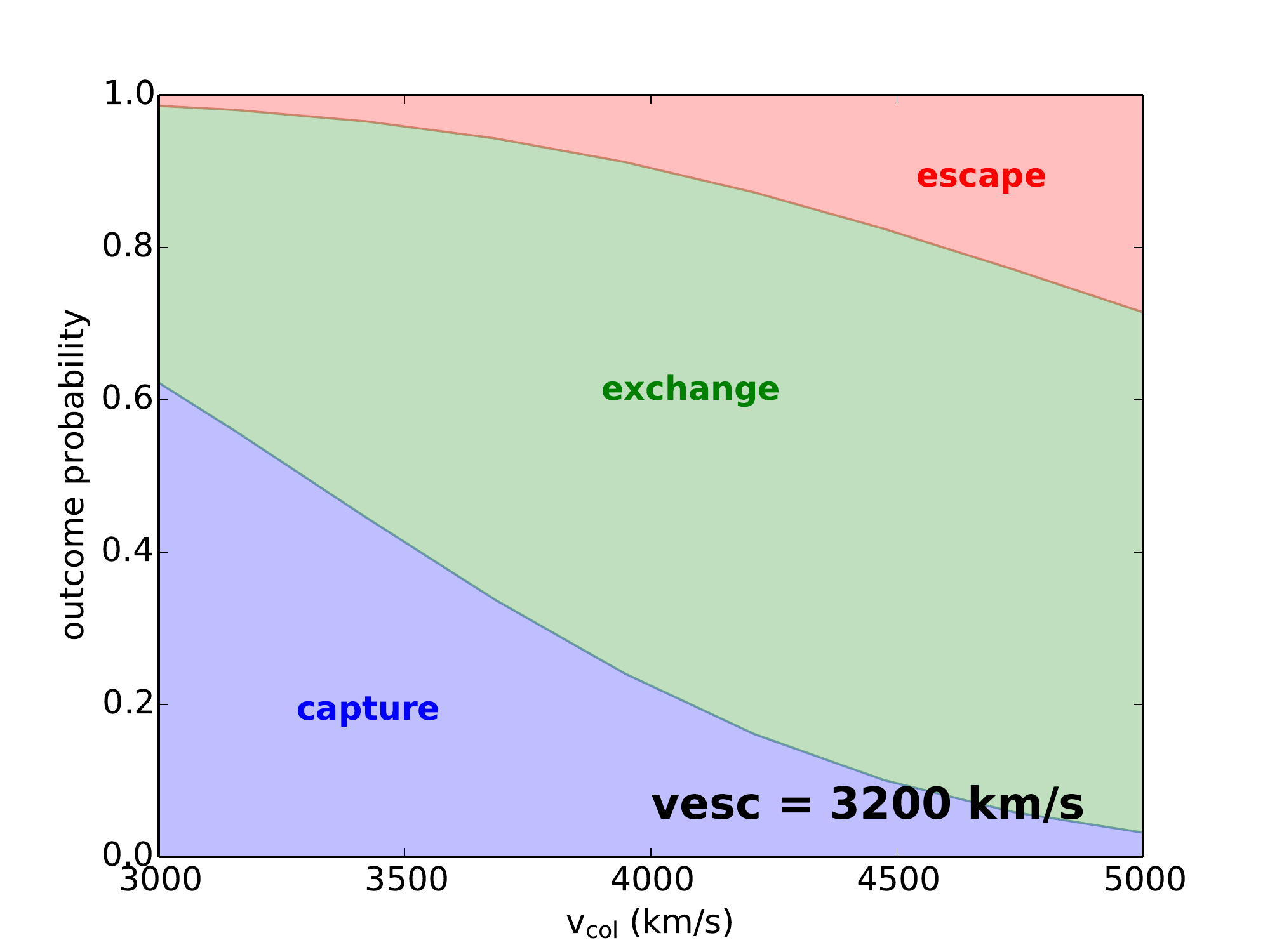}
\includegraphics[width=\columnwidth]{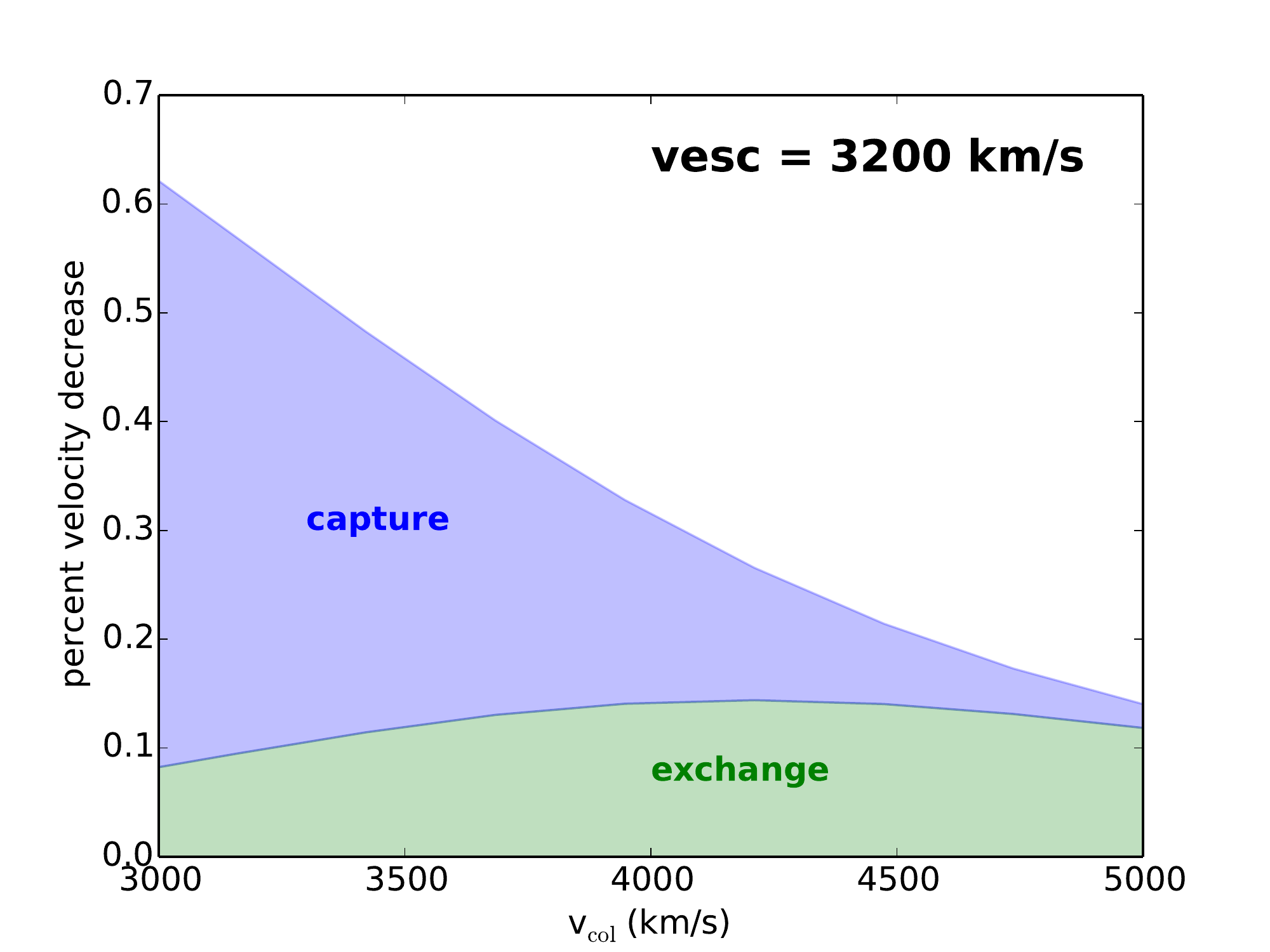}
\caption{The probability of different interaction outcomes (left) and the fractional decrease in velocity per particle, split by interaction outcome (right) as a function of collision velocity for a typical 10$^{15}$ M$_\odot$ halo with $v_\text{esc}$ = 3200 km/s.  No bulk momentum change occurs following escape interactions.}
\label{fig:pout_and_mout-1e15msun}
\end{figure*}

We revisit these assumptions in our scenario of equal-mass, cluster-sized halos, which we model with dark matter that has large-angle, velocity independent interactions.   In constructing a model for the phenomenology of offsets and merger evolution, work by \citet{2014MNRAS.437.2865K} suggests that we need to consider both the likelihood that particles are ejected from halos during an interaction between particles belonging to different halos, and the frequency of scattering between halos.

Although we only consider mergers of identical $10^{15} M_\odot$ halos here, the results can be generalized to halos of other masses (see Sec. \ref{sec:observability:offset_sizes}) and for unequal mass mergers (see Sec. \ref{sec:discussion:unEMMs}).  We show that for a cross section $\sigma_\text{SI}/m_\chi \sim 1\hbox{ cm}^2/\hbox{g}$, the phenomenology of equal-mass mergers is consistent with a drag-like picture, in contrast to the \citet{2014MNRAS.437.2865K} picture for large-angle SIDM scattering.  For larger cross sections, drag-like (Harvey et al. (2014) Drag in Fig. \ref{fig:cartoon_sidm}) and expulsive tail-like phenomenology (Kahlhoefer et al. (2014) Expulsive in Fig. \ref{fig:cartoon_sidm}) are present, complicating the modeling of offsets and merger evolution.  We show what this implies first for offsets, and then for merger evolution.

\subsubsection{Self-Interaction Kinematics}

After two dark matter particles interact, the particles' velocities are randomly scrambled (conserving linear but not angular momentum) under our assumption that interactions can be modeled as hard-sphere elastic scattering events.  If the particles are located in a dark matter halo with an escape velocity $v_\text{esc}$, three post-interaction outcomes can occur, depending on the resultant particle velocities:  both, one, or none of the interacting particles can remain bound.  Although an interaction between particles occupying the same halo will typically not unbind them, an interaction between particles from two halos colliding at high-velocity---such as occurs in a collision of galaxy clusters---can introduce significant velocity and increase the likelihood that particles become unbound.

One can compute how likely each of the three interaction outcomes are in an equal mass cluster collision with a few assumptions.  We assume that each individual halo is in dynamical equilibrium with a Maxwell-Boltzmann velocity distribution with dispersion $\sigma_v$.  Since the overwhelming majority of scatters occur in the core of each halo (where the density is the highest and is also enhanced via gravitational focusing, which pulls the pericenter of even high impact-parameter particle orbits into the cores), we use $\sigma_v$ and $v_{\mathrm{esc}}$ appropriate for cored density profiles, $\sigma_v = v_\text{esc}/4$ \citep[e.g.,][]{1994AJ....107..634T}.  If the collision of two identical halos occurs at velocity $V$ (at pericenter, when the cluster centers are coincident), we can readily calculate the probability that an interaction between two particles will result in a given pair velocities.  Given an escape velocity $v_\text{esc}$, we can derive the probability of each of the three outcomes.

The details of this calculation are outlined in Appendix \ref{apdx:poutcomes}.  In Fig. \ref{fig:poutcomes}, we show the probability of the three outcomes as a function of collision and escape velocities.  If one uses the escape velocity of an individual halo and consider interactions between its particles and particles from an incoming halo, the outcomes correspond to ``capture" (both particles bound, net acquisition of a particle), ``exchange" (one particle bound, no change in number of particles), or ``escape" (both become unbound, net loss of one particle).  To guide the eye, we have highlighted the velocity ranges that are typical for 10$^{15}$ M$_\odot$ clusters based on our simulations.  In red is the range of escape velocities measured at the centers of our clusters.  The darker red shading denotes the range in our halos with the fiducial \citet{2008MNRAS.390L..64D} concentration for 10$^{15}$ M$_\odot$ mergers, $c_\text{D08}$ = 3.3 and cross sections ranging from 1 to 10 cm$^2$/g (within this range, higher escape velocities correspond to lower cross sections).  The blue band denotes the range of collision velocities of our mergers.  The dark blue band denotes the range for our lowest velocity ($v_\text{4Mpc}$ = 0 kpc) mergers and the same range of cross sections (lower collision velocities correspond to higher cross sections).

For a typical 10$^{15}$ M$_\odot$ halo, $v_\text{esc} \approx$ 3200 km/s.  We show in Fig. \ref{fig:pout_and_mout-1e15msun} the outcome probabilities for halos with this escape velocity as a function of collision velocity.  Note that particles are overwhelmingly expected to either be captured or exchanged, even at the highest collision velocities.  Collision velocities in our cosmological simulations are $\sim$4200 km/s for which a typical two-particle interaction in our mergers will have a $\sim$11\% chance of unbinding both particles, a 20\% chance of capturing particles from the other halo, and a 69\% chance of exchange.  The velocity dependence is fairly weak: lowering (increasing) the collision velocity by 200 km/s changes the escape fraction to 9\% (15\%).

Note that in the above example, we considered whether particles were captured, exchanged, or ejected from a single halo---not from the merging system.  How do these particle kinematics affect the macroscopic kinematics of and offset formation in the merging system?  Here it is useful to treat the evolution of particles that end up bound and unbound separately.  We explore the evolution of each scenario in the following sections.

\subsubsection{Macroscopic Consequences: Drag}

The vast majority of particles undergo a capture or exchange interaction, in which one or both particles end up bound.  If an incoming particle from the opposite halo becomes bound, the halo acquires the particle's momentum, which is typically directed opposite the halo's direction of motion.  Interactions within the halo core are frequent enough to thermalize the momentum transfer, causing the halo to lose momentum.  We can thus model offsets between the galaxies and dark matter as in the frequent, small-angle regime explored by \citet{2014MNRAS.441..404H} and \citet{2014MNRAS.437.2865K, 2015MNRAS.452L..54K} even though we are modeling large-angle scatters.  Drag causes the peak of the dark matter to lag behind the galaxies', producing offsets.  It is maximal at pericenter, where the opposing current of dark matter is densest, then asymptotes to zero at apocenter.

When does drag become important?  Drag assumes the continuum approximation, which, as a rule of thumb, applies when each particle interacts with another at least once on average.  The average number of interactions per particle in the halo core during a collision is
\begin{equation}
N_\text{SI} \sim  \Gamma t  \sim n_\chi \sigma_\text{SI} v t  \sim  \frac{\sigma_\text{SI}}{m_\chi} \rho v t  \sim  \frac{\sigma_\text{SI}}{m_\chi} \Sigma.
\end{equation}
where $\Gamma$ is the rate of self-interactions, $t$ is the time during which interactions occur, $n_\chi$ and $m_\chi$ are the number density and mass of the dark matter particles, respectively, and $v$ is a characteristic interaction velocity.  Here, since most interactions will occur in the cluster core, we take $t$ to be the duration of pericenter passage, $v$ the collision velocity, and $\Sigma$ the central surface density.  For a cluster of mass 10$^{15}$ M$_\odot$, this is \citep[assuming a cored Burkert density profile appropriate for SIDM clusters from][]{2013MNRAS.430...81R}
\begin{equation}
\Sigma(r = 0) = 1 \text{ g cm}^{-2}~\left( \frac{M}{10^{15} M_\odot} \right)^{0.1}. \label{eqn:surface_density}
\end{equation}
During first pericenter passage of two 10$^{15}$ M$_\odot$ clusters, one expects each particle in the core to undergo
\begin{equation}
N_\text{SI} \approx 1 \, \text{scatter} \, \left( \frac{\sigma_\text{SI}/m_\chi}{1 \, \text{cm}^2/\text{g}} \right) \left( \frac{\Sigma}{1 \, \text{g}/\text{cm}^2} \right).
\end{equation}
Thus for a merger of 10$^{15}$ M$_\odot$ halos, the effects of drag likely become apparent when $\sigma_\text{SI}/m_\chi \gtrsim$ 1 cm$^2$/g.

How large are the offsets predicted under the drag picture?  According to the prescriptions by  \citet{2014MNRAS.441..404H} and \citet{2014MNRAS.437.2865K, 2015MNRAS.452L..54K}, if the drag force $\propto v_\text{col}^2$, offsets can be no more than
\begin{eqnarray}
	\Delta r \ll 100 \text{ kpc} \left(\frac{\sigma_\text{SI}/m_\chi}{\text{1 cm$^2$/g}}\right) \left(\frac{v_\text{col}}{4000 \text{ km/s}}\right)^{-2}.
\end{eqnarray}

\subsubsection{Macroscopic Consequences: Tails}

Particles that are unbound from their parent halo are preferentially backscattered opposite the direction the halo is moving.  This leads to the formation of a tail of particles pointing to barycenter, as noted by \citet{2014MNRAS.437.2865K}.  The tail exerts a retarding gravitational pull on both the halos and galaxies.  In an equal mass merger, the colliding halos symmetrically undergo this process.  Due to the overlap between the two resultant tails, individual tails cannot be resolved, but instead are visible as an enhancement of dark matter between the dark matter peaks.  If sufficiently many particles are backscattered, the dark matter halos can appear as a single halo for a significant amount of time after pericenter; there is a delay between the time when the galaxies and the dark matter can again be resolved into two populations.

As particles are preferentially removed from the halo cores, the halo's gravitational potential becomes shallower, encouraging further mass loss.  Unless the tails are comparable in mass to the remaining cluster, the peaks of the galaxy and dark-matter distributions will not be substantially shifted, but the centroids of the populations will.  Following the simple model for centroid offsets in Appendix D of \citet{2014MNRAS.437.2865K}, we expect maximum offsets to occur approximately a dynamical time after pericenter passage, and be of order $O(\Delta r) \approx r_\text{core} N_\text{lost}/N$ in size, where $N_\text{lost}/N$ is the fraction of core particles that escape and $r_\text{core}$ is the size of the core.  For the $10^{15} M_\odot$ halos we consider in this work, this corresponds to $10 - 400$ kpc for $1-10$ cm$^2$/g.

When do tails dominate over drag?  The fraction of a halo lost due to expulsive interactions during a cluster collision is
\begin{align}
\frac{N_\text{lost}}{N} &= 1 - \exp \left[ - N_\text{SI} P(\text{escape}) \right] \nonumber \\
&= 1 - \exp \left[ - \frac{\sigma_\text{SI}}{m_\chi} \Sigma P(\text{escape}) \right]. \label{eqn:fesc}
\end{align}
Note that this expression allows $N_{SI} > 1$, i.e. multiple scatters---a single escaping particle can eject multiple particles on its way out.  This implies that the fraction of particles that are lost from a halo increases exponentially with $\sigma_\text{SI}/m_\chi$.  If, as we have calculated above, there is a 12\% probability that a particle escapes following a self-interaction, one expects 11\%, 24\%, and 50\% of the halo to escape, assuming self-interaction cross sections $\sigma_\text{SI}/m_\chi$ = 1, 3, and 10 cm$^2$/g, respectively.  Again, note that this is escape from a single halo, not from the merger system---particles initially ejected from one halo may end up becoming acquired by the other halo.  Tails are thus likely to be negligible for cross sections of 1 cm$^2$/g, but become exponentially important as the cross section increases.  For a cross section of 10 cm$^2$/g, halos are likely to become highly disrupted upon collision.  We thus expect tails to be important for cross sections between 1 and 10 cm$^2$/g.

Eq. \ref{eqn:fesc} provides a way to predict how the relative importance of drag over tails changes for cluster collisions with different halo or merger properties.  An increase in the merger velocity, for instance, increases the escape probability, which will cause a higher fraction of the halo to be lost during collision.  As the velocity-dependence of the escape probability is relatively small, a 200 km/s change in collision velocity changes the fraction of the halo lost by only about $\sim$3\%, 6\%, and 10\% for $\sigma_\text{SI}/m_\chi$ = 1, 3, and 10 cm$^2$/g, respectively.  A similar exercise may be carried out for the halo concentration (which affects $\Sigma$ and escape probability), impact parameter ($\Sigma$), and triaxiality ($\Sigma$ and/or escape probability), among others.

\subsubsection{Merger evolution}

Missing in all treatments thus far is a description of how the overall merger evolves, beyond just the description of offsets.  We have seen that both drag and tail formation work to slow the colliding subclusters, though by different means:  drag is caused by momentum transfer between halos through the exchange of particles, whereas tails exert an additional retarding gravitational force on the clusters, just as the gas does when it coalesces into a single clump at barycenter following cluster collision \citep{2016ApJ...820...85Z}.  Therefore, we expect that self-interactions enhance the orbital decay of the colliding clusters, accelerating the merger and virialization process.  As tails and drag both become increasingly important at higher cross sections, the merger timescale should decrease.

Based on the toy model presented above, we can estimate a lower bound on the amount of momentum transfer that occurs during collision.  Appendix \ref{apdx:momentum} contains the details of these calculations.  In the right panel of Fig. \ref{fig:pout_and_mout-1e15msun}, we show the minimum fractional decrease in momentum for each of the 10$^{15}$ M$_\odot$ halos due to exchange (green) and capture (blue) interactions.  We expect an average per-particle velocity loss of at least $\Delta v(\text{capture}) \sim$ -600 km/s due to particle capture and $\Delta v(\text{exchange}) \sim$ -600 km/s due to particle exchange.  Particles that escape produce no immediate net momentum loss (e.g. due to the scrambling of particle velocities; this calculation does not account for the additional momentum loss that can be produced though subsequent tail formation).  If $N_{SI} \sim 1$, the halo momentum is reduced by $\gtrsim30\%$ at first pericenter passage, assuming a collision velocity of 4200 km/s.  Increasing (decreasing) the collision velocity by 200 km/s changes this fraction to 24\% (33\%).  Most of this decrease is due to the fact that fewer capture interactions occur at higher velocities.

We thus expect a significant fraction of the collision velocity of the merger to be lost due to self-interactions at pericenter passage, and that the fraction should increase with cross section. The magnitude of the momentum loss on the halos due to the ejected tail will also grow as a function of cross section.  For sufficiently high cross sections, the momentum loss is so high that the halo apocenters become smaller than the cluster core size, causing the halos to coalesce upon impact (see Sec. \ref{sec:offsets:sigmavel}).

\subsection{Predictions}

Since a relatively small number of SIDM-induced collisions between cluster particles lead to expulsion unless the cross section is large, we expect self-interactions to produce both drag and tails in an equal mass merger.  For $\sigma_\text{SI}/m_\chi \sim$ 1 cm$^2$/g, we expect that self-interactions should primarily induce an extra drag-type force that causes offsets between the peaks of the galaxy and dark-matter distributions.  These offsets are largest at pericenter and smallest at apocenter.  We expect that the halos lose at least 30\% of their momentum at first pericenter passage because of direct momentum exchange between halos (either through particle capture or exchange).

At higher cross sections, we expect that drag still operates, but exponentially more expulsive interactions will occur.  The exponential increase introduces a sharp cutoff of $\sim$10 cm$^2$/g above which $\gtrsim$ 50\% of the particles in each halo undergo an expulsive interaction, causing the halos to become highly disrupted and coalesce upon contact.  Between 1 to 10 cm$^2$/g, ejected particles can form dark matter tails behind each cluster, which affects both the formation of offsets and the overall evolution of the merger.  The tails exert a gravitational pull on the clusters, pulling them closer to barycenter, and hence accelerate the coalescence process.

In Sec. \ref{sec:offsets}, we use simulations to check our predictions for the general phenomenology of equal-mass mergers, and to quantify both the sizes of galaxy-dark matter offsets and the shortening of merger timescales.  We also seek to find whether there are other aspects of the phenomenology of SIDM equal-mass mergers that may lead to tighter cross section constraints.


\section{METHODS}
\label{sec:methods}

In Section \ref{sec:theory}, we discussed how self-interactions may drastically impact the dynamical evolution and undermine the validity of popular methods to constrain the self-interaction cross section in equal mass mergers.  To explore these predictions more fully, we run $N$-body dark matter-only merger simulations with self-interaction physics.  The details and validation of our modified version of GADGET-2 used to run our simulations can be found in Appendix \ref{apdx:sidm_validation}.  We treat galaxies as massive tracer particles, and do not include gas, and leave hydrodynamical simulations to future work.  In Section \ref{subsec:methods_ics} we present our choice of initial conditions, detailing how each cluster is generated and the merger parameters we adopted.  In Section \ref{subsec:methods_offsets}, we describe how we quantify the dark matter-galaxy offset.


\begin{table*}
\caption{Initial Conditions} \label{tab:ics}
\begin{tabular}{ l  c  c  c  c  c  c  c  c  c  c }\hline
\multirow{2}{*}{Simulation Name} & \multirow{2}{*}{$\frac{c_\text{DM}}{c_\text{D08}}$} & \multirow{2}{*}{$N_\text{gal}$} & \multirow{2}{*}{BCG mass} & $b$ & $v_\text{4Mpc}$ & $v_\text{infall} $ & $v_\text{col}$ & $\sigma_\text{SI} / m_\chi$ \\
& & & & (kpc) & (km/s) & (km/s) & (km/s) & (cm$^2$/g) \\ \hline \hline

\multicolumn{9}{>{\columncolor[gray]{0.8}} l }{\sc cross section} \\
df15mr1cD0000v-1cg\_pbcg\_1e5gal\_cdm & D & 57,000 & 7 $\times$ 10$^{10}$ M$_\odot$ & 0 & 0 & 1740 & 4070 & 0 \\
df15mr1cD0000v-1cg\_pbcg\_1e5gal\_1si & D & 57,000 & 7 $\times$ 10$^{10}$ M$_\odot$ & 0 & 0 & 1730 & 3980 & {\bf1} \\
df15mr1cD0000v-1cg\_pbcg\_1e5gal\_3si & D & 57,000 & 7 $\times$ 10$^{10}$ M$_\odot$ & 0 & 0 & 1720 & 3900 & {\bf3} \\
df15mr1cD0000v-1cg\_pbcg\_1e5gal\_10si & D & 57,000 & 7 $\times$ 10$^{10}$ M$_\odot$ & 0 & 0 & 1720 & 3790 & {\bf10} \\ \hline

\multicolumn{9}{>{\columncolor[gray]{0.8}} l }{\sc non-zero initial velocity} \\
df15mr1cD1000v-1cg\_pbcg\_1e5gal\_cdm & D & 57,000 & 7 $\times$ 10$^{10}$ M$_\odot$ & 0 & {\bf1000} & 2000 & 4200 & 0 \\
df15mr1cD1000v-1cg\_pbcg\_1e5gal\_1si & D & 57,000 & 7 $\times$ 10$^{10}$ M$_\odot$ & 0 & {\bf1000} & 1990 & 4100 & {\bf1} \\
df15mr1cD1000v-1cg\_pbcg\_1e5gal\_3si & D & 57,000 & 7 $\times$ 10$^{10}$ M$_\odot$ & 0 & {\bf1000} & 2000 & 4030 & {\bf3} \\
df15mr1cD1000v-1cg\_pbcg\_1e5gal\_10si & D & 57,000 & 7 $\times$ 10$^{10}$ M$_\odot$ & 0 & {\bf1000} & 1990 & 3910 & {\bf10} \\ \hline

df15mr1cD2000v-1cg\_pbcg\_1e5gal\_cdm & D & 57,000 & 7 $\times$ 10$^{10}$ M$_\odot$ & 0 & {\bf2000} & 2660 & 4410 & 0 \\
df15mr1cD2000v-1cg\_pbcg\_1e5gal\_1si & D & 57,000 & 7 $\times$ 10$^{10}$ M$_\odot$ & 0 & {\bf2000} & 2640 & 4370 & {\bf1} \\
df15mr1cD2000v-1cg\_pbcg\_1e5gal\_3si & D & 57,000 & 7 $\times$ 10$^{10}$ M$_\odot$ & 0 & {\bf2000} & 2650 & 4350 & {\bf3} \\
df15mr1cD2000v-1cg\_pbcg\_1e5gal\_10si & D & 57,000 & 7 $\times$ 10$^{10}$ M$_\odot$ & 0 & {\bf2000} & 2650 & 4270 & {\bf10} \\ \hline

df15mr1cD3000v-1cg\_pbcg\_1e5gal\_cdm & D & 57,000 & 7 $\times$ 10$^{10}$ M$_\odot$ & 0 & {\bf3000} & 3470 & 4950 & 0 \\
df15mr1cD3000v-1cg\_pbcg\_1e5gal\_1si & D & 57,000 & 7 $\times$ 10$^{10}$ M$_\odot$ & 0 & {\bf3000} & 3480 & 4920 & {\bf1} \\
df15mr1cD3000v-1cg\_pbcg\_1e5gal\_3si & D & 57,000 & 7 $\times$ 10$^{10}$ M$_\odot$ & 0 & {\bf3000} & 3480 & 4890 & {\bf3} \\
df15mr1cD3000v-1cg\_pbcg\_1e5gal\_10si & D & 57,000 & 7 $\times$ 10$^{10}$ M$_\odot$ & 0 & {\bf3000} & 3470 & 4830 & {\bf10} \\ \hline

\multicolumn{9}{>{\columncolor[gray]{0.8}} l }{\sc DM concentration} \\
df15mr1c0.5D1000v-1cg\_pbcg\_1e5gal\_1si & {\bf0.5D} & 57,000 & 7 $\times$ 10$^{10}$ M$_\odot$ & 0 & {\bf1000} & 1960 & 3490 & {\bf1} \\
df15mr1c2D1000v-1cg\_pbcg\_1e5gal\_1si & {\bf2D} & 57,000 & 7 $\times$ 10$^{10}$ M$_\odot$ & 0 & {\bf1000} & 2040 & 4940 & {\bf1} \\ \hline

df15mr1c0.5D2000v-1cg\_pbcg\_1e5gal\_10si & {\bf0.5D} & 57,000 & 7 $\times$ 10$^{10}$ M$_\odot$ & 0 & {\bf2000} & 2610 & 3780 & {\bf10} \\
df15mr1c2D2000v-1cg\_pbcg\_1e5gal\_10si & {\bf2D} & 57,000 & 7 $\times$ 10$^{10}$ M$_\odot$ & 0 & {\bf2000} & 2680 & 4960 & {\bf10} \\ \hline

\multicolumn{9}{>{\columncolor[gray]{0.8}} l }{\sc impact parameter + $v_\text{4Mpc}$ = 2000 km/s} \\
df15mr1cD1000v-1cg\_pbcg\_b500kpc\_cdm & D & 5700 & 7 $\times$ 10$^{10}$ M$_\odot$ & {\bf500} & 1000 & 2000 & 4020 & 0 \\
df15mr1cD1000v-1cg\_pbcg\_1e5gal\_b500kpc\_1si & D & 57,000 & 7 $\times$ 10$^{10}$ M$_\odot$ & {\bf500} & 1000 & 2000 & 3970 & {\bf1} \\
df15mr1cD1000v-1cg\_pbcg\_1e5gal\_b500kpc\_10si & D & 57,000 & 7 $\times$ 10$^{10}$ M$_\odot$ & {\bf500} & 1000 & 2010 & 3870 & {\bf10} \\ \hline

df15mr1cD1000v-1cg\_pbcg\_b1000kpc\_cdm & D & 5700 & 7 $\times$ 10$^{10}$ M$_\odot$ & {\bf1000} & 1000 & 2030 & 3700 & 0 \\
df15mr1cD1000v-1cg\_pbcg\_1e5gal\_b1000kpc\_1si & D & 57,000 & 7 $\times$ 10$^{10}$ M$_\odot$ & {\bf1000} & 1000 & 2030 & 3700 & {\bf1} \\
df15mr1cD1000v-1cg\_pbcg\_1e5gal\_b1000kpc\_10si & D & 57,000 & 7 $\times$ 10$^{10}$ M$_\odot$ & {\bf1000} & 1000 & 2030 & 3660 & {\bf10} \\ \hline

\multicolumn{9}{>{\columncolor[gray]{0.8}} l }{\sc impact parameter + $v_\text{4Mpc}$ = 2000 km/s} \\
df15mr1cD2000v-1cg\_pbcg\_1e5gal\_b500kpc\_cdm & D & 57,000 & 7 $\times$ 10$^{10}$ M$_\odot$ & {\bf500} & 2000 & 2660 & 4100 & 0 \\
df15mr1cD2000v-1cg\_pbcg\_1e5gal\_b500kpc\_1si & D & 57,000 & 7 $\times$ 10$^{10}$ M$_\odot$ & {\bf500} & 2000 & 2660 & 4100 & {\bf1} \\
df15mr1cD2000v-1cg\_pbcg\_1e5gal\_b500kpc\_10si & D & 57,000 & 7 $\times$ 10$^{10}$ M$_\odot$ & {\bf500} & 2000 & 2670 & 4040 & {\bf10} \\ \hline

df15mr1cD2000v-1cg\_pbcg\_1e5gal\_b1000kpc\_cdm & D & 57,000 & 7 $\times$ 10$^{10}$ M$_\odot$ & {\bf1000} & 2000 & 2680 & 3680 & 0 \\
df15mr1cD2000v-1cg\_pbcg\_1e5gal\_b1000kpc\_1si & D & 57,000 & 7 $\times$ 10$^{10}$ M$_\odot$ & {\bf1000} & 2000 & 2670 & 3740 & {\bf1} \\
df15mr1cD2000v-1cg\_pbcg\_1e5gal\_b1000kpc\_10si & D & 57,000 & 7 $\times$ 10$^{10}$ M$_\odot$ & {\bf1000} & 2000 & 2670 & 3680 & {\bf10} \\ \hline

\multicolumn{9}{>{\columncolor[gray]{0.8}} l }{\sc big BCG} \\
df15mr1cD1000v-1cg\_pbcg12\_1e5gal\_10si & D & 57,000 & {\bf10$^{12}$ M$_\odot$} & 0 & {\bf1000} & 2000 & 4100 & {\bf10} \\
df15mr1cD2000v-1cg\_pbcg12\_1e5gal\_10si & D & 57,000 & {\bf10$^{12}$ M$_\odot$} & 0 & {\bf2000} & 2660 & 4200 & {\bf10} \\ \hline \hline

\multicolumn{9}{l}{ \footnotesize Notes: Boldface font highlights key parameters that differ from the fiducal CDM simulation (shown on the first line)} \\
\multicolumn{9}{l}{ \footnotesize $c_\text{DM}$ and $c_\text{gal}$ denote the concentration of the dark matter halos and the galaxy distributions, respectively.} \\
\multicolumn{9}{l}{ \footnotesize $c_\text{D08}$ denotes the concentration prescribed by \citet{2008MNRAS.390L..64D} for 10$^{15}$ M$_\odot$ halos.} \\
\multicolumn{9}{l}{ \footnotesize $N_\text{gal}$ is the number of galaxies for each halo (thus total number of galaxies in a merger = 2 $N_\text{gal}$). } \\
\multicolumn{9}{l}{ \footnotesize $b$ is the impact parameter. } \\
\multicolumn{9}{l}{ \footnotesize $v_\text{4Mpc}$, $v_\text{infall}$, and $v_\text{col}$ denote the initial, infall, and collision velocities of the mergers. } \\
\multicolumn{9}{l}{ \footnotesize $\sigma_\text{SI}/m_\chi$ denotes the self-interaction cross section for the merger.}

\end{tabular}
\end{table*}


\subsection{Initial Conditions}
\label{subsec:methods_ics}

We modeled clusters as containing two components: galaxies and dark matter.  The clusters were generated using \textsc{HALOGEN4MUSE} \citep{2008MNRAS.386.1543Z}, modified to generate two-component spherical systems.  The dark matter halo is well fit at large radii by a NFW profile \citep{2013ApJ...765...25N, 2013ApJ...765...24N, 2016ApJ...821..116U, 2016ApJ...819...36D}
\begin{equation}
\rho(r) = \dfrac{\rho_0}{\dfrac{r}{r_s} \left( 1 + \dfrac{r}{r_s} \right)^2}
\end{equation}
where $M$ is the virial mass and $r_s$ the scale radius of the system.  We define the virial radius $r_\text{vir}$ such that the average overdensity within $r_\text{vir}$ is equal to 200 $\rho_\text{c}$, the critical density of the Universe: $M_\text{vir} = (4/3) \pi r_\text{vir}^3 (200 \rho_c)$.  The concentration $c = r_s/r_\text{vir}$ describes how centrally concentrated the distribution is.  For our fiducial simulations, we adopted a dark matter mass of 10$^{15}$ M$_\odot$ ($r_\text{vir}$ = 2.1 Mpc).  There is large scatter in the dark matter halo concentrations $c_\text{DM}$ for a given mass; we adopted median values given by the mass-halo concentration relation derived by \citet{2008MNRAS.390L..64D} for our fiducial models, which prescribes $c_\text{D08}$ = 3.3 for a 10$^{15}$ M$_\odot$ halo (and thus $r_s$ = 630 kpc).  To explore the concentration-dependence of our results, we also modeled halos with half or twice this fiducial concentration.  Each halo was represented by $6 \times 10^{6}$ dark matter particles, each with a mass of $2 \times 10^{8}$ M$_\odot$.

In galaxy clusters, the galaxy mass distribution is also well fit by an NFW profile.  We adopted a total galaxy mass of 2\% the dark matter mass \citep{2012ARA&A..50..353K} and the same concentration for both the galaxies and the dark matter \citep{2012MNRAS.423..104B}.  The galaxy distribution was modeled with $6 \times 10^4$ galaxy particles---greater by at least a factor of 10 in observed cluster mergers, but chosen to reduce noise in galaxy peak identification---each with a mass of $7 \times 10^{9}$ M$_\odot$.  As galaxies make a subdominant contribution to mass of and thus the dynamical evolution of mergers, we modeled them as collisionless ``tracer" particles.

Lastly, we added a galaxy particle ten times more massive than the other galaxy particles (e.g. $7 \times 10^{10}$ M$_\odot$) at the center of system in analogy to the brightest cluster galaxies (BCGs) observed in clusters.  BCGs can be as massive as $10^{12}$ M$_\odot$; we thus re-ran mergers with $\sigma_\text{SI}/m_\chi$ = 10 cm$^2$/g with a $10^{12}$ M$_\odot$ mass BCG particle to test how a massive BCG affects the density and potential evolution of the dark matter halo.

We evolved these single, isolated clusters for 3 Gyr to check that the CDM halos were in equilibrium and to allow cores to form in SIDM halos.  Mergers were generated by placing identical halos so that their virial radii were just touching, i.e. their halo centers were $\Delta r_x$ = 4 Mpc apart.  We explored a range of merger parameters.  We varied initial relative velocities ranging from $v_\text{4Mpc}$ = 0 to 3000 km/s and impact parameters $b$ = 0 to 1 Mpc (mergers with $b \ne 0$ were given offsets $\Delta r_y$ perpendicular to the merger axis; the distance between their halo centers were thus greater than 4 Mpc).  Initial velocities $v_\text{4Mpc}$ = 0 and 1000 km/s represent the most cosmologically realistic mergers (extending the results of \citet{2015MNRAS.448.1674J} to 10$^{15}$ M${_\odot}$ gives infall velocities---the halos' relative velocity when separated by $r_\text{vir}$ = 2100 kpc---of about 2000 km/s and an impact parameter of $b$ = 500 kpc; see also \citet{2010ApJ...718...60L} and \citet{2001ApJ...561..621R}).  We simulate velocities beyond this range in order to explore the dependence of SIDM merger outcomes on kinematic merger parameters.

We have chosen to simulate CDM and three self-interaction cross sections $\sigma_\text{SI}/m_\chi$ = 1, 3, and 10 cm$^2$/g.  The lower bound on our cross section range corresponds to the typical scale of SIDM cross section constraints in the literature \citep[e.g.,][]{2008ApJ...679.1173R, 2013MNRAS.430..105P}, and the regime where drag-type effects likely become significant.  Although some constraints are of order 0.1 cm$^2$/g \citep[e.g.][]{2015Sci...347.1462H, 2016PhRvL.116d1302K}, we have not chosen to model lower cross sections as offsets produced in such mergers are below our resolution limit.  The upper bound is chosen to bracket the cross section constraints derived from simple order of magnitude estimates from gas-DM offsets in the vein of \citet{2004ApJ...606..819M}, \citep[e.g.][]{2008ApJ...687..959B, 2012ApJ...747L..42D}, and corresponds to the cutoff at which we expect halos to be highly disrupted and coalesce upon contact (see Sec. \ref{sec:theory}).

Initial conditions are summarized in Table \ref{tab:ics}.  In addition to the initial relative velocity ($v_\text{4Mpc}$), we have provided the infall velocity $v_\text{infall}$ and the collision velocity $v_\text{col}$ (relative velocity when $\Delta r_x$ = 0) leading up to first pericenter passage.  The velocities were computed based on the dynamical evolution of the tracer BCG, which traces the evolution of the merging clusters well through first pericenter passage in all our mergers.

Our idealized, isolated mergers were evolved with a modified version of the tree-based $N$-body code \textsc{gadget}-2 that can account for SIDM physics \citep{2013MNRAS.430...81R}.  We have implemented a modification to this code to speed up the SIDM interaction calculations via adaptive SIDM smoothing lengths.  The details and validation of these modifications are in Appendix \ref{apdx:sidm_validation}.  All simulations were run on the Oakley and Ruby supercomputers at the \citet{OhioSupercomputerCenter1987}.


\subsection{Peak Finding and Offsets}
\label{subsec:methods_offsets}

Quantifying the relative displacements between a collection of extended distributions does not have a well-defined solution \citep[c.f.][Ng et al., in prep]{2016arXiv160504307R}.  Various methods have been applied to merging clusters in the literature.  In particular, different techniques are used in theoretical and observational work, and little effort has been made to translate results between the two.  In keeping with past work, we focus on identifying the location of the peaks or centroids of our distributions, and comparing peak locations to calculate offsets.  To make connections with past theoretical work \citep[e.g.][]{2014MNRAS.437.2865K} and observational work \citep[e.g.][]{2015ApJ...802...46J}, we compute the offsets based on 1D distributions (projected along the merger axis), which behaves much like a centroid-finder, and on projected 2D distributions.  We also compare the 1D and 2D results with the location of the ``tracer BCG" particle we placed at pre-merger halo centers.  The details of each of these three methods are described below.

We calculate offsets between the galaxy and dark matter distributions by comparing galaxy and dark matter peaks identified with the same method (e.g. galaxy peaks via 1D vs. dark matter peaks via 1D).  In the BCG analysis, the BCG particle's location is compared against 2D dark matter peak analyses.

We validate our peak finding methods in Appendix \ref{apdx:peak_validation}.

\subsubsection{1D Analysis}

In order to make connections with previous theoretical work \citep[such as][]{2014MNRAS.437.2865K}, we searched for peaks in 1D projections of the galaxy and dark matter mass distributions along the merger axis.  Each component (dark matter and galaxies) was analyzed separately.  For the dark matter, particle positions were binned along the merger axis, and the bin with the highest number of particles was identified as the peak.  The more sparsely sampled galaxy distributions, on the other hand, were smoothed to reduce noise via a kernel density estimate (KDE) with a gaussian kernel.  Peaks in the smoothed galaxy distributions were obtained with an optimization routine.  An initial guess at the peak location was chosen by binning the galaxy distributions.

Choosing the appropriate bin width (dark matter) and smoothing scale (galaxies) requires balancing the desire to obtain smooth distributions (obtained via large smoothing scales) and to resolve small offsets and prevent oversmoothing (small smoothing scales).  Oversmoothing is particularly a concern for the bimodal particle distributions of our merging systems; peaks separated by less than the smoothing scale may appear as a single peak.  We found that a 2 kpc bin width was sufficient for the dark matter, while a smoothing scale of 50 kpc was required for the galaxies.

To derive statistics on peak positions, we carried out a bootstrap analysis with 1000 bootstrapped particle distributions for both the galaxies and dark matter distributions.  We repeated this analysis for each individual halo in the merger.  Halo membership was assigned based on the halo each particle belonged to at the beginning of the simulation (and thus does not account for the exchange of particles between halos or expulsion during the merger).

\subsubsection{2D Analysis}

Mergers are observed in 2D projections in the plane of the sky.  Thus to make connections to observations, we also analyzed 2D projections of the galaxy and dark matter mass distributions.  Since one of our goals is to quantify the \emph{maximum} offset between the dark matter and galaxies, we will analyze our simulations as if the merger takes place entirely in the plane of the sky.  Any other orientation of the merger with respect to the plane of the sky will result in smaller measured offsets.

As the number of resolution elements increases by a factor of $N^2$ in going from 1D to 2D, the noisiness of the discretely sampled distribution and the difficulty in identifying a peak is exacerbated in 2D.  We thus smooth both the dark matter and galaxy distributions, again using a KDE estimate with a 2D gaussian smoothing kernel with a width of 100 kpc, and perform the peak search as in 1D.  No single-peak searches were performed, as a two-peak search does well.  To obtain uncertainties, we bootstrap the galaxy distribution 1000 times as before, but bootstrap the well-sampled dark matter distribution 10 times (which produces uncertainties ranging from 10-20 kpc).

We repeated the 2D analysis for each individual halo in the merger as in 1D, but with a wider 150 kpc smoothing width to account for the noisier distribution that result from halving the number of particles.

\subsubsection{The BCG Particle}

A tracer particle (with negligible mass---a factor of 10 greater than a galaxy particle) was placed at the center of each halo, mimicking the bright, massive galaxies (BCGs) often found near the centers of galaxy clusters \citep{2014ApJ...797...82L}.  The location of this tracer particle was used as another measure of the centers of the galaxy distributions.


\section{SIMULATED OFFSETS AND MERGER EVOLUTION}
\label{sec:offsets}

\begin{figure*}
\centering
\includegraphics[width=0.8\textwidth]{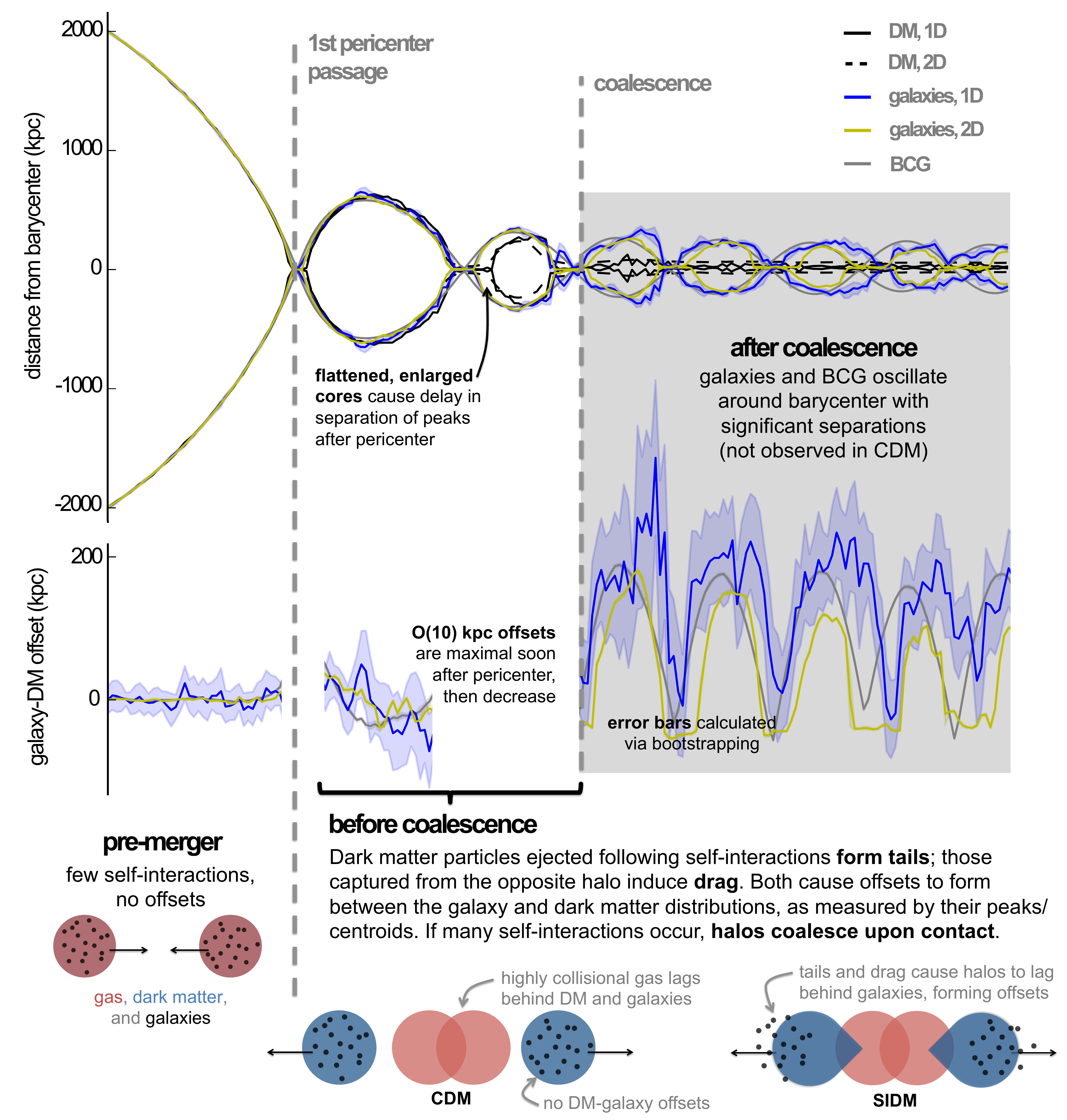} 
\caption{A prototypical equal mass SIDM merger ($\sigma_\text{SI}/m_\chi$ = 3 cm$^2$/g, $v_\text{4Mpc}$ = 1000 km/s).  In the top row, the distance of each halo from barycenter is plotted over time (the horizontal axis spans 10 Gyr), as measured by the techniques discussed in Section \ref{subsec:methods_offsets}.  Offsets between the dark matter and galaxy distributions are shown in the bottom panel.} 
\label{fig:prototype}
\end{figure*}

\begin{figure*}
\centerline{\includegraphics[width=2.5\columnwidth,trim=0cm 0.7cm 0cm 0cm,clip=true]{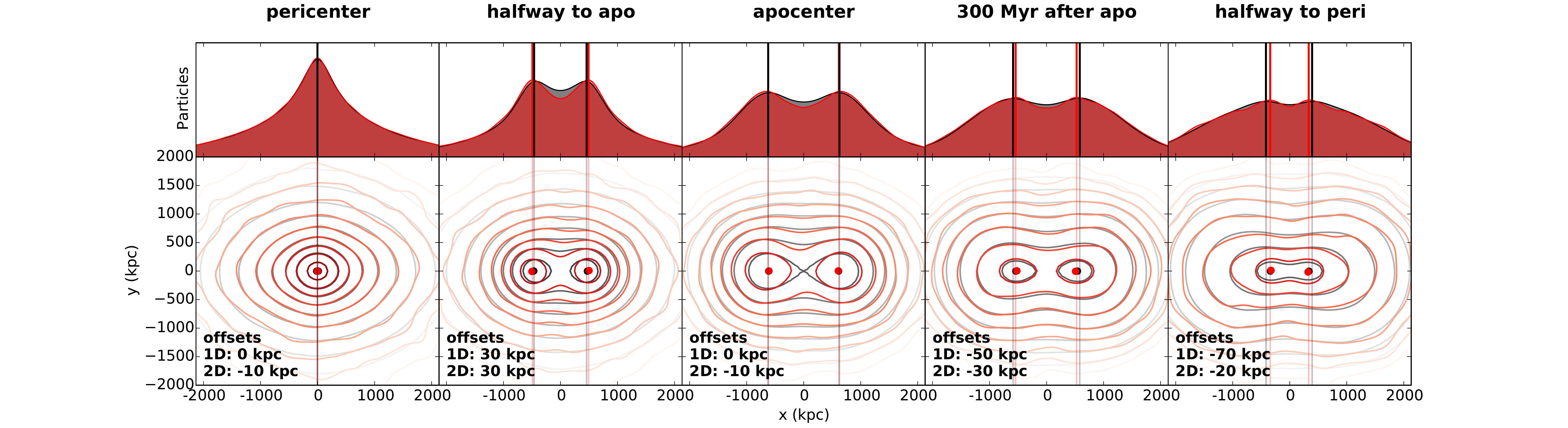}}
\centerline{\includegraphics[width=2.5\columnwidth,trim=0cm 0cm 0cm 1.5cm,clip=true]{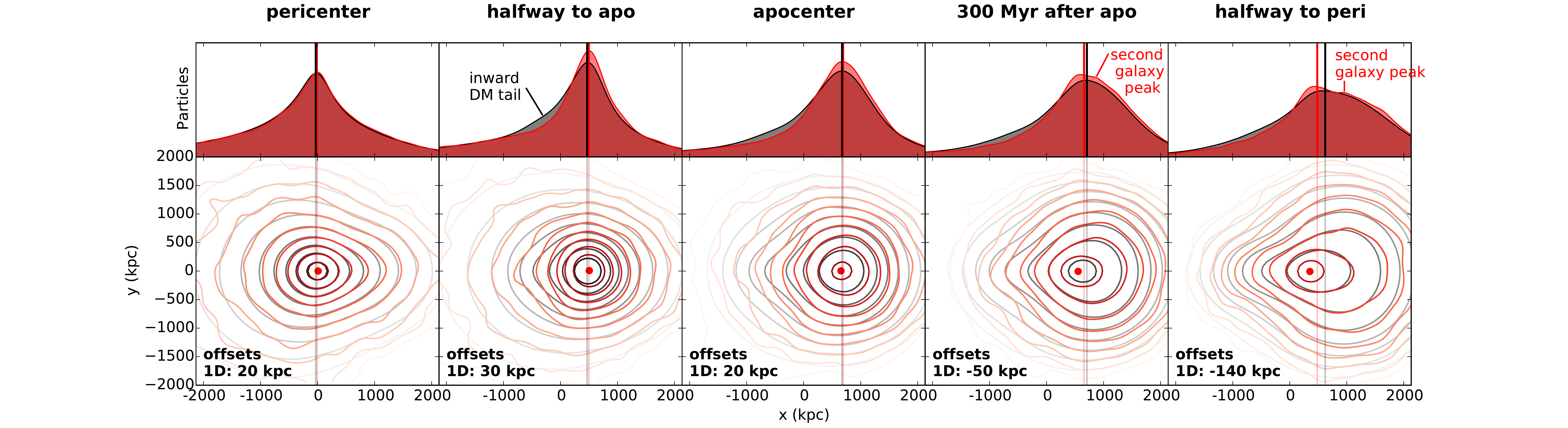}}
\caption{Evolution of dark matter and galaxies in the merger shown in Figure \ref{fig:prototype} ($v_\text{4Mpc}$ = 1000 km/s, $\sigma_\text{SI}/m_\chi$ = 3 cm$^2$/g).  The top row shows the full 2-halo dark matter and galaxy distributions, while the bottom row shows the distributions of galaxies and dark matter that were in the left halo at the beginning of the simulation.  In each row, 1D (top) and 2D (bottom) distributions of the dark matter (black) and galaxy (red) distributions are shown.  Contours in the 2D plots are logarithmically spaced; the same contour levels are used across all snapshots.  1D peaks are denoted by the vertical lines (both top and bottom rows), while 2D peaks are denoted by the points in the bottom row.} 
\label{fig:tails-1000v_3si}
\end{figure*}

In this section, we use the simulations described in Sec. \ref{sec:methods} to test the qualitative predictions of and quantitatively calibrate the equal-mass merger theory model we presented in Sec. \ref{sec:theory}.  We quantify both the overall evolution of mergers as well as the development of offsets.  We first describe the general features of an SIDM merger relative to a CDM merger in \S\ref{sec:offsets:anatomy}, and introduce nomenclature that we will use for the remainder of this paper.  We show how the orbital evolution and the magnitude of offsets depend on cross section and velocity for our fiducial values of the dark matter halo concentration and impact parameter (head-on mergers) in \S\ref{sec:offsets:sigmavel}.  In \S\ref{sec:offsets:concentration} and \S\ref{sec:offsets:impact}, we examine the effects of the dark matter halo concentration and merger impact parameter, respectively.  We leave a comparison of our simulated mergers to observations, and implied SIDM cross section constraints, to \S5 \& \S6.

\subsection{The Anatomy of an SIDM Merger}
\label{sec:offsets:anatomy}

In Figs. \ref{fig:prototype} and \ref{fig:tails-1000v_3si}, we show an equal mass merger from our suite of simulations that exhibits much of the unique phenomenology seen in SIDM mergers, one run with a self-interaction cross section of $\sigma_\text{SI}/m_\chi$ = 3 cm$^2$/g and an initial velocity $v_\text{4Mpc}$ = 1000 km/s.  As we describe below, for this choice of cross section and initial velocity, we clearly see that both drag and tail phenomena play major roles in the merger.  In Fig. \ref{fig:prototype}, peak locations of the dark matter and galaxy distributions, as measured from barycenter, are plotted on top.  The offset between the two is shown on the bottom (a positive offset denotes that the galaxies are further from barycenter than the dark matter).  We have not shown offsets when the dark matter peaks are separated by less than 600 kpc, about the halo scale radius, as overlap with the other halo biases peak locations towards barycenter (see Appendix \ref{apdx:peak_validation}).  We note that due to our choice of smoothing parameters, we can resolve offsets down to about 5 kpc (see Appendix \ref{apdx:peak_validation}).  Bootstrapped errors are denoted via the colored bands.  The main source of uncertainty (and the origin of the jitter) is due to the sparsely sampled galaxy distribution.  We note again that our simulations contain $\sim1 \times 10^5$ galaxies, which is far more than the several 10s up to a few 100 galaxies observed for most clusters.  Determining the peak of such sparsely sampled distributions---even with smoothing---is inherently noisy, and points to the need for statistical analyses of cluster mergers in order to derive more stringent SIDM constraints.

A merger can be roughly broken down into three phases based on the evolution of the dark matter halos, as illustrated by Figure \ref{fig:prototype}:  (1) a pre-merger phase, followed by (2) an intermediate pre-coalescence phase during which the dark matter halos follow oscillatory, decaying orbits until coalescence, and (3) the subsequent post-coalescence stage.  In a typical CDM merger, such as shown in Figure \ref{fig:peak_check}, the dark matter and galaxy distributions evolve identically throughout on simple oscillatory, decaying orbits until the separation between the halos becomes indistinguishable.

An SIDM merger initially evolves similarly.  The dark matter and galaxy distributions evolve identically as in CDM until first pericenter passage, when the collision of the dense halo cores give rise to many dark matter self-interactions.  Just before pericenter passage, a drag-induced offset develops in each subcluster but is impossible to observe in a real system because of the extensive overlap between the two halos.  The offsets continue to grow immediately after pericenter in part due to SIDM-induced drag, which causes the dark matter halos to lag behind galaxies.  This is evident in Fig. \ref{fig:tails-1000v_3si}, where the dark matter and galaxy distributions, normalized by the total mass, are shown in 1D (top) and 2D (bottom) projections:  the 1D galaxy peaks and 2D contours (contour levels are identical between the dark matter and galaxies) are further from barycenter than the dark matter's.  After the halos pass through each other, the drag force sharply decreases, restoring the pre-collision force equilibrium under which the dark matter-galaxy offsets are zero.  The drag-induced offsets are thus largest soon after pericenter and shrink as the system approaches apocenter.

Tails of unbound dark matter also become apparent following pericenter.  In both 1D and 2D projections, they appear as accentuated (relative to the galaxies) ``bridges" between peaks.  These tails form in both the galaxies and dark matter as a population of both are naturally unbound by the rapidly changing gravitational potential.  However, the dark matter tail is larger due to an additional population of ejected particles backscattered by self-interactions.  The preferential enhancement of the dark matter tail can pull the dark matter peaks closer to barycenter, especially for the 1D analysis, causing the dark matter-galaxy offset to be greater than it would have been due to momentum loss alone.  This effect is small for the present merger, but becomes clear in the difference between the size of the offsets in the 1- vs. 2-halo analyses for high cross section mergers (e.g. Fig. \ref{fig:tails-2000v_10si}).

Approaching apocenter, a second set of extended features forms.  In contrast to the backscattered tails discussed above that point to barycenter, outbound shells form on the opposite side of the halos.  Like the tails, the shells contain dark matter and galaxies shocked into unbound or loosely bound orbits by the changing gravitational potential.  They also contain galaxies and dark matter that have been minimally affected by SIDM---dark matter that have not undergone a self-interaction, and of course, the collisionless galaxies.  As the dark matter has been backscattered by self-interactions whereas the galaxies have not, the shells are stronger for the galaxies.  We can see this in 1D (see Fig. \ref{fig:tails-1000v_3si}, third column).  A shell of galaxy splits off and produces a bimodal galaxy distribution in the 1-halo analysis.  The 1D peaks from the shell and tail are initially of comparable size.  In the 2-halo analysis, the bimodal structure is difficult to detect; the unbound population primarily manifests as a brief outward jump of the 1D galaxy peak before the tail disperses.  The outbound shells are not bound to the dark matter halos lagging behind them and escape the system.  The 2D analysis is not strongly affected by the shell.

We note that the gravitationally unbound portion of the galaxy and dark matter tails and shells are enhanced relative to CDM.  It is easier to unbind particles from SIDM mergers; SIDM halos have cores, which have shallower potentials than CDM cusps, and their potentials become even more shallow as self-interactions transfer dark matter from halo centers to the backscattered tails.  These processes lower the energy required to become unbound in SIDM systems, promoting greater mass loss.

Apocenter is abbreviated by the inward tails of dark matter---if one could measure the dark-matter peaks for each halo individually, apocenter would occur later and the halo-halo separation would be larger.  Furthermore, galaxies in the outskirts are preferentially unbound.  The remaining galaxies lead the dark matter to barycenter and maintain negative offsets to the subsequent pericenter.

Coalescence occurs for the dark matter halos before the galaxies due to the enhanced momentum loss induced by self-interactions during pericenter passages.  The resultant halo preserves the central core of its progenitor halos.  The galaxies, on the other hand, continue to oscillate on \emph{non-decaying orbits} due to the lack of dynamical friction in the core \citep{1998MNRAS.297..517H, 2003ApJ...588L..21K, 2006MNRAS.373.1451R}.  Galaxy peaks can oscillate around barycenter with orbits of 100s of kpc---a behavior unseen in our CDM simulations.  Observing offset galaxy distributions or BCGs in unrelaxed clusters may produce strong constraints on the dark matter self-interaction cross section, a prospect that will be further explored in Section \ref{sec:betterconstraints}.

In summary, the largest offsets appear soon after pericenter, are small (at best, a few tens of kpc), and decrease, disappearing by apocenter.  While the merger discussed here illustrates the distinguishing features of SIDM mergers, this picture can change significantly given different merger and halo parameters.  As discussed in Section \ref{sec:theory}, the key factors that may lead to substantially different merger phenomenology are the number of self-interactions, and what fraction of those lead to dark-matter expulsion from each halo.  Thus one can expect the presence of offsets to be a function of a variety of merger parameters, including the self-interaction cross section, halo concentration, and impact parameter (all of which affect the number of self-interactions) and the merger velocity (which affects the expulsivity).  In the rest of this section, we explore how changes in these parameters impact the evolution of SIDM mergers and the formation of dark matter-galaxy offsets.


\subsection{Dependence on Initial Velocity}
\label{sec:offsets:sigmavel}

\begin{figure*}
\centering
\includegraphics[width=\columnwidth]{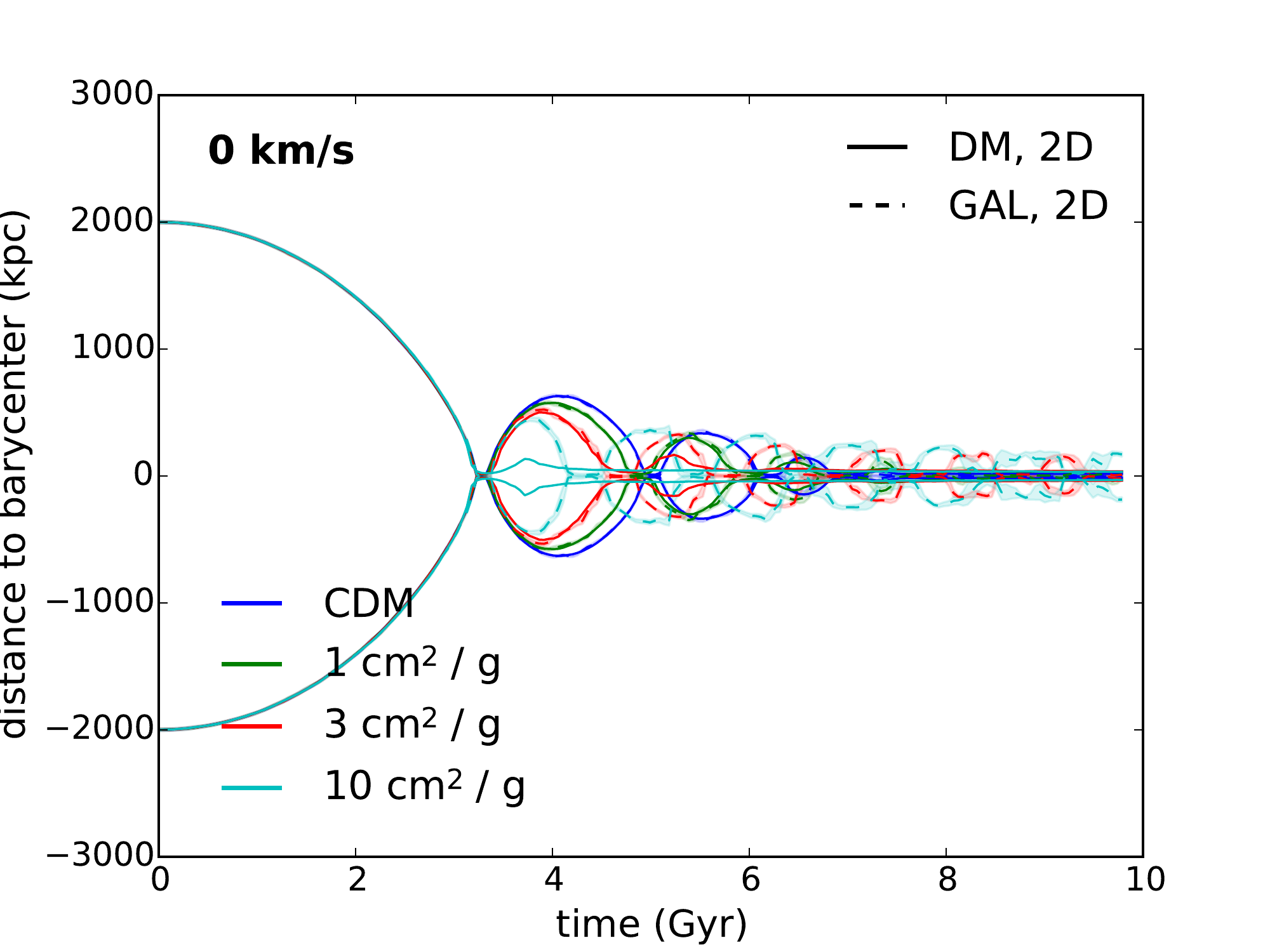}
\includegraphics[width=\columnwidth]{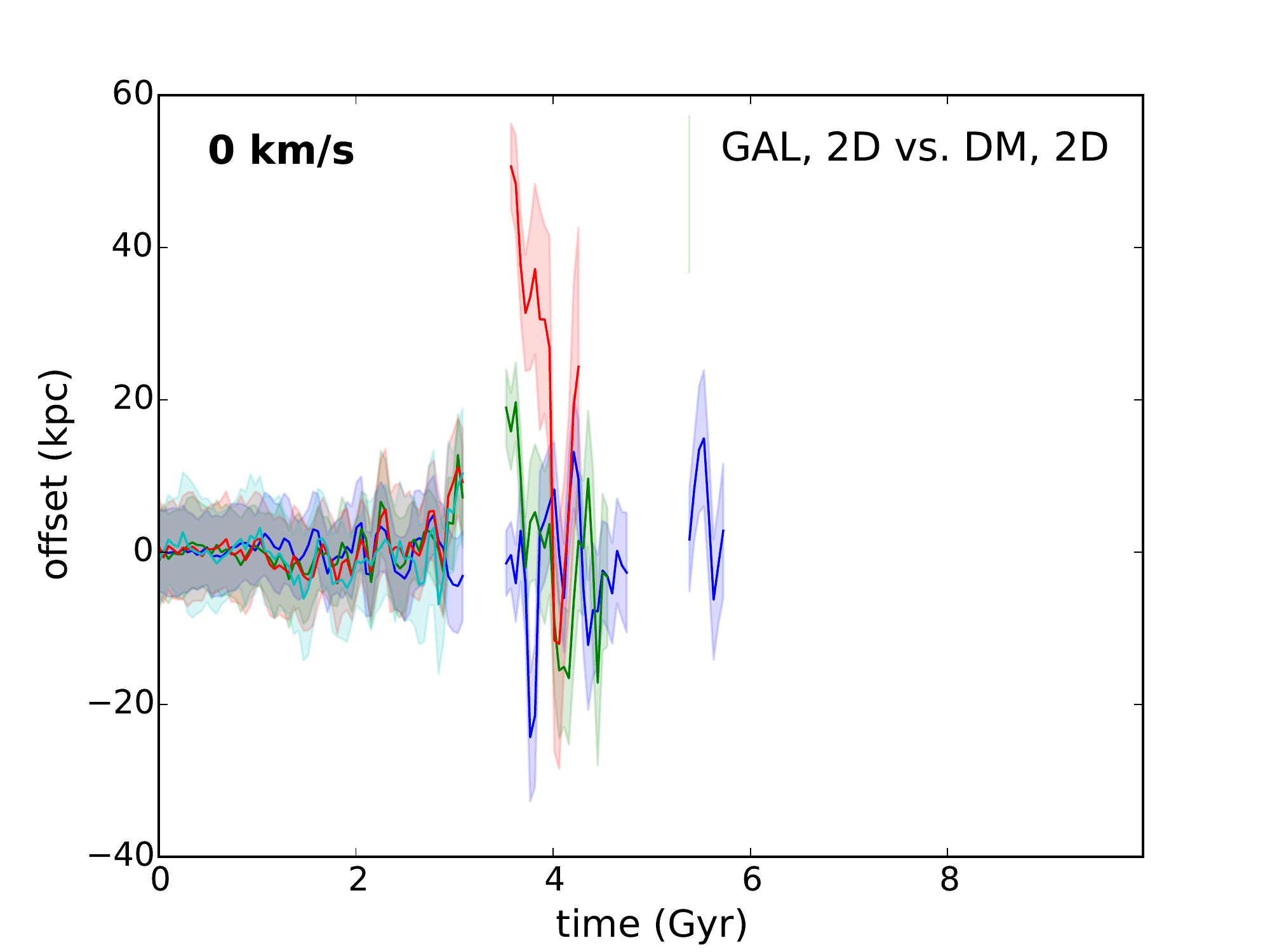} \\
\includegraphics[width=\columnwidth]{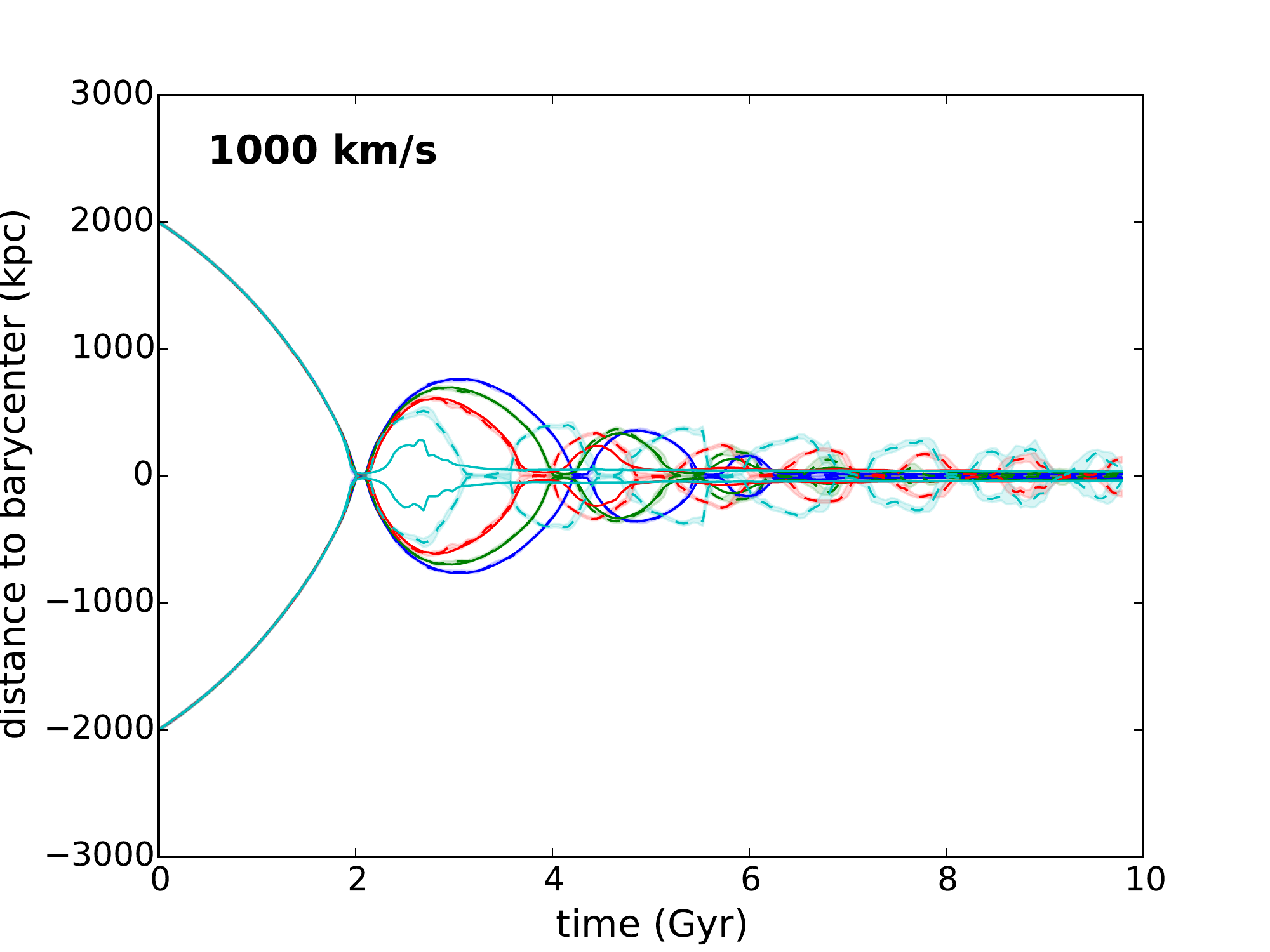}
\includegraphics[width=\columnwidth]{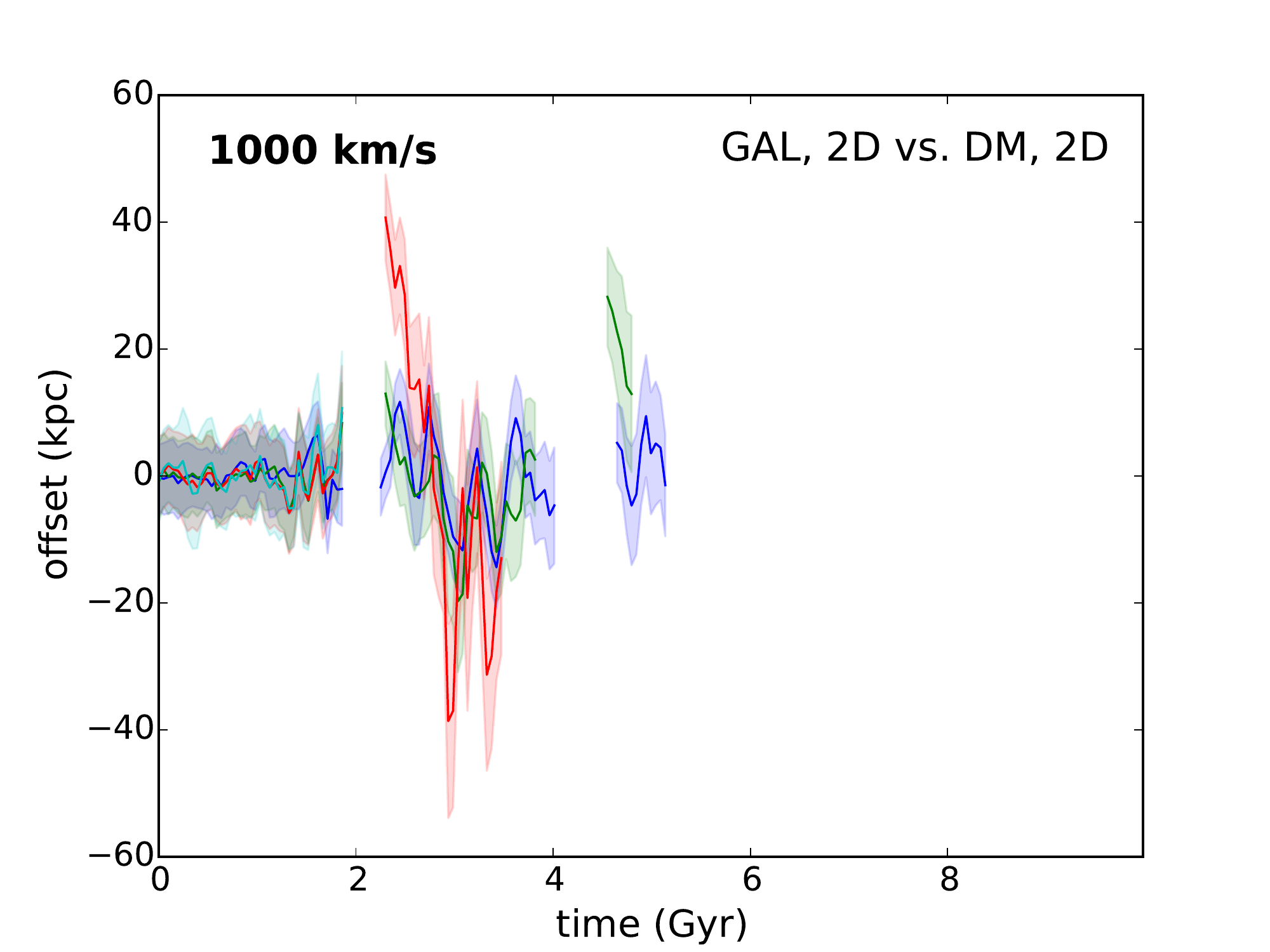} \\
\includegraphics[width=\columnwidth]{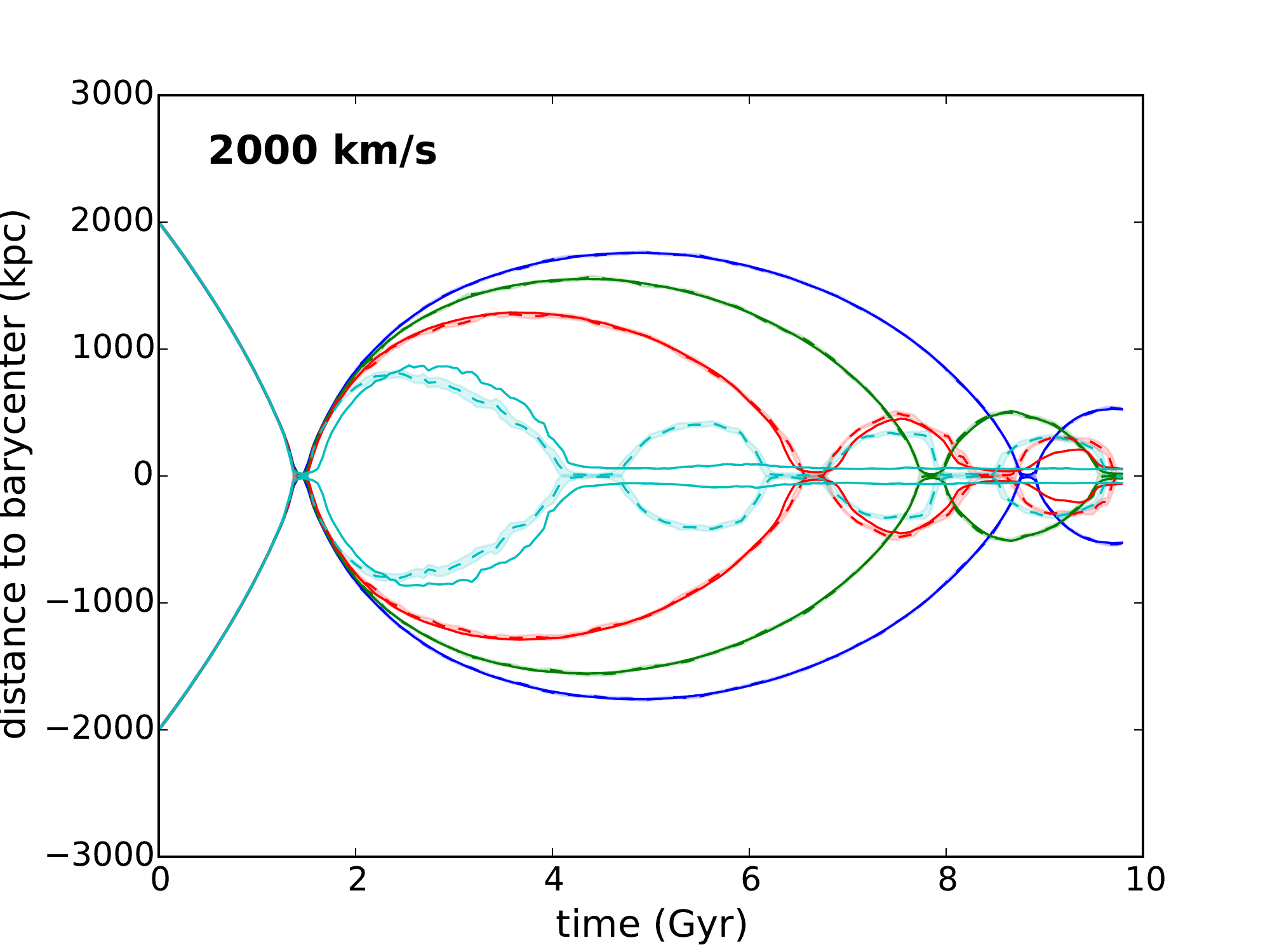}
\includegraphics[width=\columnwidth]{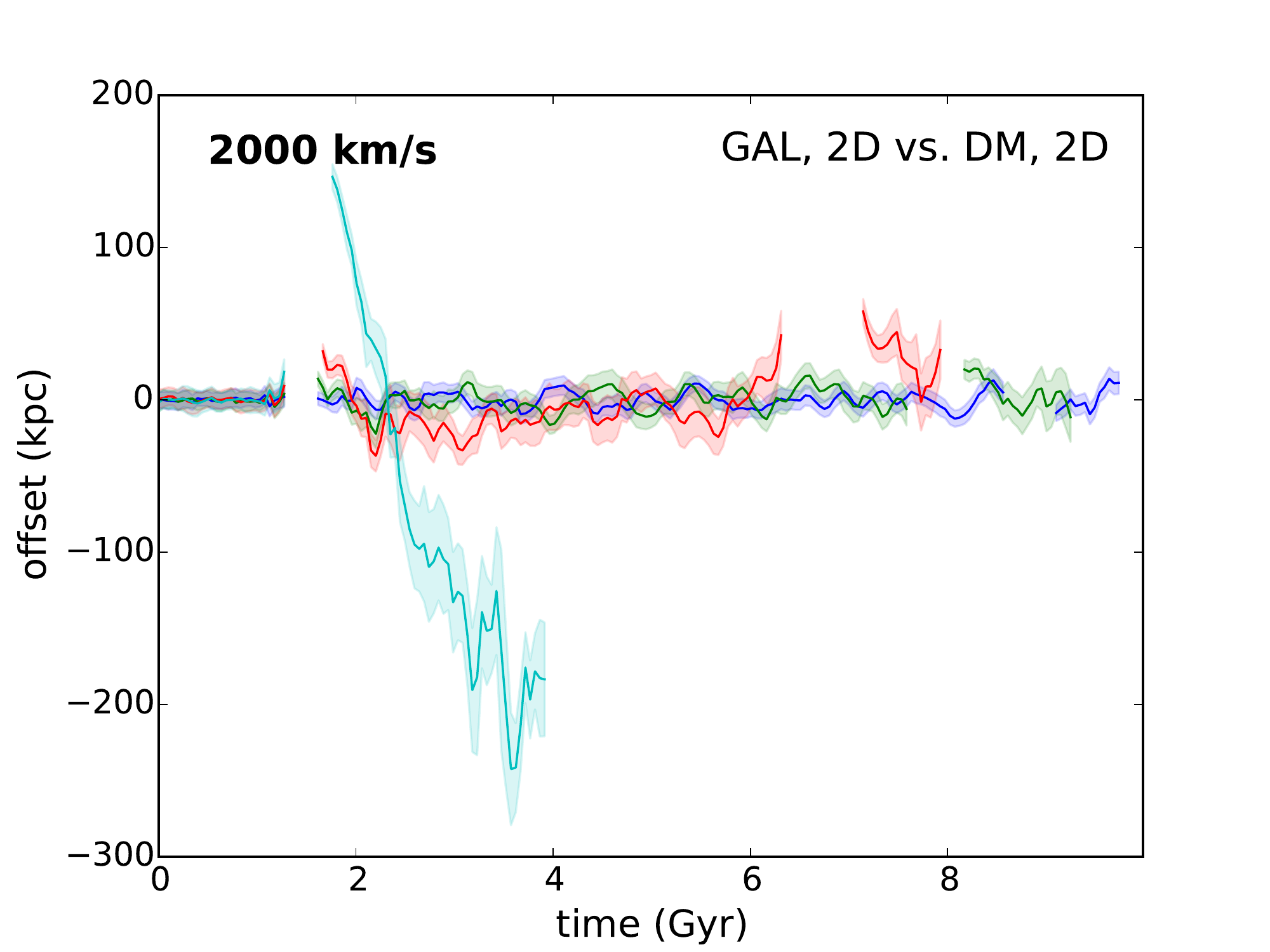} 
\caption{The dynamical evolution of (left) dark matter peaks and the galaxies and (right) dark matter-galaxy offsets for self-interaction cross sections varying from 0 (CDM) to 10 cm$^2$/g.  Rows represent different merger velocities.  From top to bottom, $v_\text{4 Mpc}$ = 0, 1000, and 2000 km/s.  Self-interactions can dramatically shorten the lifetime of the merger, speeding virialization, even with cross-sections as low as 1 cm$^2$/g.  The shortening of the merger timescale depends primarily on the self-interaction cross section and merger kinematics.  For high enough cross sections and low merger velocities, halos merge upon contact.}
\label{fig:rc-v0_dmm_2D}
\end{figure*}

\begin{figure*}
\centering
\includegraphics[width=\columnwidth]{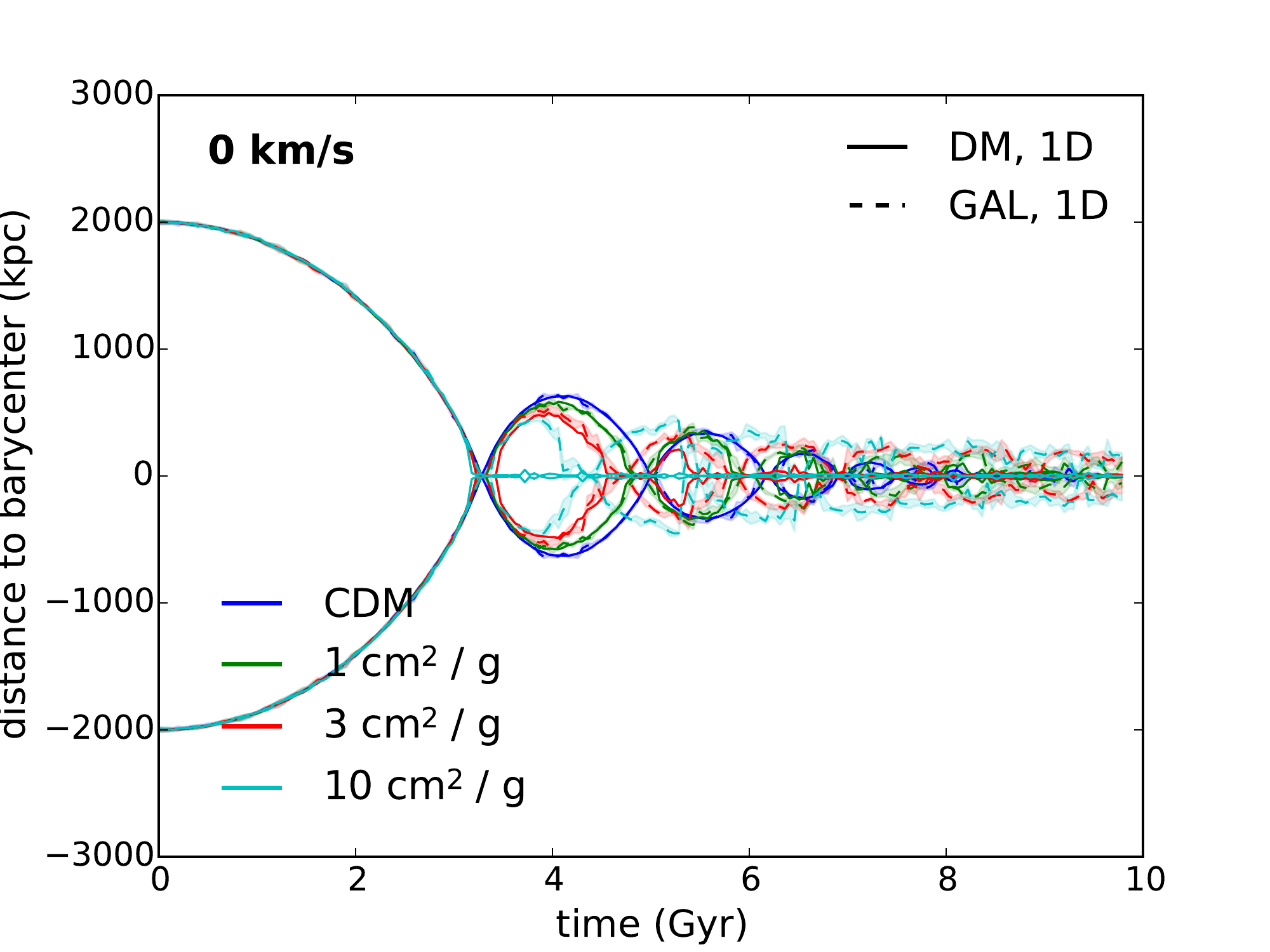}
\includegraphics[width=\columnwidth]{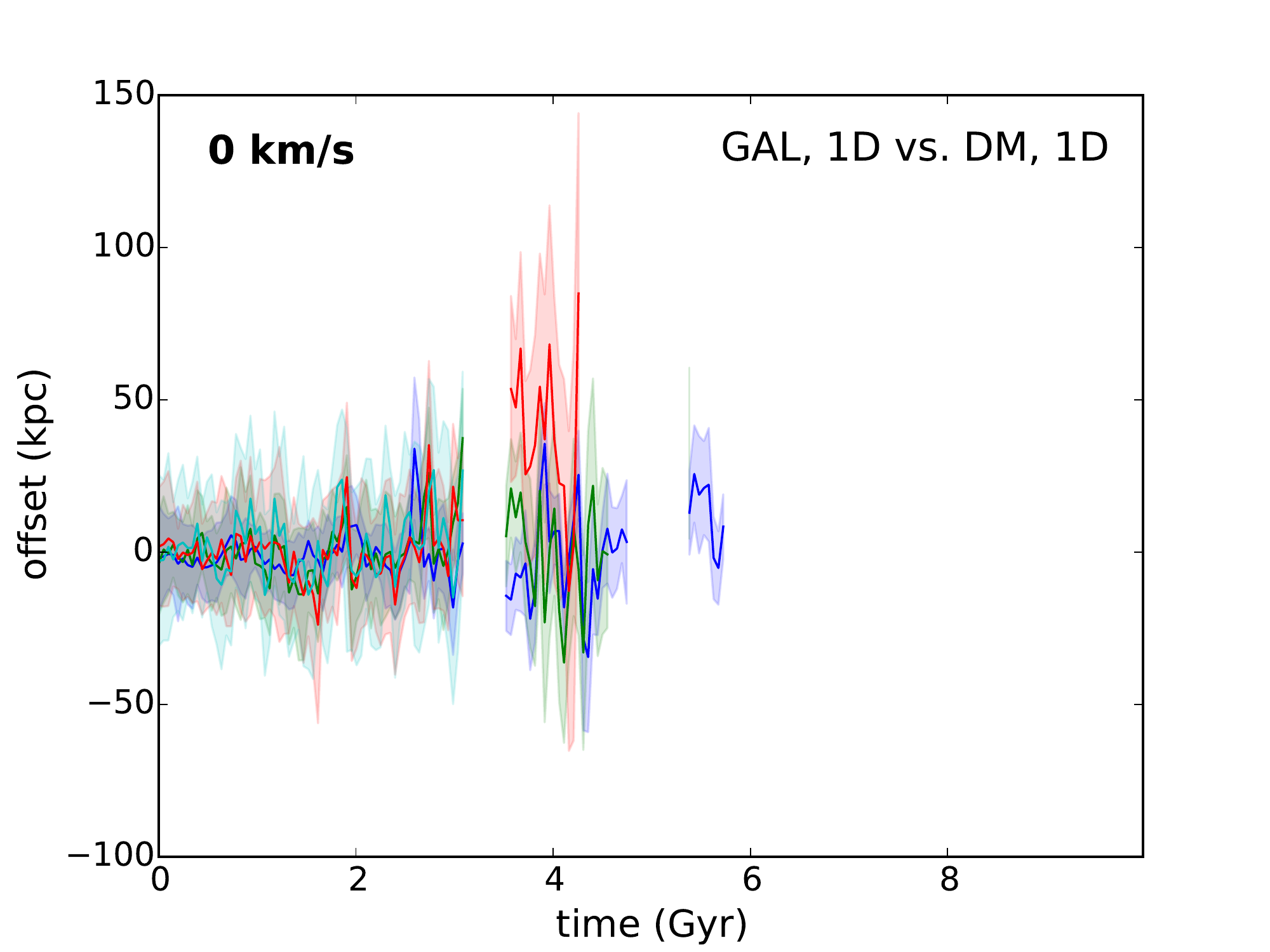} \\
\includegraphics[width=\columnwidth]{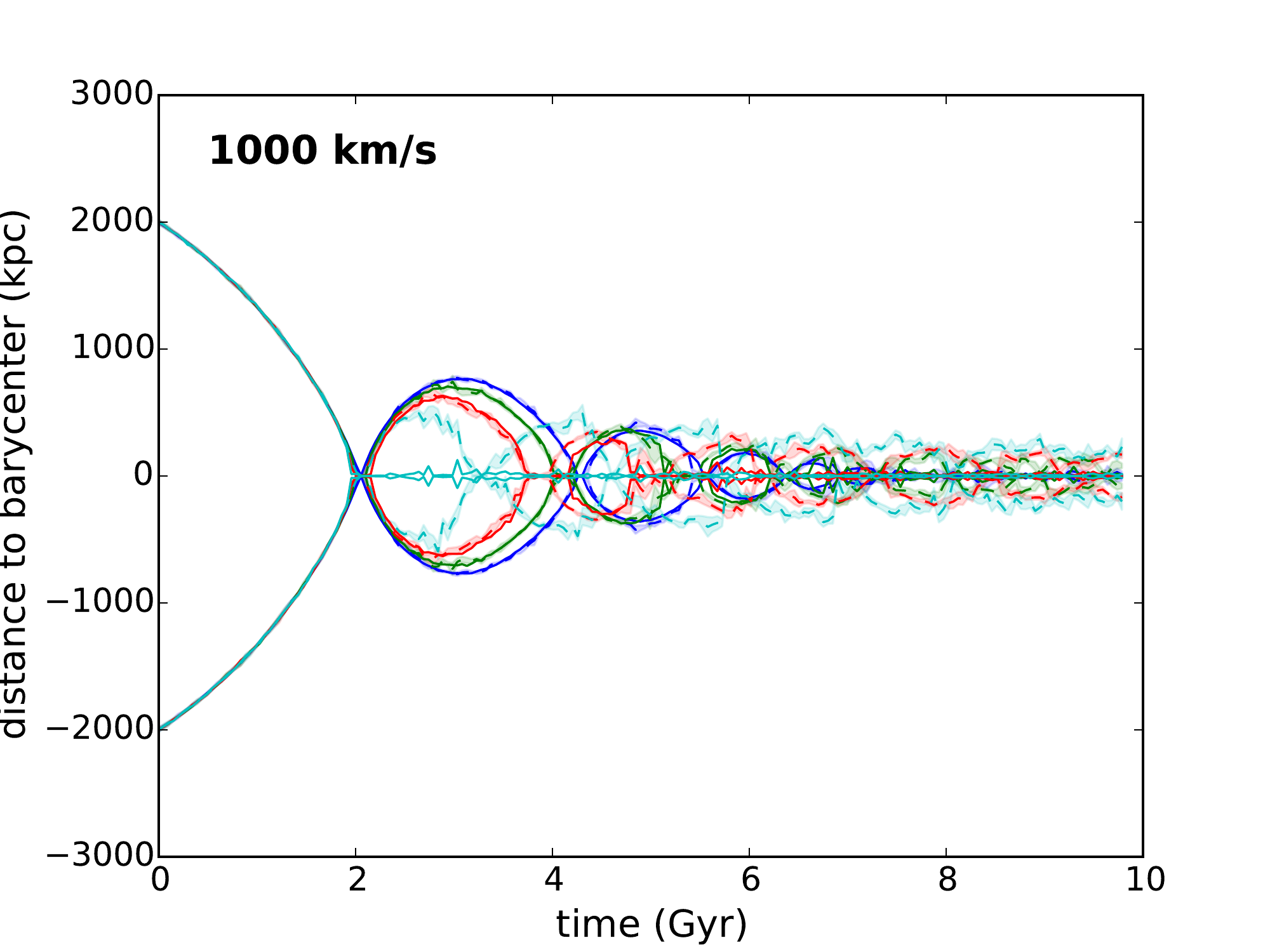}
\includegraphics[width=\columnwidth]{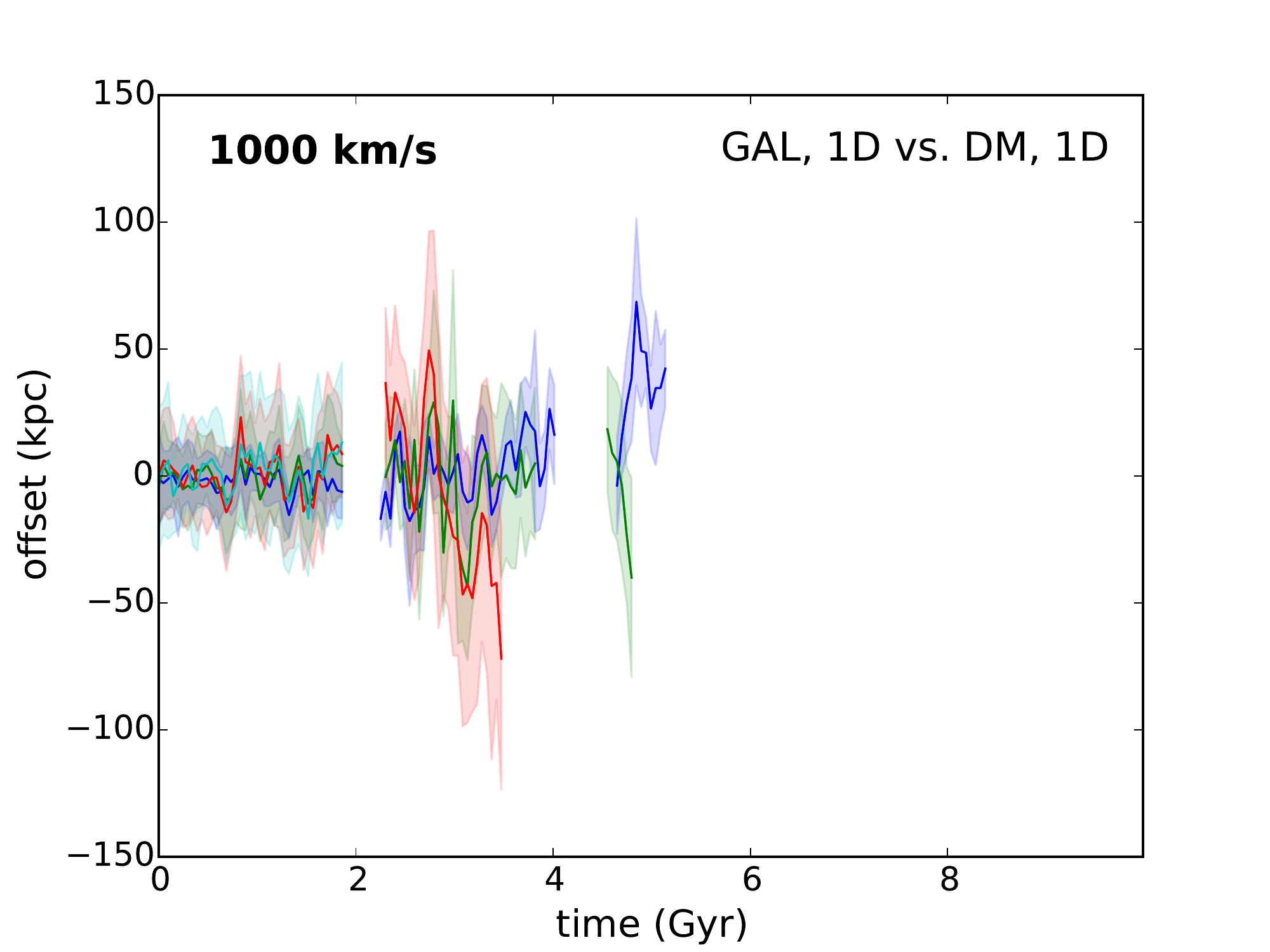} \\
\includegraphics[width=\columnwidth]{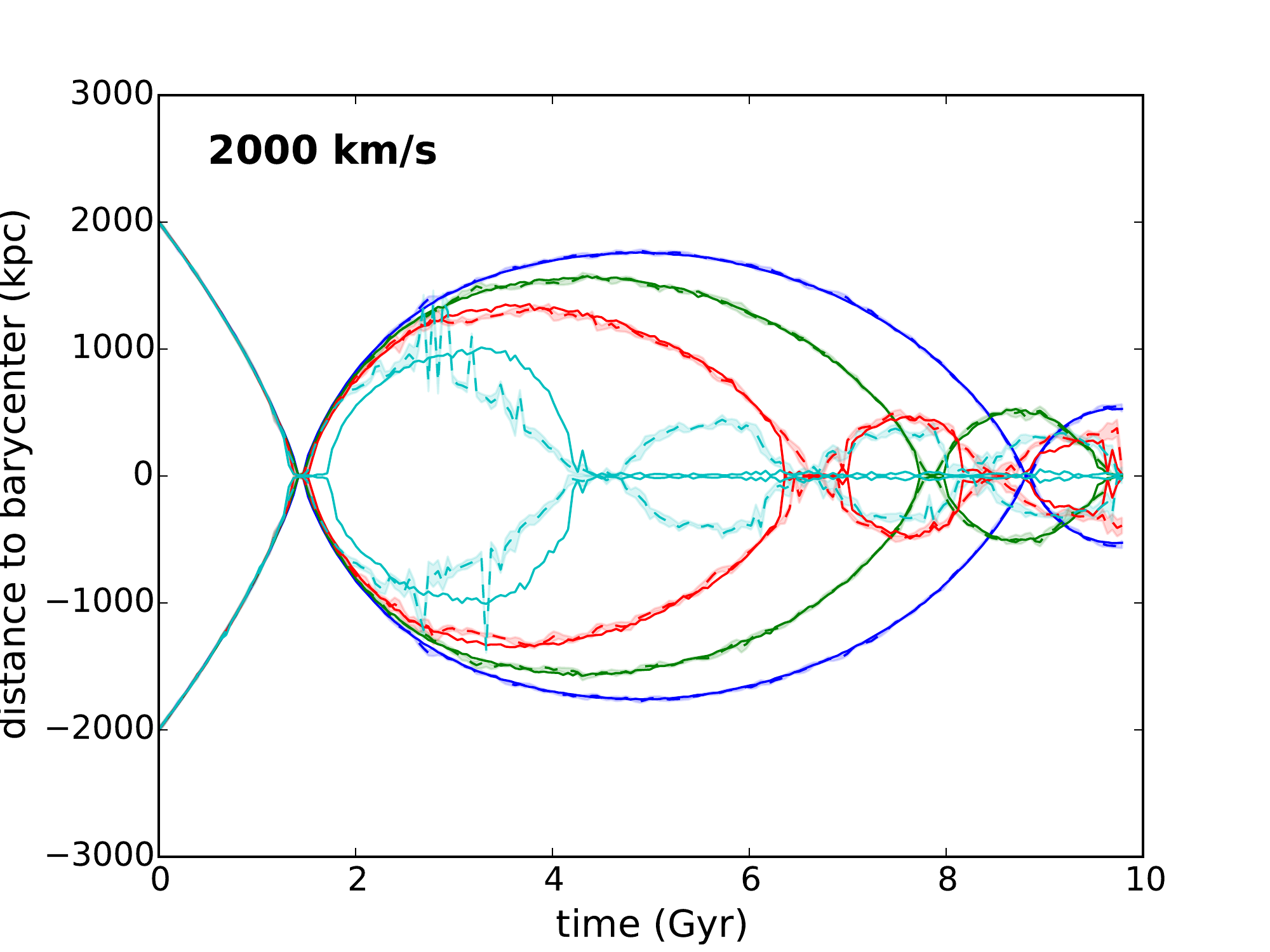}
\includegraphics[width=\columnwidth]{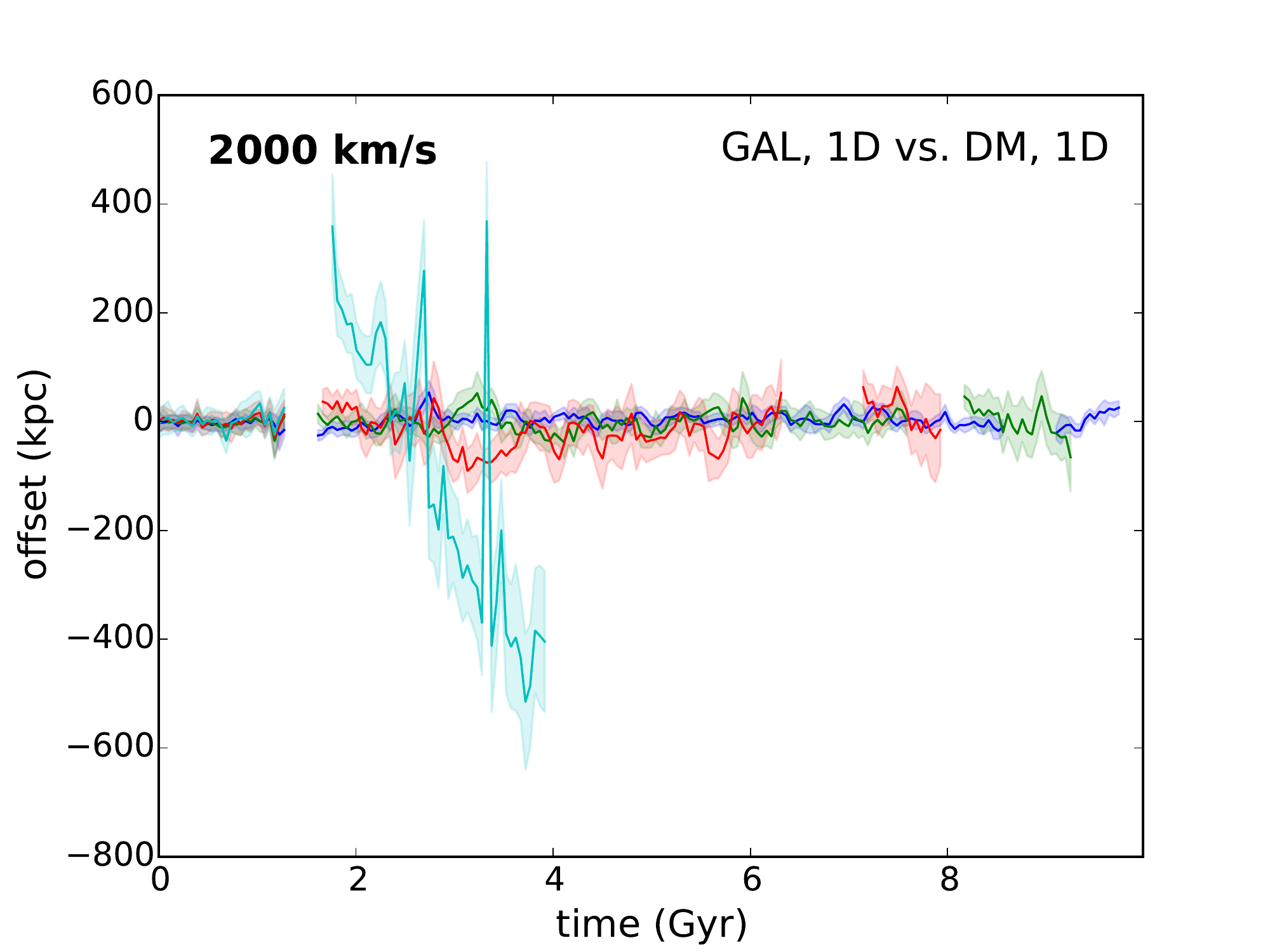} 
\caption{Same as Figure \ref{fig:rc-v0_dmm_2D} except for the 1D analysis.  For high enough cross sections and low merger velocities, halos merge upon contact.  Note that significant dark matter-galaxy peak offsets appears to develop only with relatively high cross sections and merger velocities.}
\label{fig:rc-v0_dmm_1D}
\end{figure*}

\begin{figure*}
\centerline{\includegraphics[width=2.25\columnwidth,trim=0cm 0.7cm 0cm 0cm,clip=true]{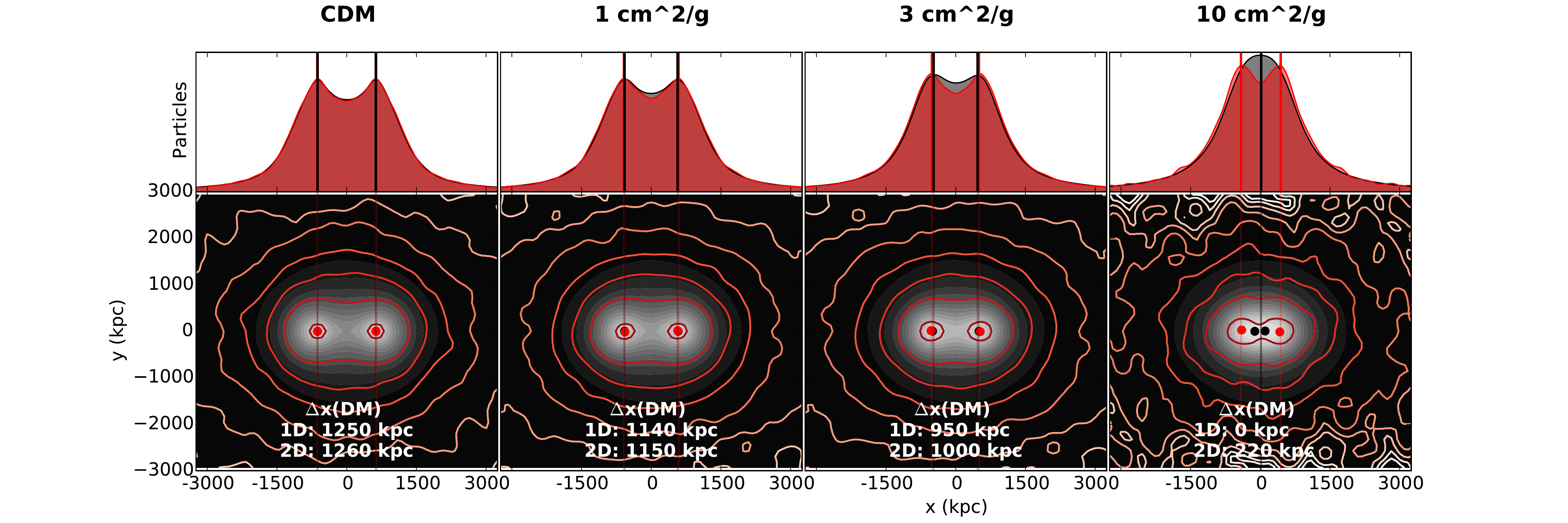}}
\centerline{\includegraphics[width=2.25\columnwidth,trim=0cm 0cm 0cm 1.5cm,clip=true]{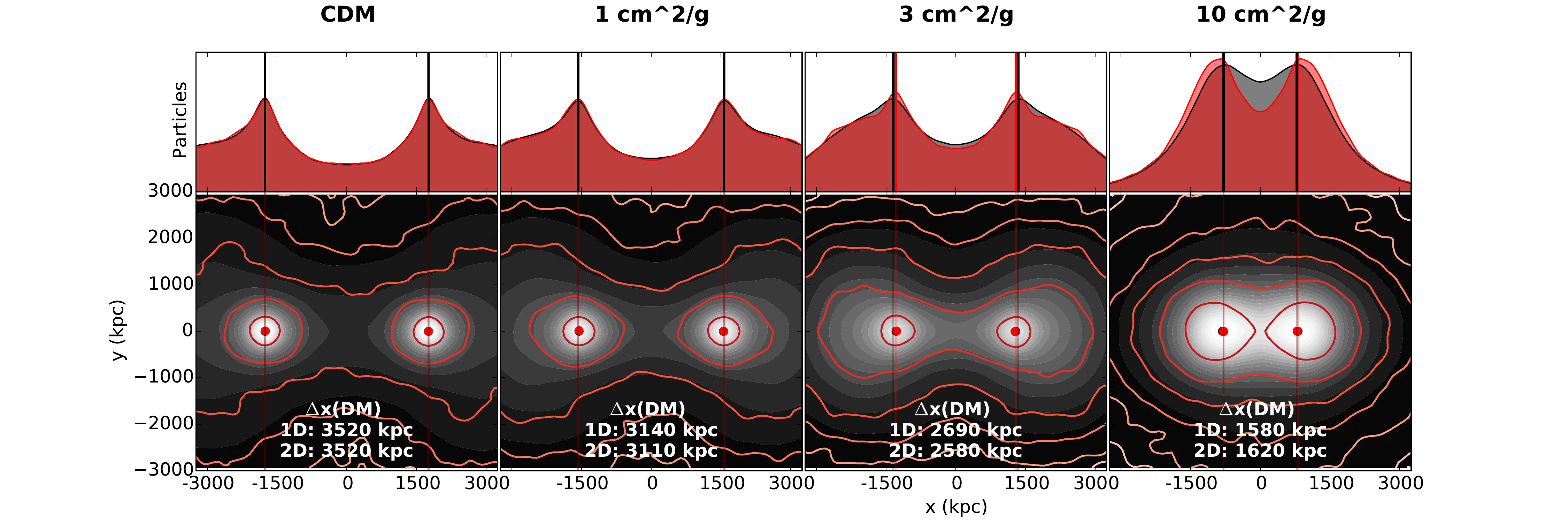}}
\caption{Evolution of dark matter and galaxy distributions after collision.  Shown are the 1D and 2D distributions at apocenter for mergers with $v_\text{4Mpc}$ = 0 (top row) and 2000 km/s (bottom row) for cross sections (from left to right) $\sigma_\text{SI}/m_\chi$ = 0 (CDM), 1, 3, and 10 cm$^2$/g.  The $v_\text{4Mpc}$ = 0 km/s, 10 cm$^2$/g merger coalesced upon contact, and thus has no apocenter; the snapshot corresponding to the apocenter of the galaxy peak trajectories is shown instead.  The 2D color scale has been inverted to highlight the outbound shells.}  
\label{fig:tails-apo}
\end{figure*}

The single-most important parameter controlling SIDM merger evolution, aside from the self-interaction cross section, is the merger velocity.  The velocities we explore here ($v_\text{4Mpc}$ = 0 - 2000 km/s, or $v_\text{infall} \approx$ 1700 - 2700 km/s) span the range of cosmologically typical values \citep{2015MNRAS.448.1674J}.  Mergers with $v_\text{4Mpc}$ = 1000 km/s ($v_\text{infall} \approx$ 2000 km/s) represent the most typical case.  To illustrate the effects of ejection-induced tails, we present results from both our 1D and 2D analyses, which we show in Figs. \ref{fig:rc-v0_dmm_2D} and \ref{fig:rc-v0_dmm_1D}, respectively.  In both figures, we show the evolution of peaks on the left; dark matter peaks are denoted by solid lines, while that of the galaxies are dashed.  Offsets between the two are shown on the right, except when the dark matter halos are within about a scale radius of each other, when the two halos bias each other significantly (see Appendix \ref{apdx:peak_validation}).

In our lowest velocity mergers ($v_\text{4Mpc}$ = 0 km/s; $v_\text{infall} \approx$ 1700 - 1750 km/s), shown on the top row of both figures, collision velocities range from $\sim$3800 - 4100 km/s, a regime in which we expect most self-interactions to result in the capture and exchange of dark matter particles and thus the greatest drag-type phenomology near pericenter.  As expected, no significant offsets appear in the CDM merger.  In the SIDM mergers, however, small, positive (the galaxies lead the dark matter) offsets of 20 and 50 kpc appear for the 1 and 3 cm$^2$/g mergers, respectively, although the offsets for the 1 cm$^2$/g case are only marginally statistically significant.  The 10 cm$^2$/g merger coalesces upon impact due to the sheer number of momentum-loss inducing self-interactions that occurs.  Offsets are similar in magnitude between the 1D and 2D analyses, indicating that tails are not significant.  This is expected, as only a small fraction of self-interactions are expulsive (the escape probability per interaction is $\sim$11\%, see \S\ref{sec:theory}).  1D and 2D maps of the dark matter and galaxy distributions at apocenter reveal the lack of outbound shells, which are shown in the top panel of Figure \ref{fig:tails-apo}.  In the 1D projections, the wings of the mass distributions in all mergers are similarly steeply falling.  In the 2D projections, the galaxy and dark matter contours are roughly circular, modulo effects due to halo overlap.  The inward tail of backscattered dark matter becomes stronger with higher cross section, as seen in the thickening bridge between dark matter peaks with increasing cross section.

The two additional, observable effects on merger evolution produced by enhanced momentum loss also become clear.  Firstly, the apocentric separation of the halos is reduced on the order of hundreds of kpc.  As expected, this effect is more pronounced for mergers with high cross sections.  
Secondly, merger timescales are shortened; the time it takes to reach first apocenter is reduced on order of several tens of Myr.  


This picture changes with higher merger velocities.  The mergers enter a regime in which we expect self-interactions to result in fewer momentum-reducing exchange and capture of particles and instead result in significant expulsion.  The following rows of Figure \ref{fig:rc-v0_dmm_2D} show mergers with initial velocities $v_\text{4Mpc} = $ 1000 km/s (middle row, $v_\text{col} \approx$ 3910-4200 km/s) and 2000 km/s (bottom row, $v_\text{col} \approx$ 4270-4410 km/s).  Increasing the initial velocity to 3000 km/s produced ``fly-by" cluster collisions after which the clusters remained unbound to each other and did not merge.

Offsets appear for all SIDM mergers except for those in which the dark matter halos merge upon contact.  Again, just after pericenter passage, offsets are positive as the galaxies lead the dark matter due to SIDM-enhanced momentum loss.  Note, however, that higher velocity mergers exhibit smaller initial offsets---about 20 kpc for $v_\text{4Mpc}$ = 2000 km/s vs. 50 kpc for $v_\text{4Mpc}$ = 0 km/s with 3 cm$^2$/g---consistent with our expectation that self-interactions are less likely to result in momentum loss at higher collision velocities.  Though the maximum offset size is smaller in higher velocity mergers, offsets are longer lasting.

Tails are stronger at higher velocities.  In the bottom row of Fig. \ref{fig:tails-apo} are shown the 1D and 2D maps of mergers with $v_\text{4Mpc}$ = 2000 km/s at apocenter.  It is clear that the outbound shells of the galaxies and dark matter are much stronger than in the $v_\text{4Mpc}$ = 0 km/s case---the 2D isodensity contours are more lobe-shaped and extended along the merger axis.  In the 1D maps, one can see that at higher cross sections, the outbound shell of galaxies more strongly rivals the peak of bound galaxies; the larger lags and shallower potentials of the dark matter halos at higher cross sections allow more galaxies to escape at apocenter.

Note too, that the outbound shells of dark matter becomes increasingly smooth (vs. peaked) with increasing cross section as more of the dark matter has been redirected into the backscattered tail and outbound shells.  This makes it easier for the shell to bias the peak offsets to larger barycentric distances, especially in the 1D analyses.


\begin{figure*}
\centerline{\includegraphics[width=2.5\columnwidth,trim=0cm 0.7cm 0cm 0cm,clip=true]{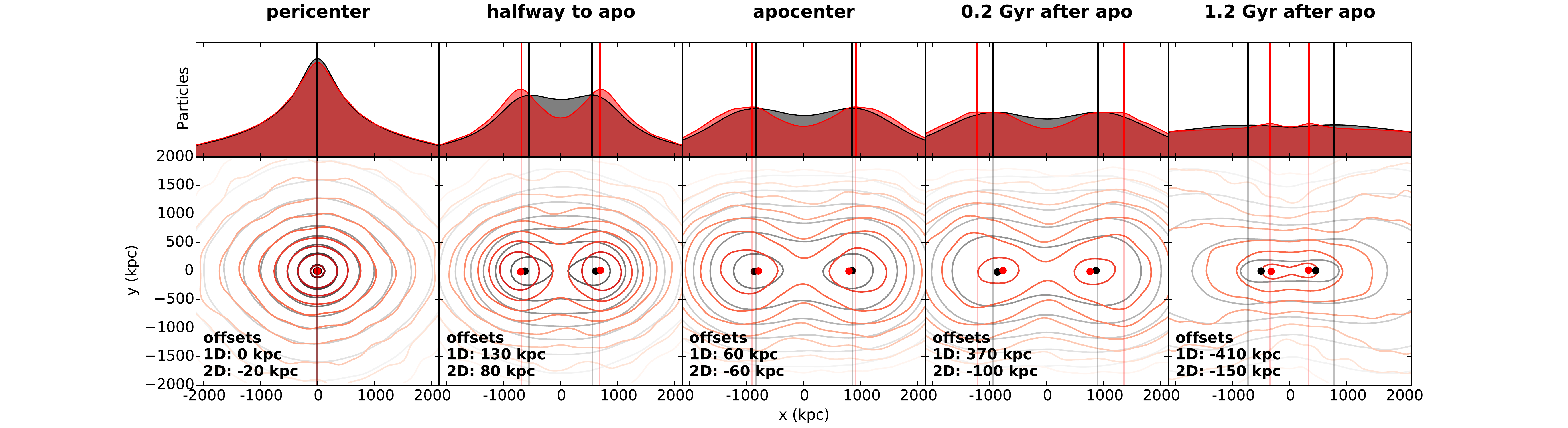}}
\centerline{\includegraphics[width=2.5\columnwidth,trim=0cm 0cm 0cm 1.5cm,clip=true]{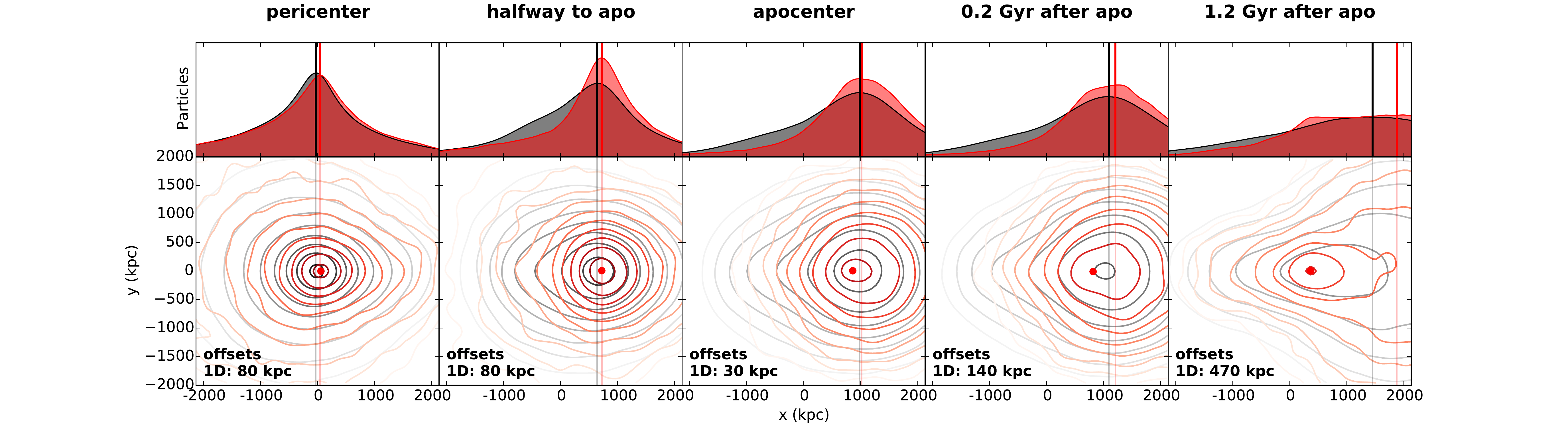}}
\caption{Development of heavy tails at high cross sections.  Panels are the same as in Fig. \ref{fig:tails-1000v_3si} except shown is the 10 cm$^2$/g merger with $v_\text{4Mpc}$ = 2000 km/s.  In the 2D plots, the DM surface density color scale is linear, while the red galaxy contours are logarithmically spaced; the same contour levels are used across all snapshots.  1D peaks are denoted by the vertical lines (both top and bottom rows), while 2D peaks are denoted by the points in the bottom row.} 
\label{fig:tails-2000v_10si}
\end{figure*}

The 10 cm$^2$/g mergers exhibit substantially different phenomenology.  As predicted in Sec. \ref{sec:theory}, the halos merge upon contact unless the initial velocity is about 2000 km/s.  In the 2000 km/s merger, the presence of a very large core ($r_\text{core} \approx$ 690 kpc $> r_s$) causes the dark matter distribution to look like a singly-peaked profile for a few 100 Myr after pericenter before resolving into two peaks.  Examining the 1D and 2D distributions, we find substantial tails that lead to offsets that differ by 200 kpc.  Figure \ref{fig:tails-2000v_10si} shows the evolution of the distributions from first pericenter passage.  As before, the galaxies initially lead the dark matter following pericenter passage, and offsets appear just after pericenter passage.  Following apocenter, large outbound shells of galaxies unbound by the changing gravitational potential separate from the bound galaxies, which is particularly prominent for the one-halo analyses.  Notably, the peak caused by the tail of unbound galaxies \emph{exceeds} the peak of bound galaxies in the 1D analysis, and only gradually disperses.  The large outbound tail veils the population of galaxies that remain bound, making robust peak identification difficult (note the increased jitter following apocenter, particularly in the 1D analysis), and biases peaks away from barycenter.  About 1 Gyr before second pericenter, the tails disperse sufficiently enough that the small population of bound galaxies are revealed near barycenter.  At second pericenter, the dark matter halos merge, while the bound galaxies follow oscillatory orbits in the broad core of the coalesced dark matter halos, much like lower cross section mergers.  The resultant dark matter halo appears nearly disrupted; following apocenter, the dark matter distribution becomes very flat.  The broad core presents a difficulty for weak lensing shear observations, which recover gradients in the matter distribution.  A full lensing analysis including strong lensing and/or magnification can, however, establish the existence of a broad core, for example, in MACS J1149 \citep{2009ApJ...703L.132Z}.

Lastly, we note that SIDM reduces the apocentric separations and merger timescales to a greater degree in higher velocity mergers.  For the 10 cm$^2$/g merger with $v_\text{4Mpc}$ = 2000 km/s, the apocentric separation is smaller by a factor of 2 and merger timescale is 2/3 that of CDM.


\subsection{Dependence on Halo Concentration}
\label{sec:offsets:concentration}

\begin{figure*}
\centering
\includegraphics[width=\columnwidth]{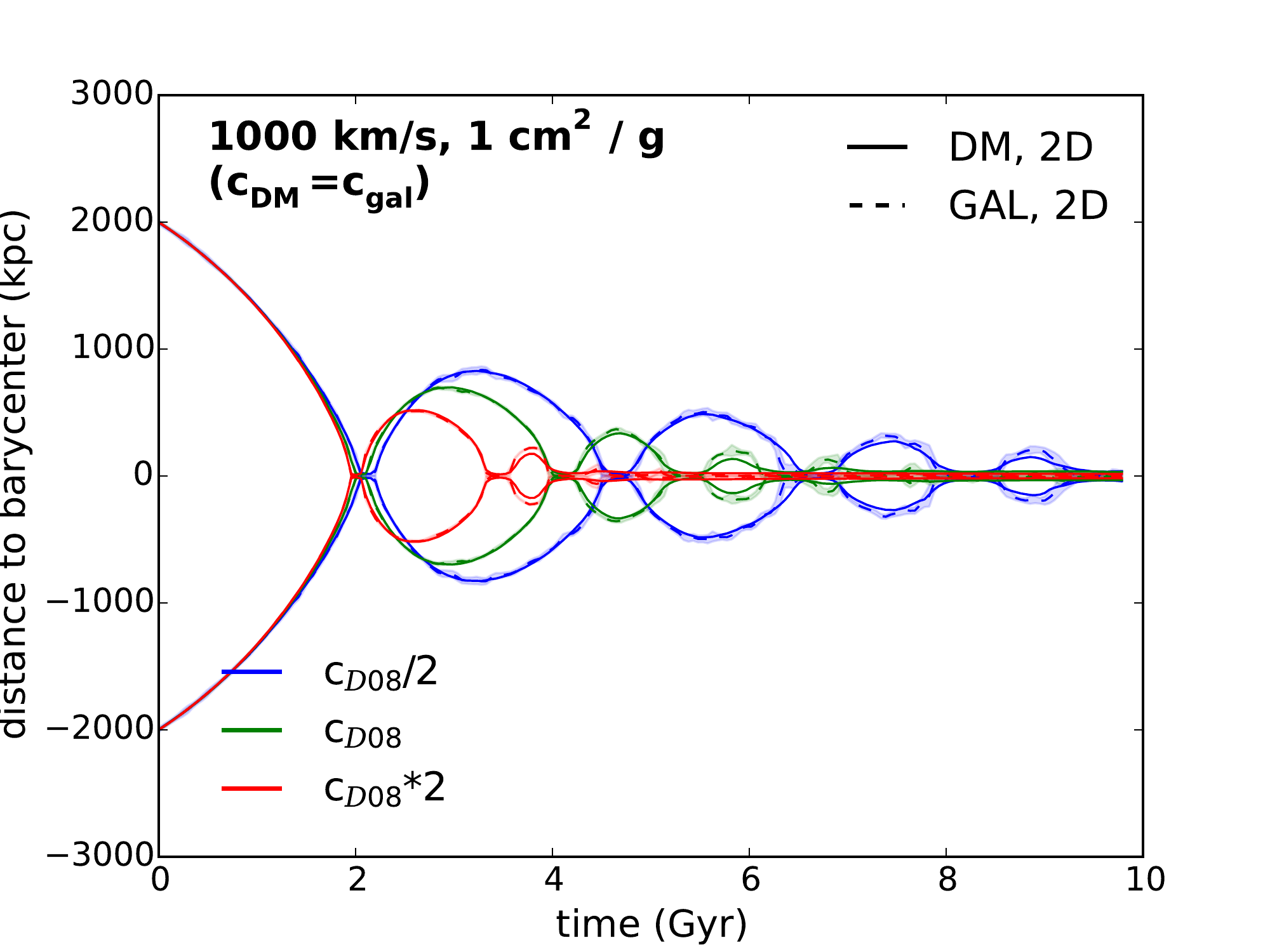}
\includegraphics[width=\columnwidth]{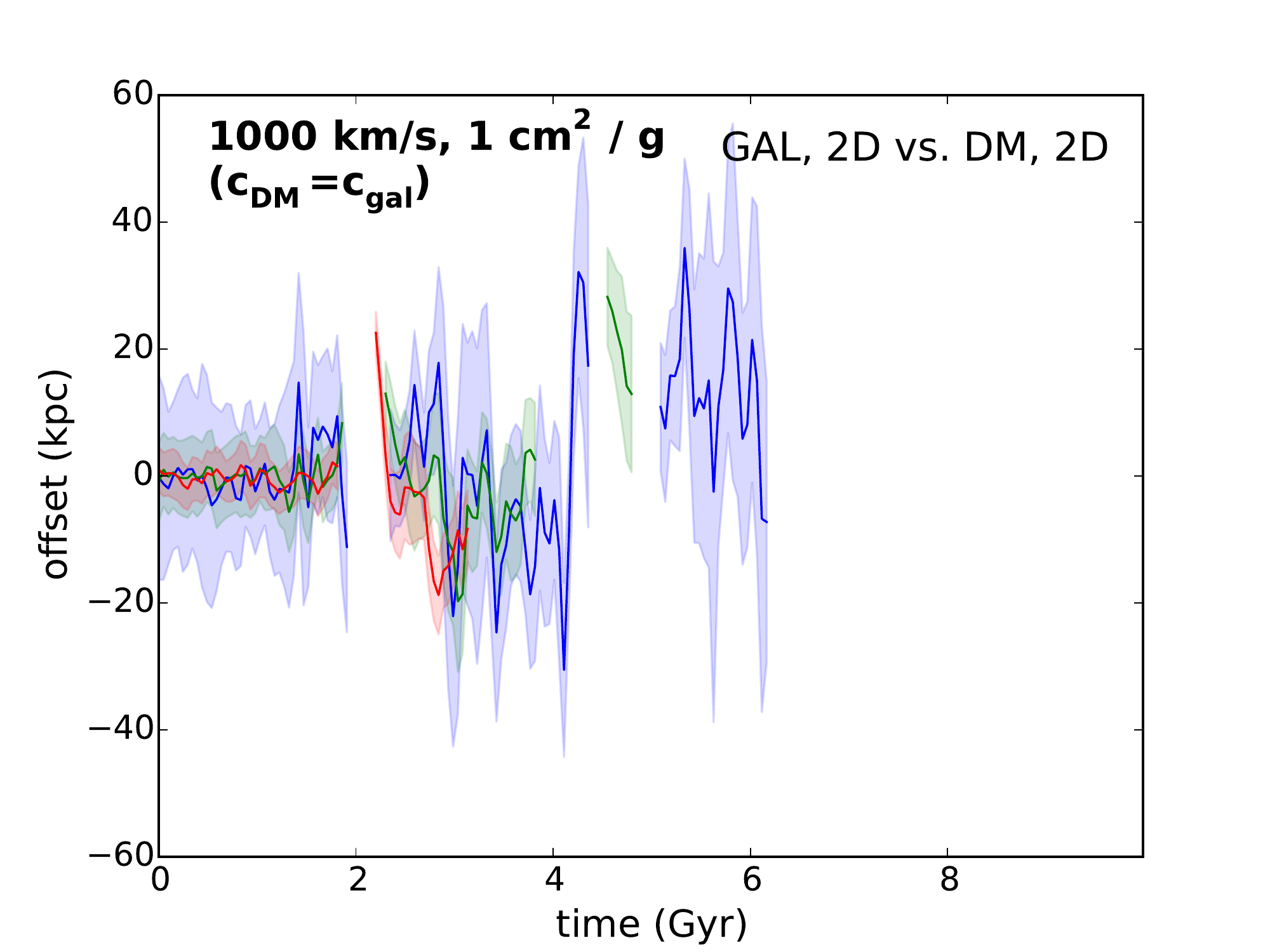} \\
\includegraphics[width=\columnwidth]{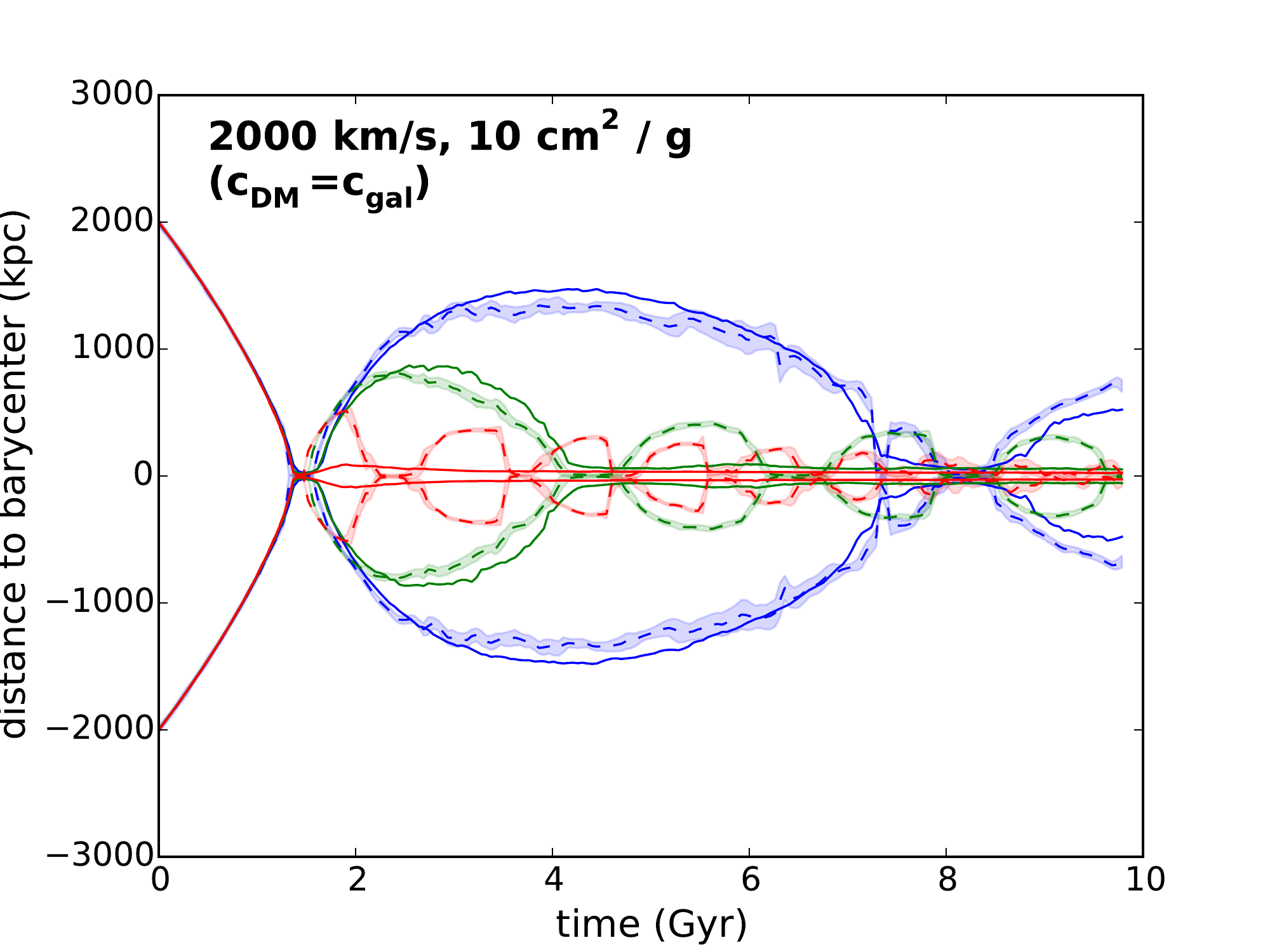}
\includegraphics[width=\columnwidth]{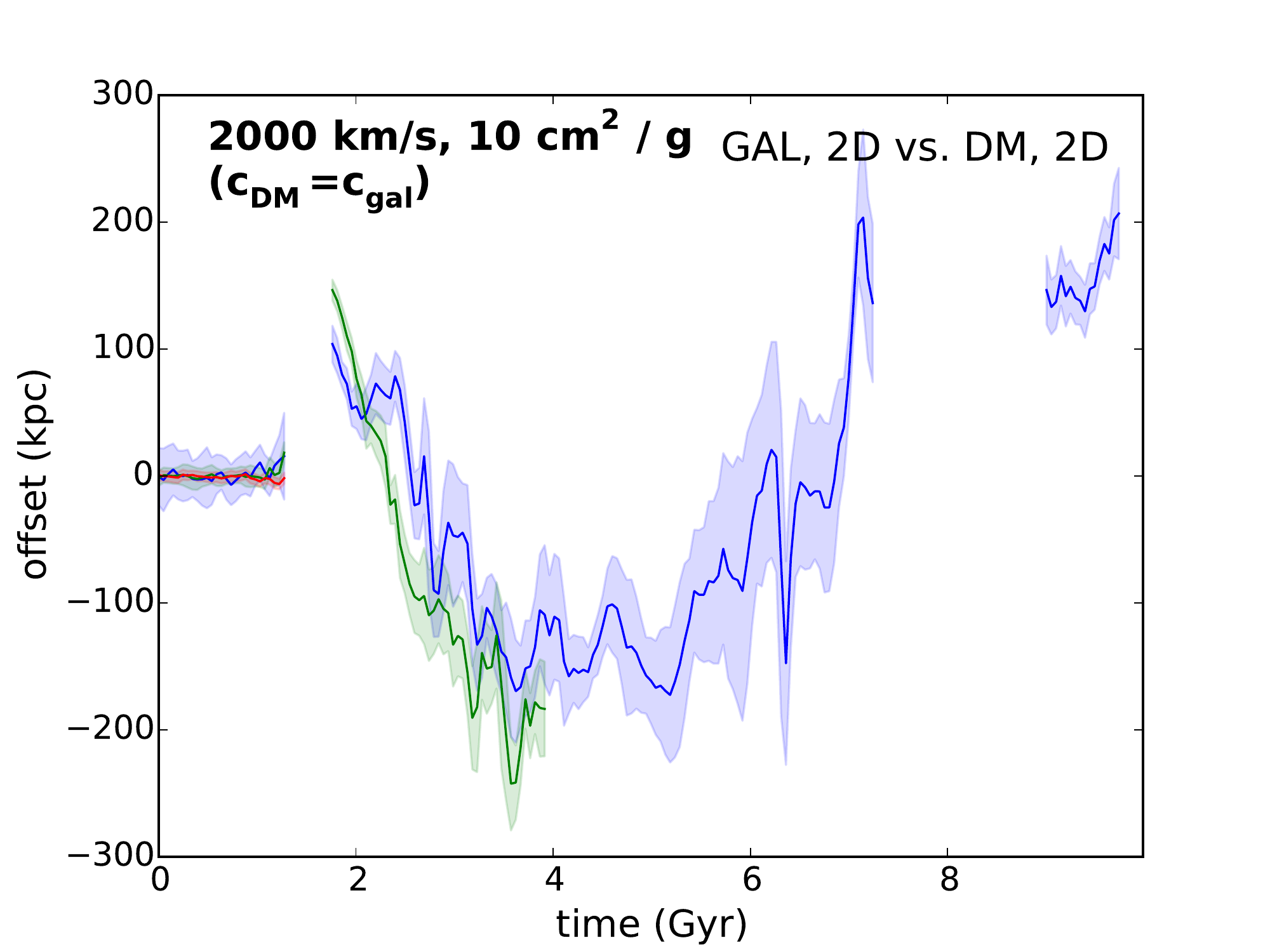} \\
\caption{Dependence on the concentration of the dark matter halos.  As before, the evolution of dark matter peaks and the galaxies are shown on left and the dark matter-galaxy offsets on the right for SIDM mergers with the most significant offsets.  The top row is of a merger with a cross-section of 1 cm$^2$/g and $v_\text{4 Mpc}$ = 1000 km/s, and the bottom row is of a merger with cross-section 10 cm$^2$/g and $v_\text{4 Mpc}$ = 2000 km/s.  Halo concentrations were allowed to vary from half to twice the fiducial concentration (obtained from \citet{2008MNRAS.390L..64D}).  }
\label{fig:rc-cDM}
\end{figure*}

The concentration of the matter distribution affects offset formation by changing two key properties of the merger:  the column of material that a dark matter particle at the core of a cluster passes through (thereby affecting the number of self-interactions that occur), and the binding energy of the dark matter particles (affecting the expulsivity of a given particle collision).  While higher concentrations increases the column of material and thus increases the number of self-interactions that occur, particles are more strongly bound to their parent halos.  The dark matter is thus less likely to become unbound post-interaction and undergo more momentum-loss inducing capture or exchange interactions.  Thus one expects that halos with higher concentrations will exhibit larger offsets immediately after pericenter passage, weaker tails, and shorter merger timescales.  Conversely, mergers with lower concentrations, in which particles are less strongly bound to their parent halos, are more likely to be ejected following a self-interaction.  More massive tails form and the bulk momentum of the halos is not as strongly affected.  This allows the large tails to persist for several Gyr.

We tested these predictions by modifying the concentrations (of both the dark matter and galaxy distributions) in the two mergers with the largest traditional offsets for each set of mergers with $v_\text{4Mpc}$ = 1000 and 2000 km/s:  one with an initial velocity of $v_\text{4Mpc}$ = 1000 km/s and a cross section of 1 cm$^2$/g, and the merger with an initial velocity $v_\text{4Mpc}$ = 2000 km/s and a cross section of 10 cm$^2$/g.  We varied the concentration by a factor of two in both directions (again evolving each halo in isolation for 3 Gyr before merging identical halos together) for each of these mergers.  The peak evolution for each set of mergers is shown in Figure \ref{fig:rc-cDM}.

As expected, there is evidence of greater momentum loss in mergers with higher concentrations.  For mergers with 1 cm$^2$/g and $v_\text{4Mpc}$ = 1000 km/s, offsets soon after pericenter are about 22, 10, and 0 kpc at concentrations that are twice, equal, and half of the fiducial concentration, respectively.  Timescales are a dramatic function of concentration:  the time between pericenters, for instance, are shortened by order a Gyr, with the highest concentration case having the shortest timescales.  Mergers with 10 cm$^2$/g and $v_\text{4Mpc}$ = 2000 km/s have offsets decrease in size for decreasing concentration, while the time between pericenter passages nearly doubles.  The high-concentration halos coalesce on impact.  In contrast, a half fiducial concentration produces a large upturn in offsets occurs approaching second pericenter, becoming positive (e.g. the dark matter catches up with and overtakes the galaxies).  This is likely due to large inward dark matter tails, which pulls the dark matter peaks towards barycenter, combined with a very flat potential, which allow the tails to play a larger role than in mergers of more concentrated halos.

Note that the errors on the peaks and offsets are much larger at lower concentrations.  The matter distribution is much flatter at lower concentrations, which makes precise peak-finding difficult, particularly for the sparsely sampled galaxy distribution.

Thus we find that the concentration of the dark halos plays a major role in the evolution of the merger and in the maximum size and evolution of the offsets.  For fixed initial velocity and cross section, high-concentration halos have shorter merger times and larger offsets, and more closely follow predictions from the drag picture than do low-concentration halos.



\subsection{Dependence on Impact Parameter}
\label{sec:offsets:impact}

\begin{figure*}
\centering
\includegraphics[width=\columnwidth]{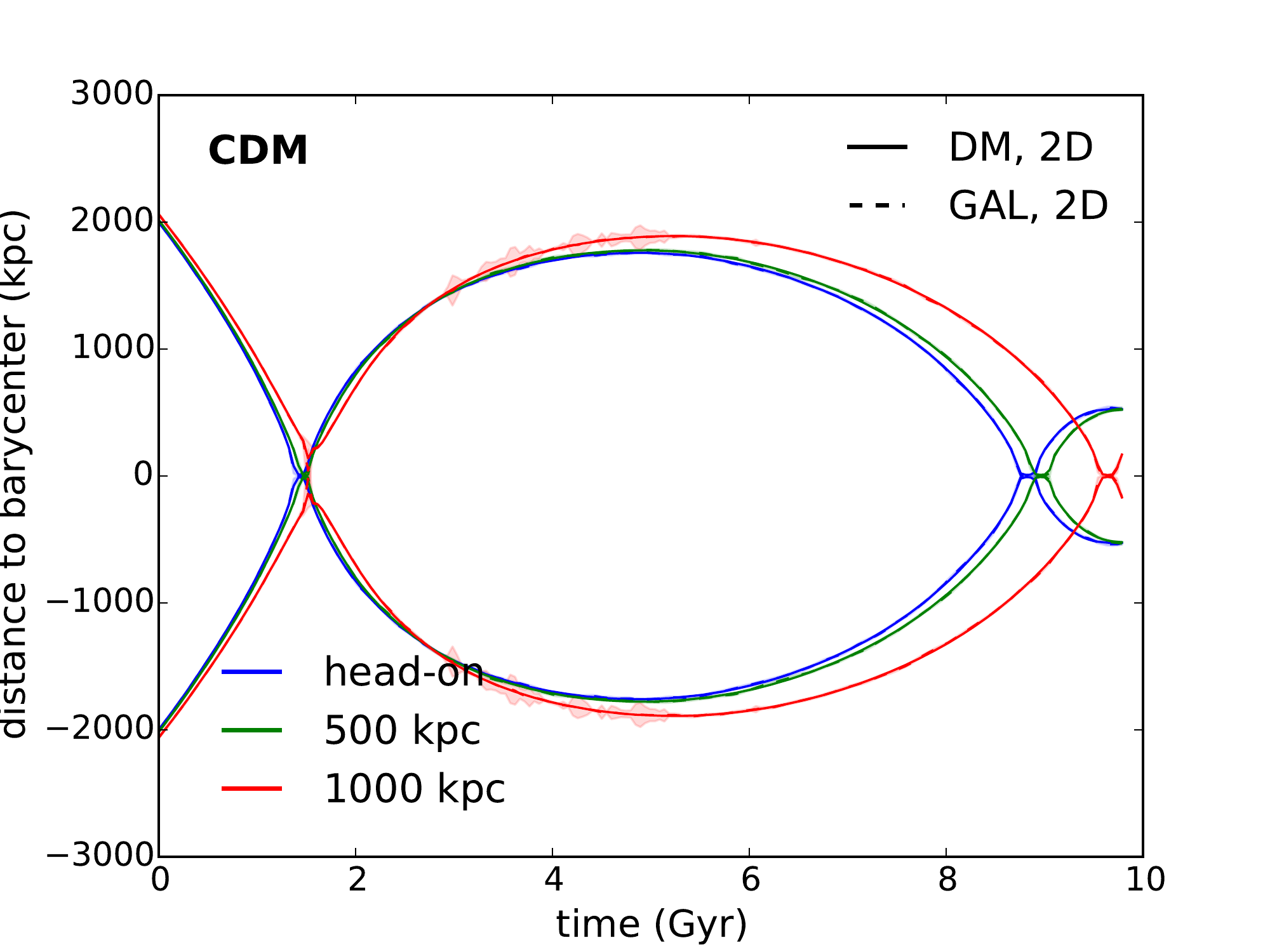}
\includegraphics[width=\columnwidth]{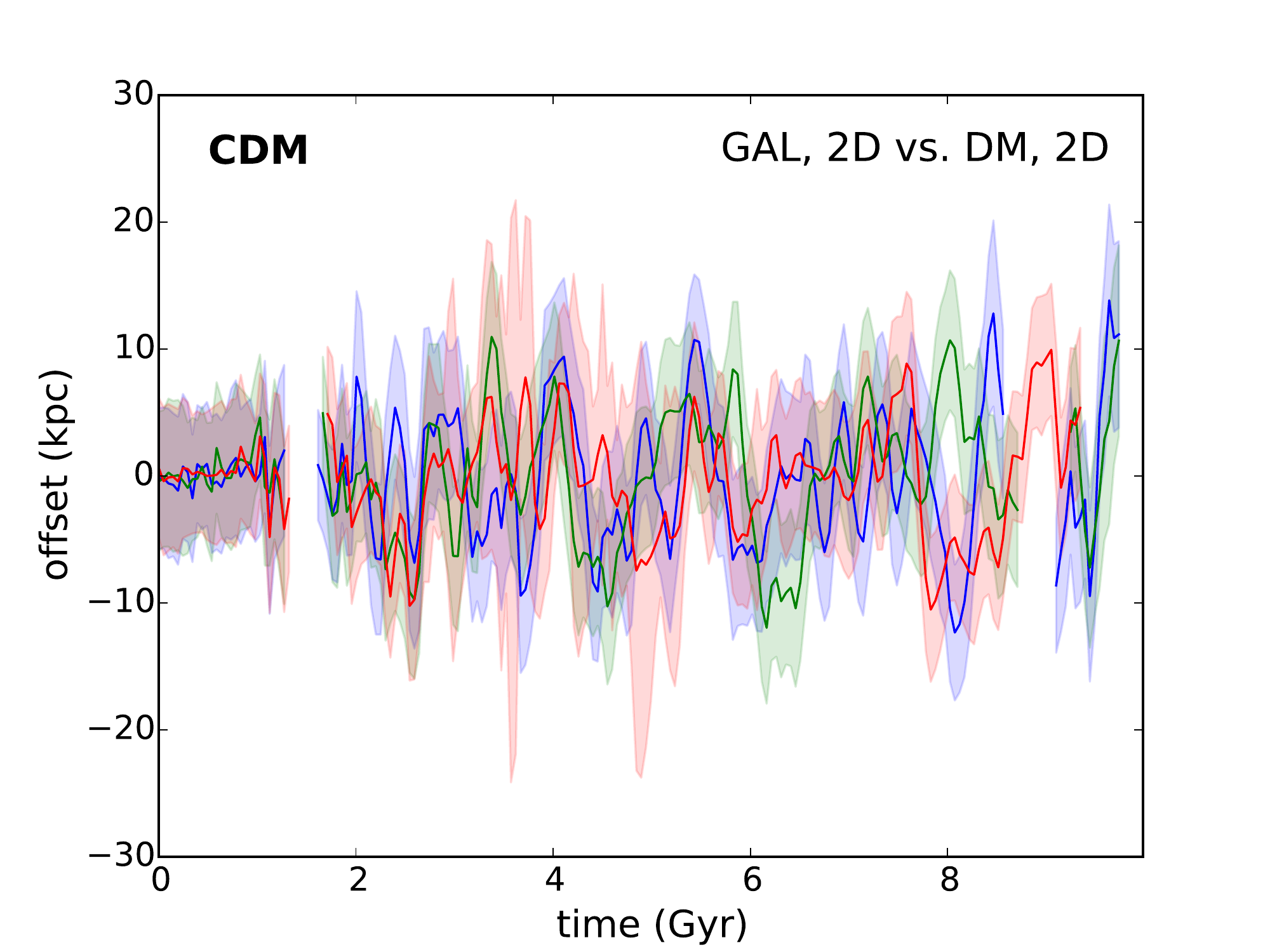} \\
\includegraphics[width=\columnwidth]{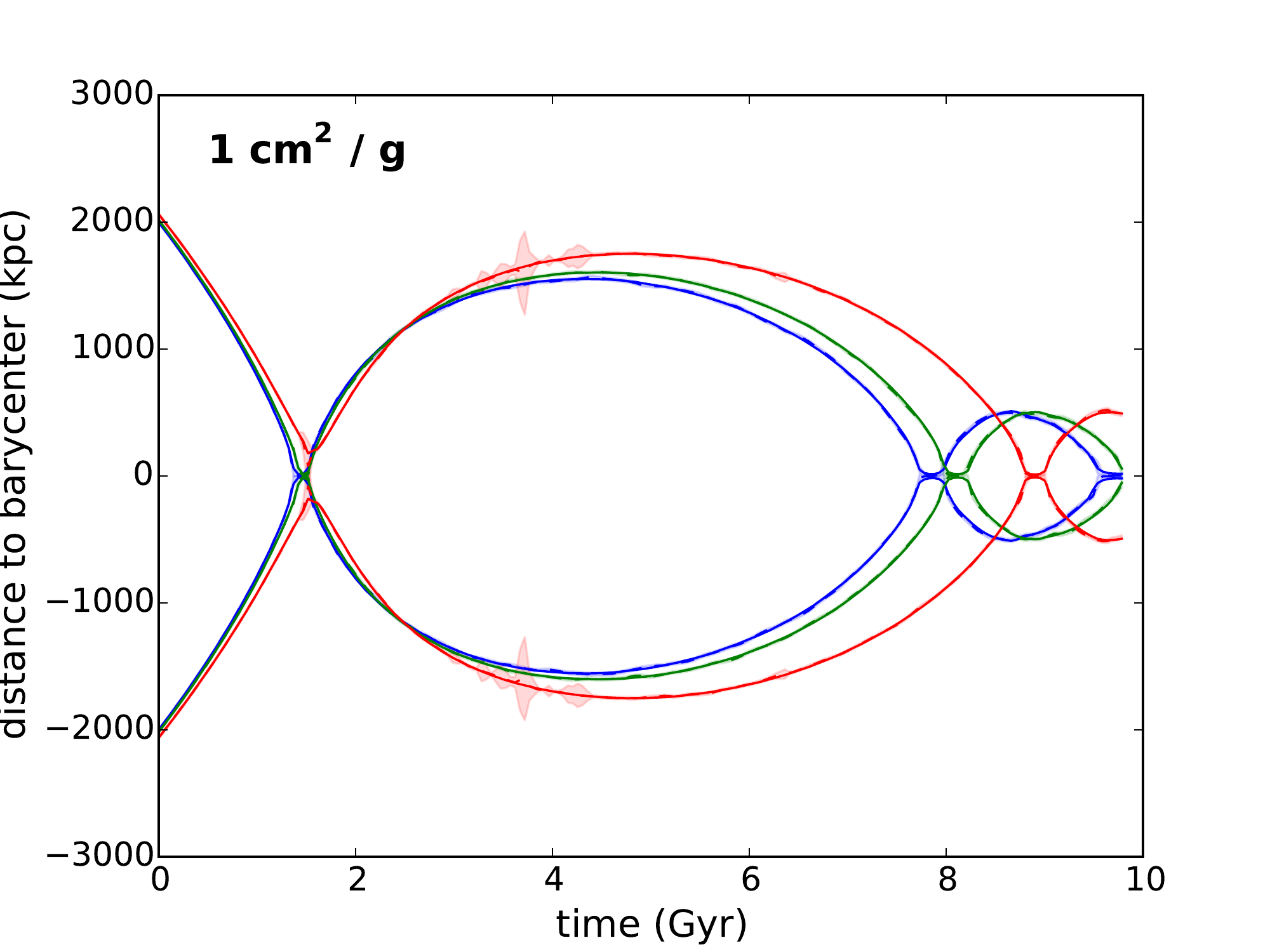}
\includegraphics[width=\columnwidth]{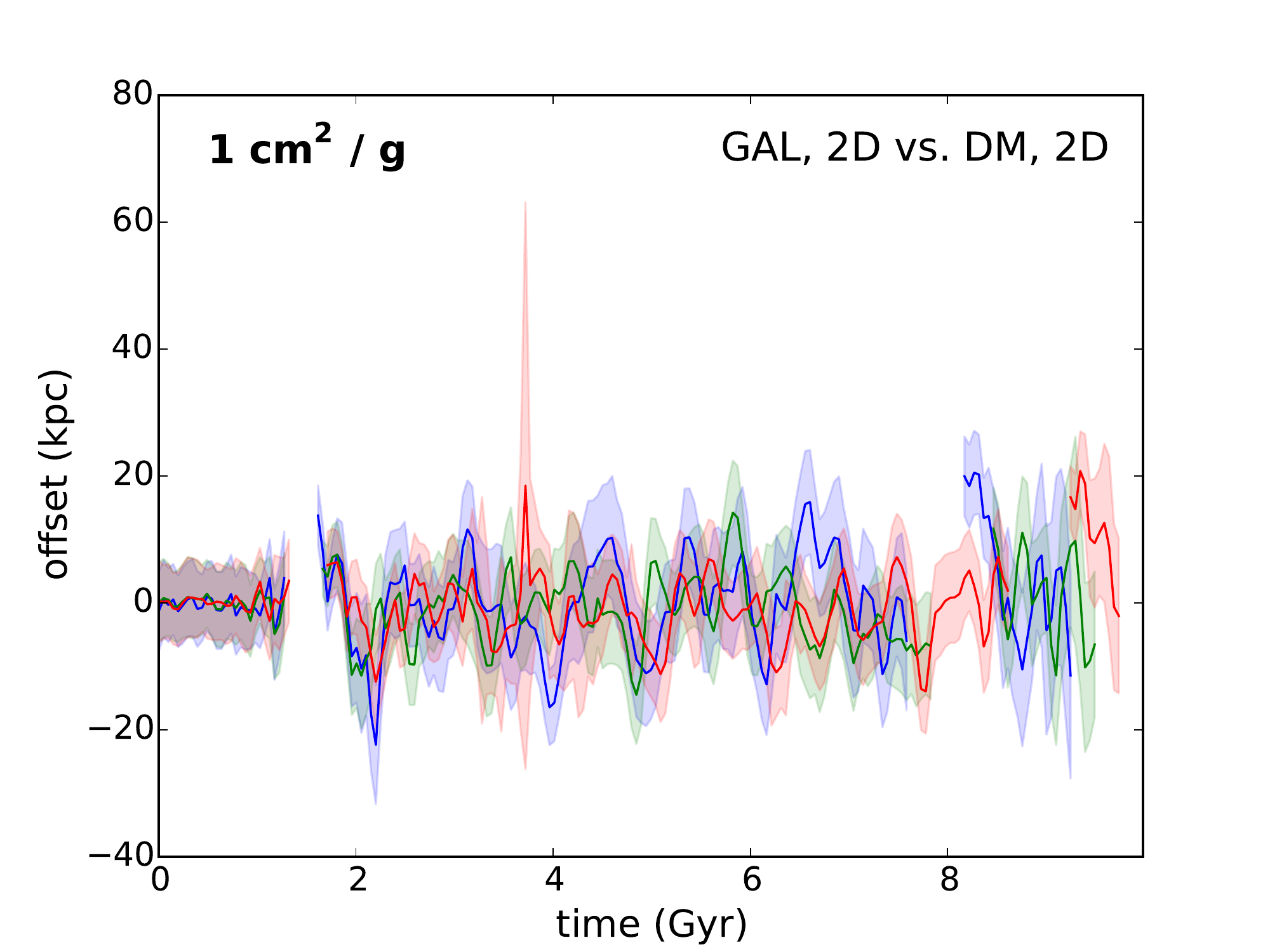} \\
\includegraphics[width=\columnwidth]{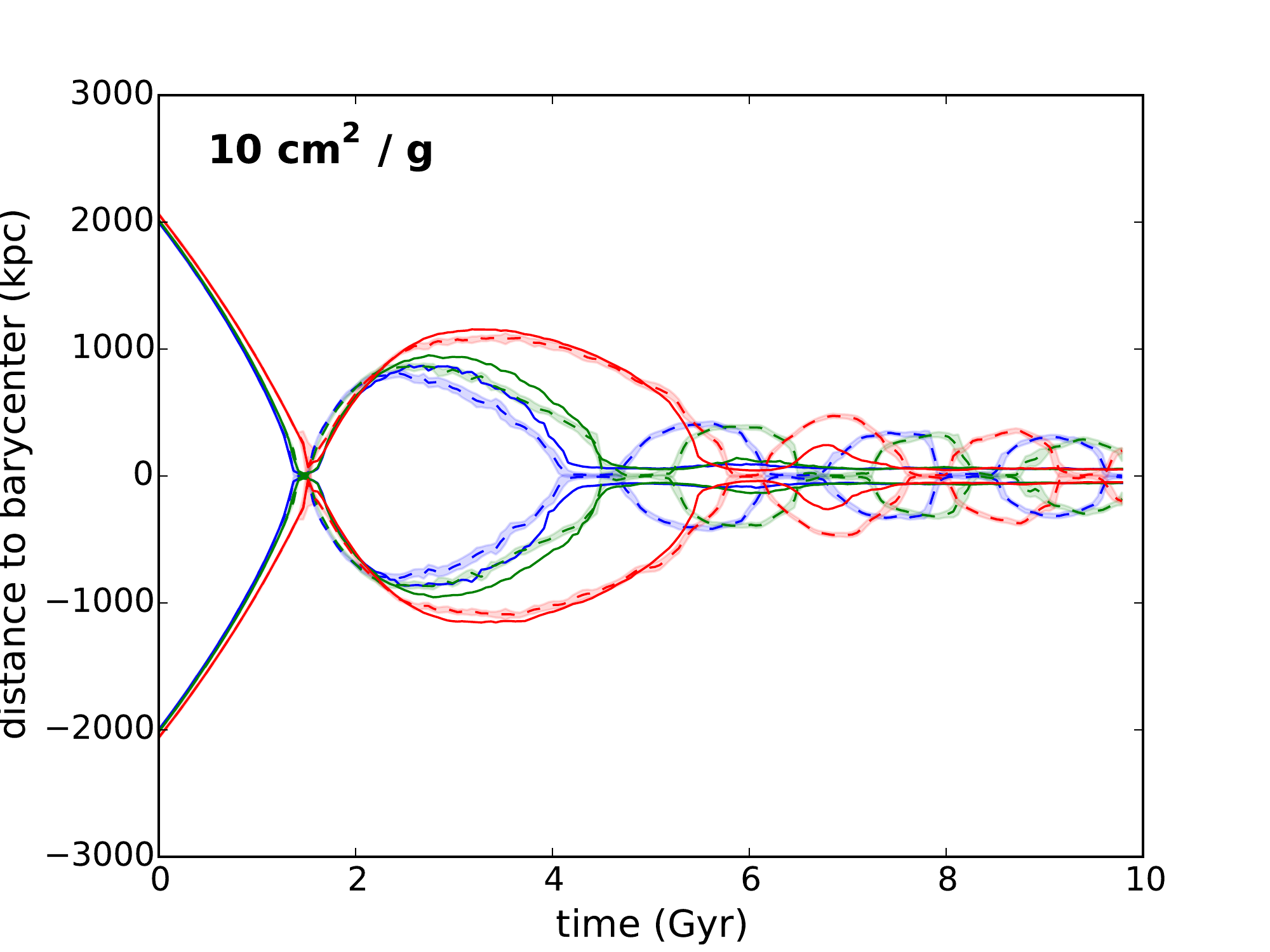}
\includegraphics[width=\columnwidth]{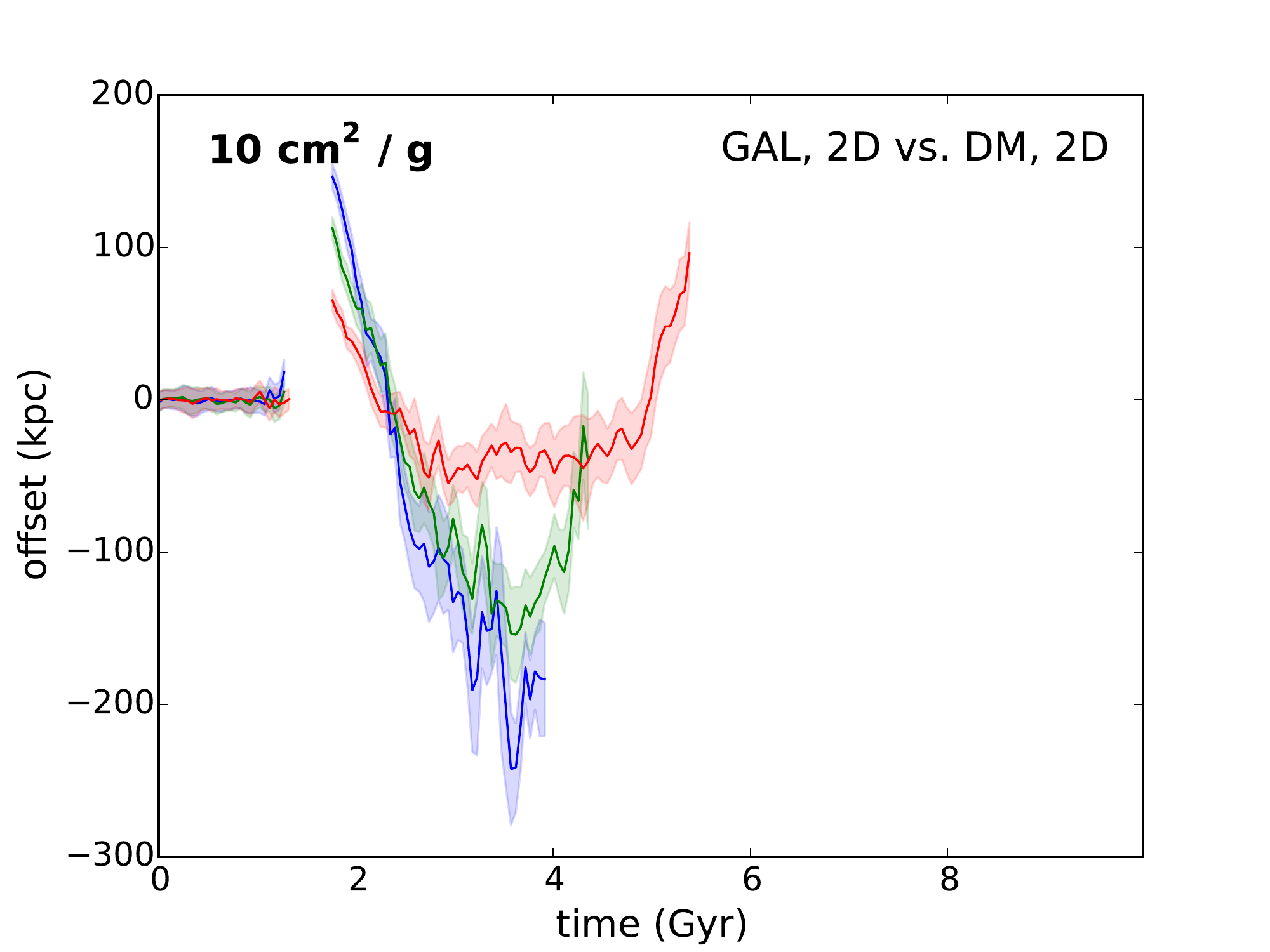} \\
\caption{Dependence on impact parameter of the dark matter halos.  As before, the evolution of dark matter peaks and the galaxies are shown on left and the dark matter-galaxy offsets on the right for mergers with $v_\text{4Mpc}$ = 2000 km/s.  Each row represents a different cross section.  From top to bottom: CDM, $\sigma_\text{SI}/m_\chi$ of 1 cm$^2$/g, and $\sigma_\text{SI}/m_\chi$ of 10 cm$^2$/g.  The impact parameter was allowed to vary from 0 to 1000 kpc (in comparison, the halo scale radius $r_s$ = 600 kpc, and core sizes are 90 and 200 kpc for $\sigma_\text{SI}/m_\chi$ = 1 and 10 cm$^2$/g, respectively.}
\label{fig:rc-b}
\end{figure*}

Mergers are not necessarily head-on.  Increasing the impact parameter $b$ decreases the column through which the dark matter halos pass (and severely so if $b > r_s$), as well as increasing the merger timescale.  While the former decreases the number of self-interactions that occur, the latter increases the offset lifetime.  We illustrate this for a subsample of our mergers, shown in Figure \ref{fig:rc-b}, which were run with initial velocities $v_\text{4Mpc}$ = 2000 km/s and self-interaction cross sections of 0, 1, and 10 cm$^2$/g (top, middle, and bottom rows, respectively).  For each set of mergers, we shifted the halos perpendicular to the merger axis so that $b$ = 0, 500 ($\approx r_s$), and 1000 ($> r_s$) kpc.

For the 1 cm$^2$/g mergers, which produced a 10 kpc offset in a head-on collision, no offsets formed with an impact parameter comparable to the scale radius (500 kpc).  In comparison, mergers with a cross section of 10 cm$^2$/g still exhibited offsets even with the largest impact parameter ($b \approx 2r_s$).  However, the maximum magnitude of the offsets decreased by factors of several but lasted longer than in the head-on collision.


\subsection{Summary}

\begin{figure*}
\centering
\includegraphics[width=2\columnwidth]{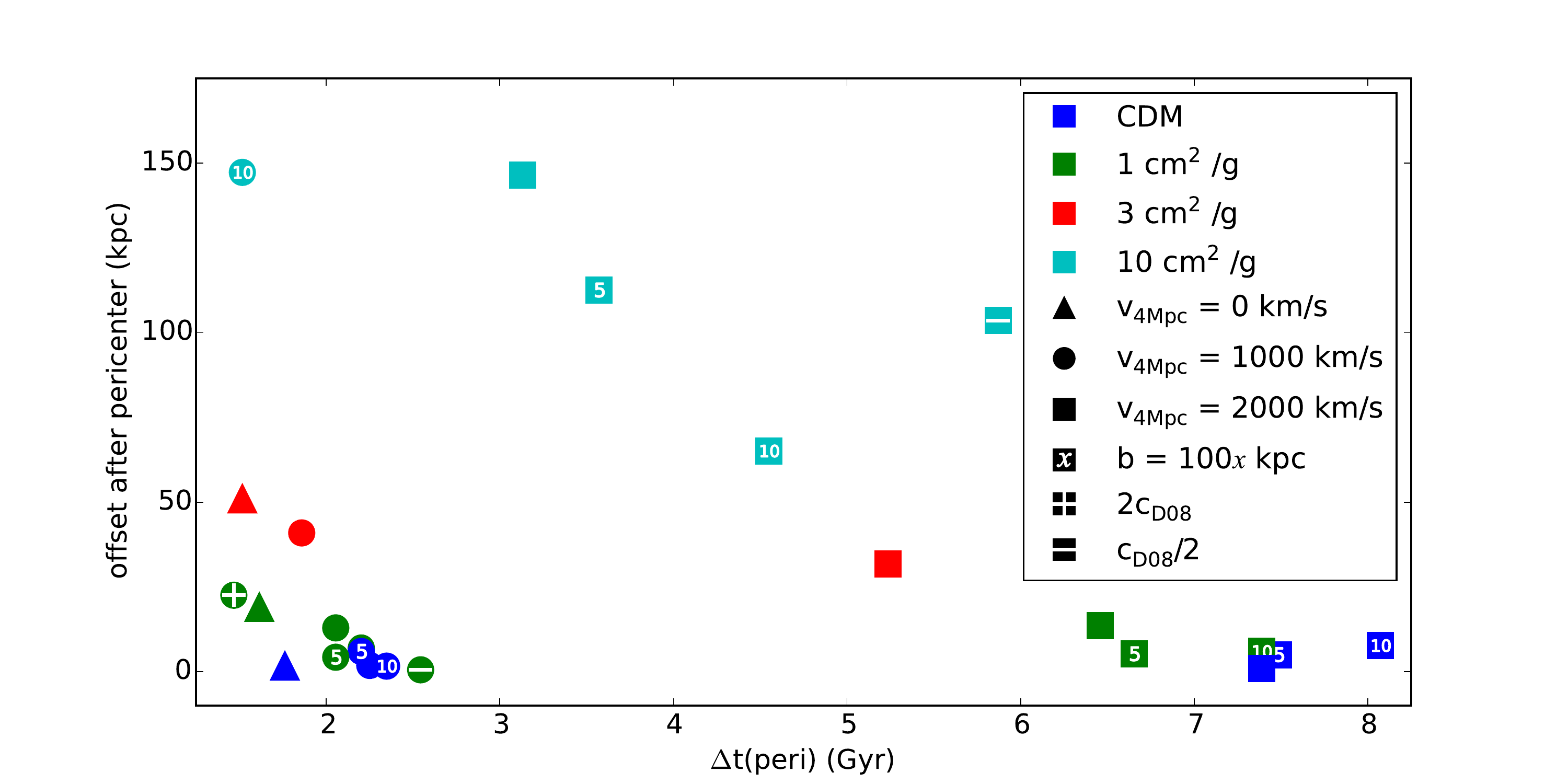}
\caption{Offsets just after pericenter passage---typically the maximum offset---for all mergers, plotted over the time between first and second pericenter (a proxy of the lifetime of offsets), calculated based on our 2D analysis.  Offsets for mergers that coalesced upon contact are not shown.  Colors denote the cross section, while shapes denote the inital velocity $v_{\text{4Mpc}}$.  Additional mark-ups denote changes in concentration (a `+' (`-') for twice (half) the fiducial concentration, and a number denoting a non-zero impact parameter).  Cosmological mergers correspond to $v_{4Mpc} \approx$ 1000 km/s, $c_\text{DM} \approx c_\text{D08}$, and small impact parameters, e.g. those in the lower left corner, which have a maximum offset size between 15-40 kpc for 10$^{15}$ M$_\odot$ mergers.}
\label{fig:summary-offset_vs_time}
\end{figure*}

In Fig. \ref{fig:summary-offset_vs_time}, we show for each of our simulations the offset size just after pericenter passage---the maximum offset in all simulations with the exception of the high velocity, high cross section merger ($v_\text{4Mpc}$ = 2000 km/s, $\sigma_\text{SI}/m_\chi$ = 10 cm$^2$/g). We do not plot cases where immediate coalescence occurred.  Offsets are plotted over the time between first and second pericenter, which we use as a proxy to the duration of offsets.

Clearly, the size of dark matter-galaxy offsets is not a simple function of cross section.  While offsets are most sensitive to the self-interaction cross section, they can change dramatically under different merger conditions.  For a cross section of 1 cm$^2$/g, the approximate limit currently claimed in the literature, we find never find offsets greater than $\sim$20 kpc.  This is achieved in head-on collisions with low merger velocities and/or high halo concentrations---all of which encourage greater momentum loss at pericenter passage.  Offsets vanish if the halo concentration is decreased by a factor of 2 or if an impact parameter the size of the cluster's scale radius is introduced.

We see that offsets form in only a narrow range of cross sections.  At higher cross sections, offsets are larger and can form across a larger range of merger conditions.  For a cross section of 3 cm$^2$/g, the largest offsets are $\sim$50 kpc.  At a cross section of 10 cm$^2$/g, halos coalesce upon contact unless the merger velocity to be unusually high.  The maximum possible offset before completely disrupting the halos is of order 100 kpc.

Offsets that do form are transient.  The conditions that produce the largest offsets also hasten merger lifetimes, causing the offsets to be short lived; there is a trade off between offset sizes and lifetimes.  In addition, the offset sizes we have predicted are only so large if the merger occurs in the plane of the sky.  Observing the largest offsets we predict requires observing a merger at exactly at the right time (just after pericenter passage) and with exactly the right orientation with respect to the line of sight.


\section{OBSERVED OFFSETS}
\label{sec:observability}

In light of the findings in Sec. \ref{sec:offsets}, we here revisit the case for self-interactions in observed equal mass mergers.  In Sec. \ref{sec:observability:offset_sizes}, we review the cosmological ranges in merger and halo parameters, from which we infer an expected range of offsets.  We then compare these predictions with those measured in observed equal mass mergers and discuss their implications in Sec. \ref{sec:observability:observed_mergers}.


\subsection{Expected Offsets}
\label{sec:observability:offset_sizes}

In the previous section, we modeled mergers across a large range of parameters to study how offset formation and size depend on merger parameters.  Here, we discuss what are realistic, cosmologically favored ranges for merger parameters and their implications for the range of expected offset sizes.

{\bf Merger velocities}.  For 10$^{15}$ M$_\odot$ mergers, \citet{2015MNRAS.448.1674J} predicts infall velocities of $\sim$2000 km/s, or collision velocities of $\sim$4100 km/s (akin to our $v_\text{4Mpc}$ = 1000 km/s simulations).  Presuming that clusters follow the median mass-concentration relation and are head-on, we expect offsets to be at most $\sim$20 kpc for $\sigma_\text{SI}/m_\chi$ = 1 cm$^2$/g and at most $\sim$50 kpc for $\sigma_\text{SI}/m_\chi$ = 3 cm$^2$/g.  In this regime, cross sections of 10 cm$^2$/g cause immediate coalescence (including for impact parameters as large as 1000 kpc).  Thus by virtue of observing well-separated, post-collision clusters, we can immediately rule out cross sections $\gtrsim$ 10 cm$^2$/g.

{\bf Scatter in the halo concentration distribution}.  We have seen that offset and merger evolution are strongly dependent on halo concentration.  There is a large scatter in the mass-concentration distribution:  \citet{2008MNRAS.390L..64D} find $\sigma$(log$_{10}c_{200}$) = 0.15 (or a factor of about 1.4) for their sample of clusters at $z \sim$ 0, which translate for 10$^{15}$ M$_\odot$ mergers with median $c$ = 3.3 to a range of 2.3 - 4.7.  In our simulations, we varied $c$ by a factor of 2, and found that the offset changed by $\pm$10 kpc for our $v_\text{4Mpc}$ = 1000 km/s, $\sigma_\text{SI}/m_\chi$ merger.  Thus we expect the scatter in the concentration to change the size of offsets by $\pm$5 kpc for 1 cm$^2$/g (and likely only slightly more for 3 cm$^2$/g).

{\bf Halo masses}.  The halo mass function sharply decreases with mass at cluster scales---we expect many more mergers of 10$^{14}$ M$_\odot$ halos than 10$^{15}$ M$_\odot$ halos.  What sort of offsets should we expect in these mergers?  The toy model described in Sec. \ref{sec:theory} can again provide guiding estimates.  Recall that the probability of escape following a self-interaction is dependent on the collision and escape velocities of the halos.  \citet{2015MNRAS.448.1674J} found that mergers of clusters of 10$^{14}$ M$_\odot$ have an infall velocity that scales as
\begin{equation*}
v_\text{in} \sim 1.1 \sqrt{ \frac{ G M }{ R }} \sim M^{1/3},
\end{equation*}
where $M$ and $R$ are the virial mass and radius (i.e. $M_\text{200}$ and $R_\text{200}$).  Estimating that orbital and collision velocities scale similarly, we have that the collision velocity scales as
\begin{equation}
v_\text{col} \sim 4000 ~ \text{km/s} ~ \left( \frac{M}{10^{15} M_\odot} \right)^{1/3}.
\end{equation}
The escape velocity scales similarly with mass,
\begin{equation}
v_\text{esc} \sim \sqrt{\frac{ G M }{ R }} \sim 3100 ~ \text{km/s} ~ \left( \frac{M}{10^{15} M_\odot} \right)^{1/3}.
\end{equation}
Thus equal mass mergers of different mass scales will roughly trace a 1:1 diagonal line in the two left-most panels of Fig. \ref{fig:poutcomes}.  This implies that regardless of the mass of an equal mass mergers, a two-particle collision will have the same likelihood of unbinding both particles.

What changes between mergers of different mass scales is their surface density, which determines the number of self-interactions that occur.  Lower-mass clusters have lower surface densities, which decreases the number of self-interactions that occur.  The central surface density of a cluster with a cored Burkert density profile (given in Eq. \ref{eqn:surface_density}) scales with the mass, but only weakly.  This implies that \emph{offset formation is a largely self-similar process in equal mass mergers across all mass scales, only slightly enhanced in more massive mergers}.  For example, the central surface density of a 10$^{14}$ M$_\odot$ halo is about 0.8 g cm$^{-2}$, which by Eq. \ref{eqn:fesc} we expect 9\%, 20\%, and 43\% of the halo to escape following a single pericenter passage for $\sigma_\text{SI}/m_\chi$ = 1, 3, and 10 cm$^2$/g, respectively, as compared to 12\%, 24\%, and 50\% for 10$^{15}$ M$_\odot$ halos.

However, we expect that offsets in less massive mergers are smaller.  Both the drag and tail physics scale with mass.  In \citet{2014MNRAS.441..404H}'s drag model, the drag force scales as a positive power of the merger speed, while in \citet{2014MNRAS.437.2865K, 2015MNRAS.452L..54K}'s tail model centroid-based offsets scale with the core radius of the dark matter halo.  Both of these dependences are smaller for less massive halos.  Thus while equal mass mergers of all masses are similarly likely to form offsets, they become increasingly difficult to observe at low masses.  The strongest offset-based limits will come from the most massive mergers.

{\bf Direction of offset}.  One long-standing idea concerning galaxy-DM offsets in merging clusters is that SIDM should cause the galaxies to appear to lead the DM relative to the merging cluster barycenter.  However, in Sec. \ref{sec:offsets}, we find that this only occurs near pericenter (both leaving and, if nonzero, approaching).  In fact, the dark matter peak tends to lie further from barycenter than the galaxy peak (e.g. negative offsets) for at least half the orbit as the merger approaches and returns from apocenter.  Therefore, modulo observational selection effects, in a sample of equal mass merging clusters that span all merger phases, we expect the sign of the offset between dark matter and galaxies with respect to barycenter to be approximately equally distributed.  However, some of the current methods for discovering mergers favor systems that have just undergone pericenter passage.  X-ray emission and radio relics, for instance, select for younger mergers \citep{2013ApJ...765...21S}.  This may cause a prevalence of positive offsets in observational samples of mergers.  The merger phase of some observed on-going mergers, however is unclear.  For such mergers, determining the merger phase may provide valuable discriminating power between models.


\subsection{Observed Equal Mass Mergers}
\label{sec:observability:observed_mergers}

\begin{table*}
\caption{Observed Mergers: Parameters and Offsets}\label{tab:obs_mergers}
\begin{tabular}{ l  r  l  c  c  c  c  c  c  l } \hline

\multirow{2}{*}{name} & \multicolumn{2}{c}{$M_i / M_\odot$} & \multirow{2}{*}{$\frac{M_1}{M_2}$} & \multirow{2}{*}{$N_\text{gal}$} & halo-halo & DM-galaxy & DM-galaxy & DM-BCG & \multirow{2}{*}{references} \\

& \multicolumn{2}{c}{($h_{70}^{-1} 10^{15} $M$_\odot$)} & & & separation & offset ($\Sigma_\text{gal}$) & offset ($L$) & offset \\ \hline \hline

CIZA J2242.8+5301 & N & 1.10$^{+0.37}_{-0.32}$ & \multirow{2}{*}{1.1} & \multirow{2}{*}{218} & \multirow{2}{*}{1060 $\pm$ 110 kpc} & 160 $\pm$ 130 kpc & 320 $\pm$ 185 kpc & 55 $\pm$ 95 kpc & \multirow{2}{*}{Jee et al. 2015} \\

(Sausage Cluster) & S & 0.98$^{+0.38}_{-0.25}$ && & & 220 $\pm$ 240 kpc & 240 $\pm$ 145 kpc & 85 $\pm$ 130 kpc \\ \hline

\multirow{2}{*}{El Gordo} & NW & 1.38 $\pm$ 0.22 & \multirow{2}{*}{1.8} & \multirow{2}{*}{89} & \multirow{2}{*}{740 $\pm$ 70 kpc} & 100 kpc & 15 $\pm$ 60 kpc & \multirow{2}{*}{---} & \multirow{2}{*}{Jee et al. 2014} \\

 & SE & 0.78 $\pm$ 0.2 & & & & 400 kpc & 15 $\pm$ 75 kpc & & \\ \hline

DLSCL J0916.2+2951 & S & $0.31^{+0.12}_{-0.079}$ & \multirow{2}{*}{1.8} & \multirow{2}{*}{210} & \multirow{2}{*}{1000$^{+110}_{-140}$ kpc} & 130 $\pm$ 70 kpc & \multirow{2}{*}{---} & \multirow{2}{*}{---} & Dawson et al. 2012 \\

(Musket Ball Cluster) & N & $0.17^{+0.20}_{-0.072}$ & & & & -50 $\pm$ 90 kpc & & & Dawson et al. 2013 \\ \hline

MACS J0025.4-1222 & NW & $0.26^{+0.05}_{-0.14} ~ \ddagger$ & \multirow{2}{*}{1.0} & \multirow{2}{*}{$\lesssim$200} & \multirow{2}{*}{520 $\pm$ 115 kpc} & \multirow{2}{*}{---} & 30 $\pm$ 120 kpc & \multirow{2}{*}{---} & \multirow{2}{*}{Bradac et al. 2008} \\

(Baby Bullet) & SE & $0.25^{+0.10}_{-0.17} ~ \ddagger$ & & & & & 45 $\pm$ 150 kpc & \\ \hline \hline

\multicolumn{10}{p{\linewidth}}{\footnotesize Halo-halo separations were calculated between mass peaks.} \\

\multicolumn{10}{p{\linewidth}}{\footnotesize Quoted halo masses are based on WL measurements of $M_{200}$, unless marked by a $\ddagger$, in which case the given mass is M($<$ 300 kpc) in an aperture centered around a BCG.}

\end{tabular}
\end{table*}

We now turn to a discussion of observed equal mass mergers.  While several tens of merging clusters have been discovered, only a handful have weak and/or strong lensing-derived mass maps and spectroscopically or photometrically identified cluster members, which are prerequisite for measuring offsets.  In Table \ref{tab:obs_mergers}, we have listed the subset of equal mass and near equal mass binary mergers ($M_1/M_2 < 2$) on which such detailed studies have been performed.  All clusters are massive ($M_{200} \gtrsim 10^{14}$ M$_\odot$) and classified as ``dissociative" mergers, i.e. those that have two well-separated dark matter halos and have gas that is offset from the dark halos as well as the galaxies, which is a clear indicator that a violent cluster collision has recently occured.  Most of the clusters have additional merger signatures (e.g. radio relics) that indicate that the merger axis likely lies near the plane of the sky.  The sample probes a large range in merger phase, from El Gordo, which has subclusters separated by $\sim 1 r_s$, followed by the Baby Bullet and the Sausage, both separated by $\sim 1.5 r_s$, and then the Musket Ball, at $\sim 2.6 r_s$.

The table provides halo and merger parameters as well as measurements of the dark matter-galaxy offset.  Note that we have not tabulated measurements of offset dependencies such as collision velocity, impact parameter, and halo concentration.  These values have not been systematically measured for all of these mergers; attempts to do so have yielded conflicting results.  However, reasonable ranges for most of the parameters can be derived.  There are no independent measurements of the concentration of the mass/galaxy distributions in clusters.  The number of confirmed cluster members and the noise in weak lensing mass maps are too low to permit a fit to their distributions with the concentration as a free parameter.  Thus the median mass-concentration relation is used instead.  Given that we do not expect the scatter in the mass-concentration relation to strongly affect the size of offsets, we consider our simulations with the fiducial (median) concentration to be a sufficient comparison.  Impact parameters are expected to be low due to the observed x-ray morphologies, which are substantially offset from the galaxies and dark matter towards barycenter, which can only be caused by a head-on or nearly head-on collision.  For mergers in which radio relics have been observed, this is further supported by the symmetry of their radio relics' brightness distributions.

We note that for galaxy peaks we use the observed number density peaks (as opposed to luminosity density peaks, which is beyond the scope of this work) in order to make a direct comparison to our simulations.  All mergers for which such a measurement has been made have at least one significant offset.  These offsets range from 100-400 kpc---far exceeding our predictions.  The offsets cannot be explained by any of our cosmologically favored simulated mergers; the largest of these offsets cannot be explained by any of our simulated mergers, including those with unrealistic parameters.  The smallest offsets in this range also require both a high cross section $\gtrsim$ 10 cm$^2$/g and cosmologically unfavored high collision velocities.  Additional investigation will be required to interpret the discrepancy between the simulated and observed offsets properly. For example, the largest offset (in the southeast subcluster of El Gordo) has no uncertainty quoted, so we are unable to determine whether this offset is even in tension with our simulations.

Foreground/background contamination by non-member galaxies may be artificially biasing galaxy peaks.  To overcome this issue, observers have calculated offsets based on galaxy {\it luminosity} peaks rather than galaxy number density peaks.  This is driven by the fact that luminous galaxies are more tightly clustered and more likely to be spectroscopically confirmed cluster members.  In addition, observers perform other cuts to reduce foreground/background contamination.  Red sequence cuts are commonly used, and simulating this would also require a level of detail beyond the scope of the current simulations.  Photometric redshift cuts are in principle more neutral with respect to galaxy member properties, but even these will favor luminous red galaxies to some extent, as luminous red galaxies are much more likely to have well-constrained photometric redshifts.

A further issue is that real clusters are embedded in large-scale structure, and there are likely to be filaments connecting to the merging clusters more or less along the merger axis. Red-sequence luminosity maps are likely to be much more robust than number density maps against the effects of these filaments. We note that the El Gordo southeast number density offset is indeed in the direction expected from this effect.

Although all these issues are worth noting, ultimately shot noise may be one of the most important factors in determining the size of the observed offsets.  We used 6000 galaxies per subcluster and found that typical offset uncertainties were at most $\sim$20 kpc with our 2D analysis.  Fewer galaxies are used in a typical observational analysis, depending on the strictness of the cuts. At the strict end, requiring spectroscopic member confirmation typically yields $\sim$100 members per subcluster. Extrapolating by $\sqrt{n}$ then yields $\sim$160 kpc uncertainties, implying that offsets of $\sim$200 kpc would not be in substantial tension with zero.  Observers can increase the number of galaxies by factors of several by employing photometric cuts or a concerted spectroscopic study, but they may not be able to reach 6000 galaxies per subcluster without inviting substantial foreground/background contamination.  Thus, the uncertainty in number density peak location is unlikely to improve to better than several tens of kpc---too large to detect the offsets we find here unless a large ensemble of clusters is used.

The uncertainty in luminosity density peak location---preferred by observers---may be substantially smaller, but we are as yet unable to compare these directly to simulations (Ng et al., in prep).  We have listed such offsets for the sample of observed equal and near equal mass mergers in Table \ref{tab:obs_mergers}.  The results are again mixed.  Both the El Gordo and Baby Bullet mergers exhibit smaller $O$(10) kpc offsets that are not significant.  The Sausage Cluster, on the other hand, still exhibits large $O$(100) kpc offsets---in fact, offsets in both subclusters are significant and are larger than number density-based offsets.

Nevertheless, our conclusion is that some of the observed number density-based offsets are significantly larger than those we expect for a SIDM hypothesis, most likely due to observational uncertainties.  For the systems in which offsets are consistent with zero, the uncertainties are much larger ($\sim$100 kpc) than the maximum size of offsets we expect near current SIDM cross section limits ($\sim$10 kpc for 1 cm$^2$/g).  A dramatic reduction in observational systematics is required to achieve competitive SIDM cross section constraints with offsets.


\section{BCG MISCENTERING IN MERGED REMNANTS}
\label{sec:bcg_miscentering}

\begin{figure}
\centering
\includegraphics[width=\columnwidth]{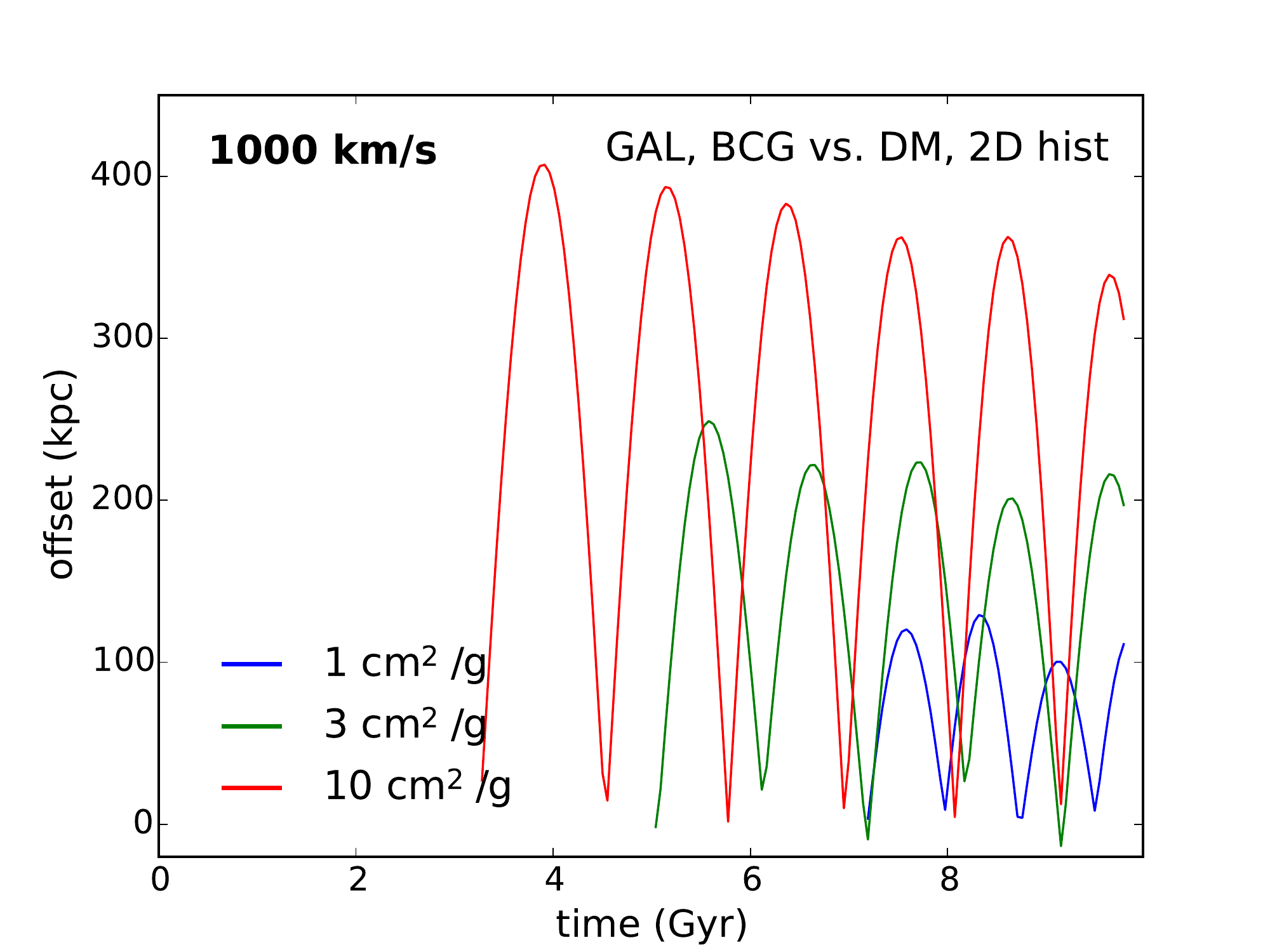}
\caption{BCG oscillations post-coalescence.  Shown are BCG-DM offsets in a head-on merger with a cosmological velocity $v_\text{4Mpc}$ = 1000 km/s and the fiducial concentration $c_\text{DM}$ = $c_\text{D08}$, modeled with various cross sections.  The BCG oscillations are about 100, 200, and 300 kpc for $\sigma_\text{SI}/m_\chi$ = 1, 3, and 10 cm$^2$/g.}
\label{fig:bcg_oscillations}
\end{figure}

\begin{figure}
\centering
\includegraphics[width=\columnwidth]{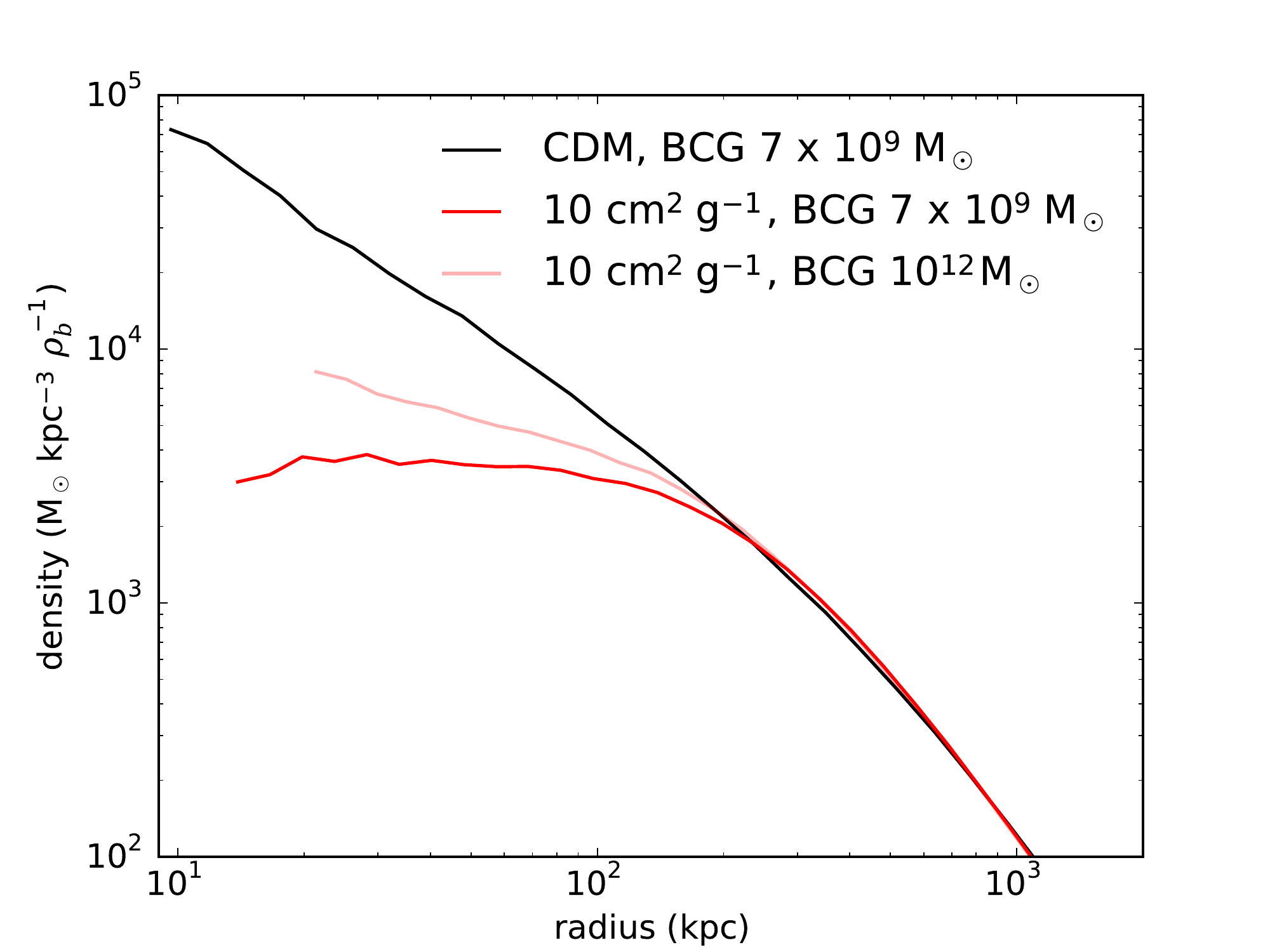}
\caption{The density profile of a SIDM halo (10 cm$^2$/g) with a massive, 10$^{12}$ M$_\odot$ BCG (pink), compared to that of a SIDM halo with the fiducial 7 $\times 10^7$ M$_\odot$ BCG (red) and a CDM halo (black), also with the lower fiducial BCG mass.}
\label{fig:bigBCG-density}
\end{figure}

\begin{figure*}
\centering
\includegraphics[width=\columnwidth]{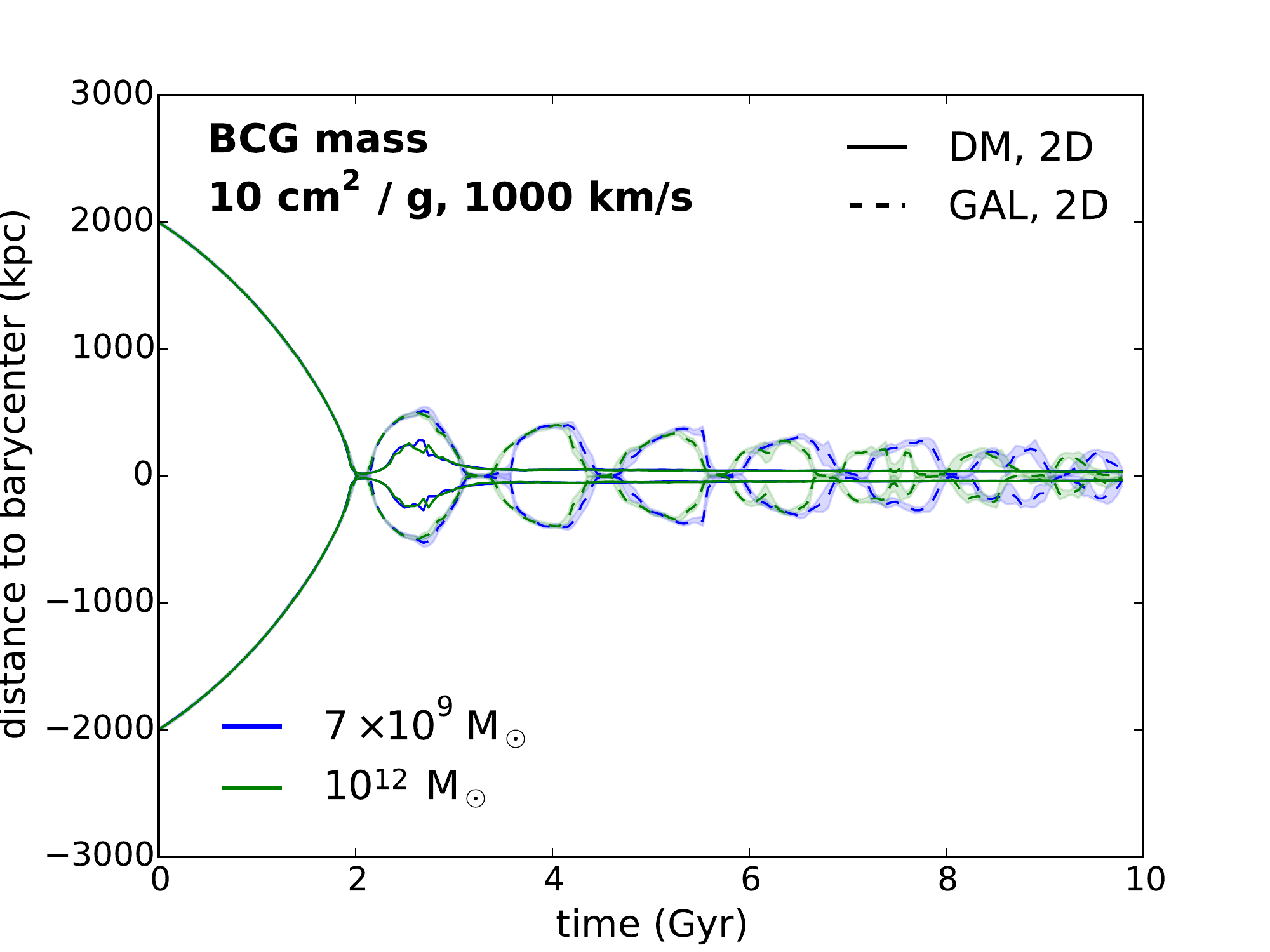}
\includegraphics[width=\columnwidth]{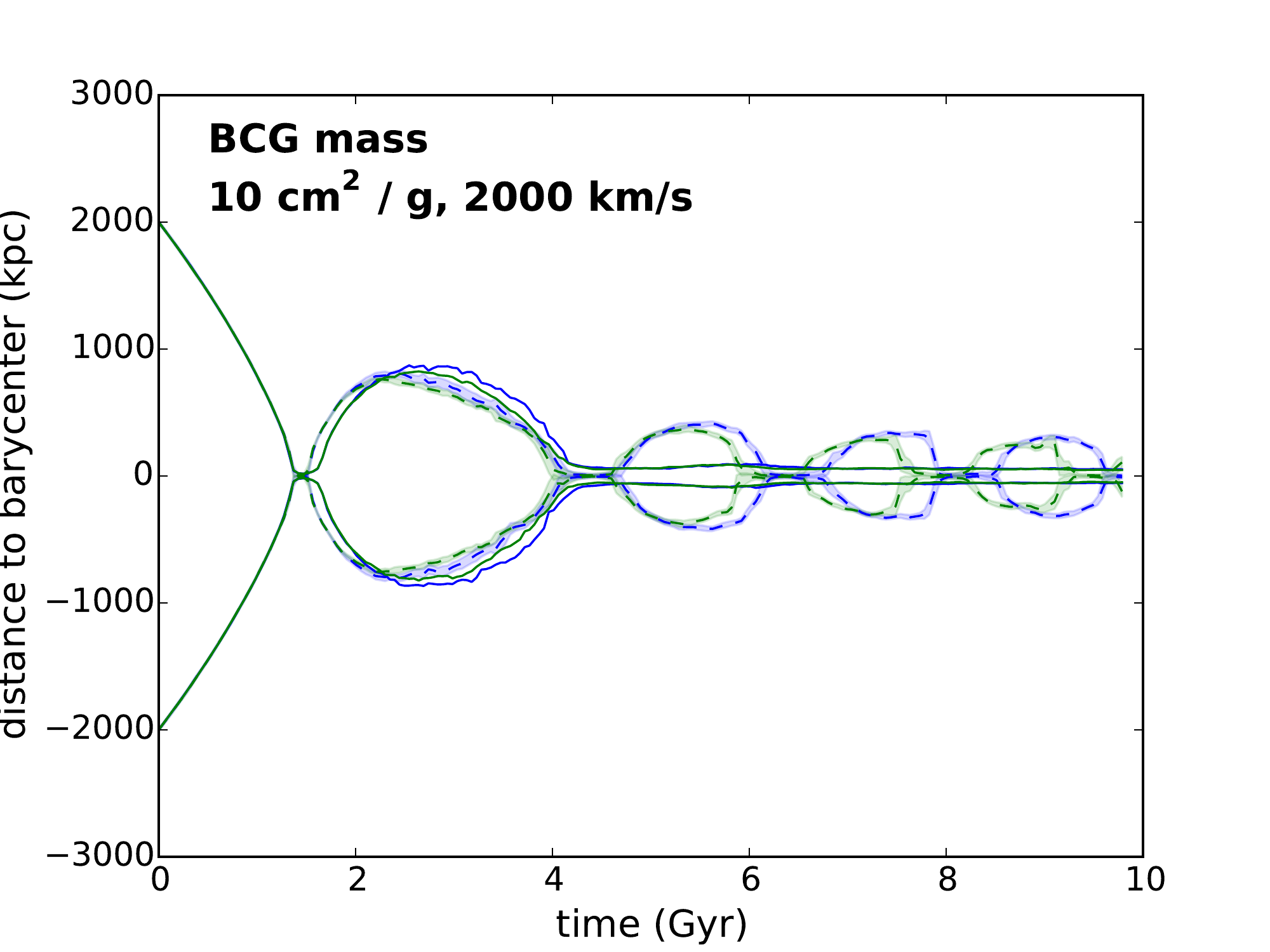}
\caption{Dependence of galaxy offsets on the mass of the BCG.  The evolution of dark matter and galaxies peaks are shown on for a 10 cm$^2$/g merger with initial velocity $v_\text{4Mpc}$ = 1000 km/s (left) and 2000 km/s (right).  Two BCG masses have been simulated, the fiducial 7 $\times 10^9$ M$_\odot$ BCG, and 10$^{12}$ M$_\odot$, the approximate range of observed BCG masses.}
\label{fig:rc-bigBCG}
\end{figure*}

While the existence of observable galaxy-dark matter peak offsets is not a generic prediction of SIDM mergers, we have seen that self-interactions can dramatically change the evolution of a merger.  In particular, we have noted that the galaxy and BCG distributions oscillate following coalescence of SIDM halos (see Fig. \ref{fig:prototype}), a behavior unseen in CDM mergers.  In this section, we explore the possibility of exploiting these oscillations to produce constraints on $\sigma_\text{SI}/m_\chi$.  If one assumes that similar behavior occurs in unequal mass mergers, the observed miscentering of BCGs in clusters require $\sigma_\text{SI}/m_\chi \lesssim$ 0.1 cm$^2$/g---the tightest constraint yet.

\subsection{Predictions}

Clusters retain information about $\sigma_\text{SI}/m_\chi$ in the dynamics of their galaxy distributions after coalescence.  In CDM mergers, galaxy distributions and the dark matter halos they are embedded in coalesce simultaneously.  After the systems relax to equilibrium, the galaxies are coincident with the dark-matter distribution, and a central galaxy typically lies at the deepest part of the cluster's potential well \citep[][Ng et al., in prep.]{2014ApJ...786...79M}.  However, in SIDM mergers (even with cross sections as low as 1 cm$^2$/g), in which self-interactions accelerate the coalescence of the dark matter halos, the galaxies continue to oscillate around barycenter.  We see this in all the simulations we presented in Sec. \ref{sec:offsets} (Figs. \ref{fig:rc-v0_dmm_2D}, \ref{fig:rc-cDM}, \ref{fig:rc-b}).  Such systems would be observed as clusters with relaxed dark matter and X-ray gas populations but with miscentered central galaxies (e.g., the BCG) and/or galaxy distributions.  The orbits of the BCGs, in particular, \emph{decay minimally}---they oscillate with a characteristic amplitude about the size of the core of the coalesced dark matter halos, in which the density gradient is zero, rendering nonexistent the dynamical frictional force responsible for orbital decay \citep{2006MNRAS.373.1451R}.  Fig. \ref{fig:bcg_oscillations} shows BCG-DM offsets following coalescence for the cosmological $v_\text{4Mpc}$ = 1000 km/s, head-on, $c_\text{DM} = c_\text{D08}$ merger, modeled with different self-interaction cross sections (e.g. middle row of Fig. \ref{fig:rc-v0_dmm_2D}).  The BCG's orbital amplitude is $\sim$100, 200, and 300 kpc for self-interaction cross sections of 1, 3, and 10 cm$^2$/g, respectively, of order $\sim 0.1 R_\text{vir}$.  The stability/persistence of dark matter cores implies that this ``core sloshing" can be sustained indefinitely, even after the cluster relaxes; in Fig. \ref{fig:bcg_oscillations}, one can see that the orbits do not decay by more than about 15\% for several Gyr after coalescence.  This naturally raises two questions: how robust are our predictions of core sloshing?  And can observations of the offset between BGCs and other measures of the center of a relaxed halo be used to constrain the SIDM cross section?

In Sec. \ref{sec:offsets}, we showed core sloshing for simulations in which galaxies each had the same, small mass.  However, BCGs can be up to several $10^{12} M_\odot$---and can thus have a dynamically important role at the centers of clusters \citep{2013ApJ...765...25N}.  In Fig. \ref{fig:bigBCG-density}, we show how the density profile of a DM halo is affected if a massive BCG is dropped into the center of an equilibrium self-gravitating halo from our initial conditions, which is then allowed to evolve for 3 Gyr.  We find that the BCG distorts the halo density profile.  As explained by \citet{2014PhRvL.113b1302K}, this is because SIDM acts as a fluid settling into hydrostatic equilibrium in an external potential, in this case sourced by the BCG.  However, when we collide two of these halos, we find that a broad SIDM core consistent with our low-mass BCG simulations forms, and the BCG orbits decay only slowly.  Thus, we expect that the core sloshing phenomenon is a robust prediction of the remnants of equal-mass mergers of SIDM halos, with the amplitude of the BCG excursions from the halo centers tracing the core size.  Likewise, the galaxies bound to the core also slosh along the merger axis, so that both the BGC and the peak of the galaxy distribution are typically significantly offset from the halo center.

In order to determine if core sloshing persists in more realistic mergers with gas, substructure, unequal mass ratios, and complex merger histories, further numerical work is required.

\subsection{Observations}

We can use observations of galaxy clusters to set SIDM cross section constraints based on the excursions of BCGs and the galaxy populations as a whole from the center of the halos.  There are a number of recent studies on the centering of galaxy clusters, to explore what baryonic tracer most closely traces the center of the dark-matter halo for stacked weak lensing analyses \citep{2012ApJ...757....2G}, to explain the small-scale clustering of galaxies \citep{2011MNRAS.410..417S, 2015MNRAS.446..578G, 2015MNRAS.452..998H}, and to cross-calibrate Sunyaev-Zeldovich observations of clusters with X-ray gas properties and optical richness \citep{2013ApJ...767...38S, 2015MNRAS.454.2305S}.  What these studies show is that the BCG is, on average, highly aligned with the center of the halo.  In a survey of all nearby Abell clusters, \citet{2014ApJ...797...82L} find that the median offset between BCGs and the peak of the X-ray flux is only 10 kpc.  There is a long tail to high separations that \citet{2014ApJ...797...82L} attribute to ongoing merging activity \citep{2014ApJ...786...79M}.  While Sunyaev-Zeldovich clusters show a broader distribution in offsets between BCGs and the X-ray gas, \citet{2016MNRAS.457.4515R} show that the more relaxed clusters have smaller offsets between the two than the cluster population as a whole.  Interestingly, \citet{2012ApJ...757....2G} find that the strongest cluster weak lensing signal comes from identifying the bright galaxy closest to the X-ray flux peak as the cluster center.  This choice beats out the X-ray peak, and various measures based on the ensemble galaxy population (luminosity-weighted centroid, stellar-mass-weighted centroid, etc.) as the choice of halo center.

All of these lines of evidence suggest that BCGs are nearly coincident with the halo center for relaxed clusters in the real Universe, that most BGCs are offset by at most a few tens of kpc.  While we have not yet simulated cross sections below 1 cm$^2$/g, nor have we determined if this sloshing mode is a generic feature of mergers with a large mass ratio between subclusters, we can estimate what order-of-magnitude limit we might be able to obtain on the SIDM cross section if the trends from our simulations hold for those other systems.  In our simulations, the BCG and the peak of the galaxy distribution spend most of their time away from the exact cluster center near apocenter, which is approximately the size of the core.  Thus, we estimate the cross section that corresponds to a core size $\lesssim 10 $ kpc.  For our $10^{15}M_\odot$ mergers, the merger remnant has a core size of $\sim 100$ kpc for $\sigma_\text{SI}/m_\chi = 1\hbox{ cm}^2/\hbox{g}$.  The core size is set by the per-particle scattering rate $\Gamma \sim (\sigma_\text{SI}/m_\chi) \rho v \gtrsim H$, where $H$ is the Hubble rate, $\rho$ is the CDM density profile of a halo, and $v$ is the typical relative speed between particles \citep{2013MNRAS.430...81R}.  For $\sigma_\text{SI}/m_\chi \lesssim 1 \hbox{ cm}^2/\hbox{g}$, the core lies within the NFW scale radius, where $\rho \propto r^{-1}$.  Note that $\rho \propto r^{-2}$ for NFW profiles in the region in which cores form for $\sigma_\text{SI}/m_\chi \approx 1-10$ cm$^2$/g, explaining the relative amplitudes of oscillation in Fig. \ref{fig:bcg_oscillations}.  Thus, if the core size of the merger remnant in our simulations ($M\approx 2\times 10^{15} M_\odot$) is $\sim 100$ kpc for 1 cm$^2/\hbox{g}$, then we expect a core of 10 kpc to correspond to $\sim 0.1 \hbox{ cm}^2/\hbox{g}$.  For velocity-independent SIDM models, the core size increases with halo mass, thus we expect the strongest SIDM constraints to come from the largest relaxed clusters.

As noted above, further work is required in order to determine if core sloshing persists in more realistic mergers with gas, substructure, unequal mass ratios, and complex merger histories.  If core sloshing remains a robust prediction of SIDM, a better statistical analysis of BCG and galaxy miscentering may lead to the tightest constraints yet on velocity-independent SIDM.


\section{REPURPOSING MERGER OBSERVATIONS: BETTER CONSTRAINTS}
\label{sec:betterconstraints}

While the existence of observable galaxy-dark matter peak offsets is not a generic prediction of SIDM mergers, we have seen that self-interactions can dramatically change the evolution of a merger.  Such differences can provide alternative---and perhaps stronger---means to constrain the self-interaction cross section.  We discuss observables in ongoing mergers whose relative merit need to be explored in more detail in future work.


\subsection{Apparent Merger Fraction}
\label{subsec:merger_frac}

One key feature of equal mass SIDM mergers is the shortening of the merger timescale with even small cross sections, and that---if merger and halo parameters encourage self-scatters to result in momentum loss (e.g. low merger velocities, high concentrations, and low impact parameter)---the timescale can be shortened to the point that the dark matter halos coalesce upon contact.  In the limit of high self-interaction cross sections, one could imagine that all mergers---accounting for the intrinsic scatter in merger and halo conditions---would result in immediate coalesence, and we would never observe an evidently post-collision merger with well-separated dark matter halos.  Thus an (admittedly weak) upper bound can be placed on the self-interaction cross section simply by observing clusters evidently in the process of merging in which their dark matter halos have not yet merged.  Our results suggest that $\sigma_\text{SI}/m_\chi >$ 10 cm$^2$/g are ruled out, given that we do observe equal mass mergers with well-separated dark matter halos.

At lower, astrophysically allowed cross sections for which we still expect to see post-collision, ongoing mergers with well-separated dark matter halos, we may still be able to make use of the fact that systems in the process of merging should be rarer in SIDM cosmologies.  For the $v_\text{4Mpc}$ = 1000 km/s mergers, we find that the merger timescale is shorted by $\sim$10\%, $\sim$20\%, and $\sim$70\% for $\sigma_\text{SI}/m_\chi$ = 1, 3, and 10 cm$^2$/g compared to CDM (these numbers also scale with $v_\text{4Mpc}$).  Assuming that in SIDM cosmologies the merger rate is unchanged, these numbers give the relative likelihood one would see a merger under these cross sections.  Based on these timescales, combined with cosmological distributions of merger and halo parameters, one can estimate the number of apparent binary mergers that one  expects to see in a sample of clusters for a given self-interaction cross section.  However, we note that more realistic estimates of the apparent merger fraction must account for gas dynamics, which has been shown to dramatically modify estimates of merger timescales \citep{2016ApJ...820...85Z}.  We recommend that cosmological hydrodynamical simulations be undertaken to determine the relative incidence rates of mergers as a function of SIDM cross section.


\subsection{Halo-Halo Separation}

In addition to shortening the merging timescale, the enhanced momentum loss caused by self-interactions reduces separations at apocenter.  In Sec. \ref{sec:offsets:sigmavel}, our simulations show that under changes in $\sigma_\text{SI}/m_\chi$, the apocentric separation can vary on order of a few 100s of kpc for a given merger velocity and up to on order a Mpc for the highest velocity mergers.  We note, however, that apocentric separation is strongly controlled not just by cross section and velocity but also the halo concentration, which alone can also push the apocentric separation out by as much as a factor of 2.  For our $v_\text{4Mpc}$ = 1000 km/s mergers with the fiducial concentration, the first apocenter is shortened by $\sim$10\%, $\sim$20\%, and 100\% (i.e. halos never separate post-collision) for $\sigma_\text{SI}/m_\chi$ = 1, 3, and 10 cm$^2$/g.

At minimum, projected separations between halo centers provides a lower bound on SIDM cross sections.  The largest separation between observed near-equal mass mergers is the Sausage cluster, which shows a peak-to-peak separation of $\sim 2.6 r_s$ (corresponding to $\sim$1 Mpc; see Table \ref{tab:obs_mergers}).  Conservatively, this implies that cross sections greater than 10 cm$^2/$g can be excluded.  To obtain tighter constraints, a statistical study of halo-halo separations in cosmological simulations, weighted by the probability of discovering systems during merger (either by X-ray offsets or radio relics), may provide a constraint on the self-interaction cross section.


\subsection{Mass-to-Light Ratios}
\label{subsec:MLRs}

\begin{figure*}
\centerline{
\includegraphics[width=0.75\columnwidth,trim=5mm 1.5cm 2cm 0cm,clip=true]{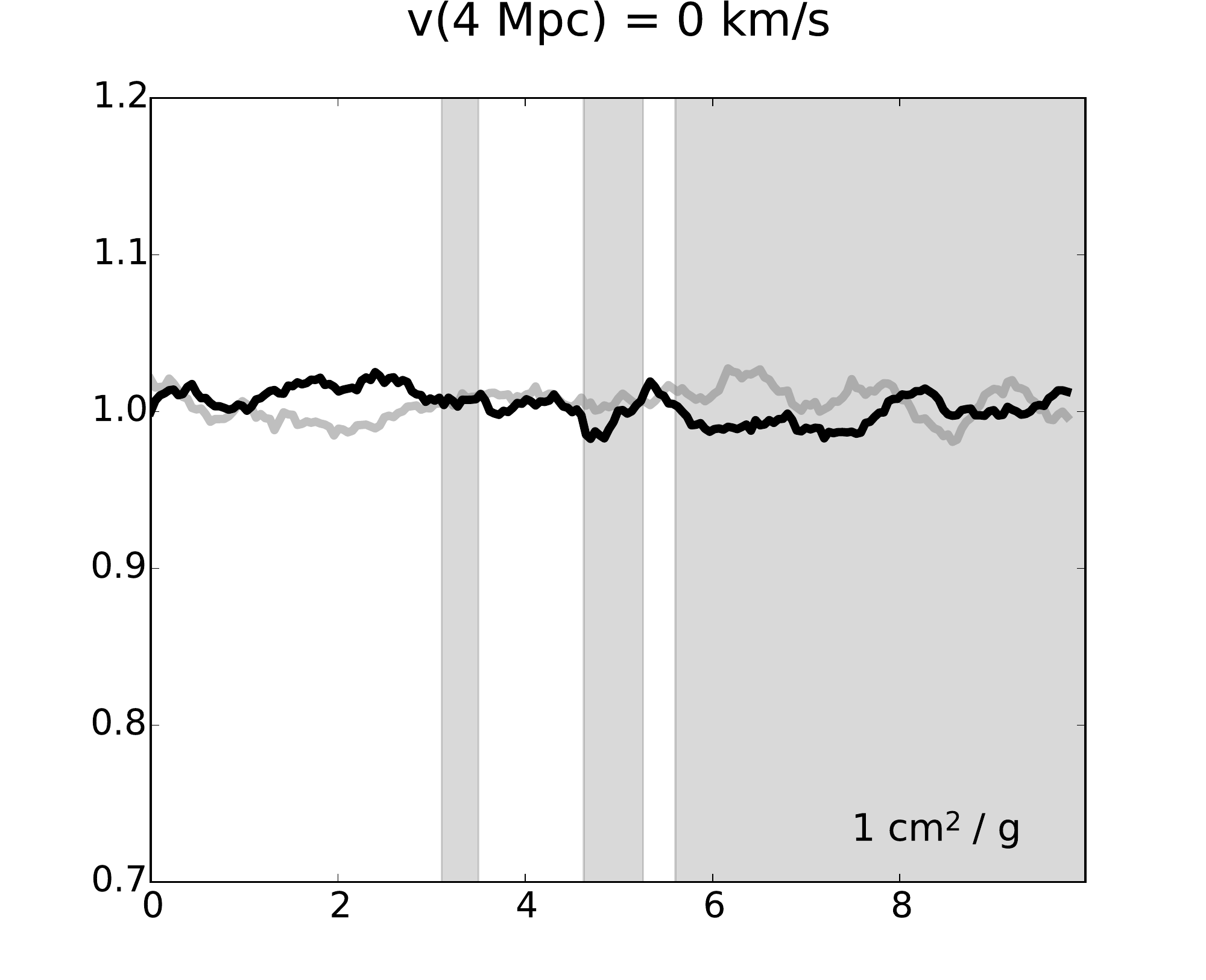}
\includegraphics[width=0.75\columnwidth,trim=1cm 1.5cm 2cm 0cm,clip=true]{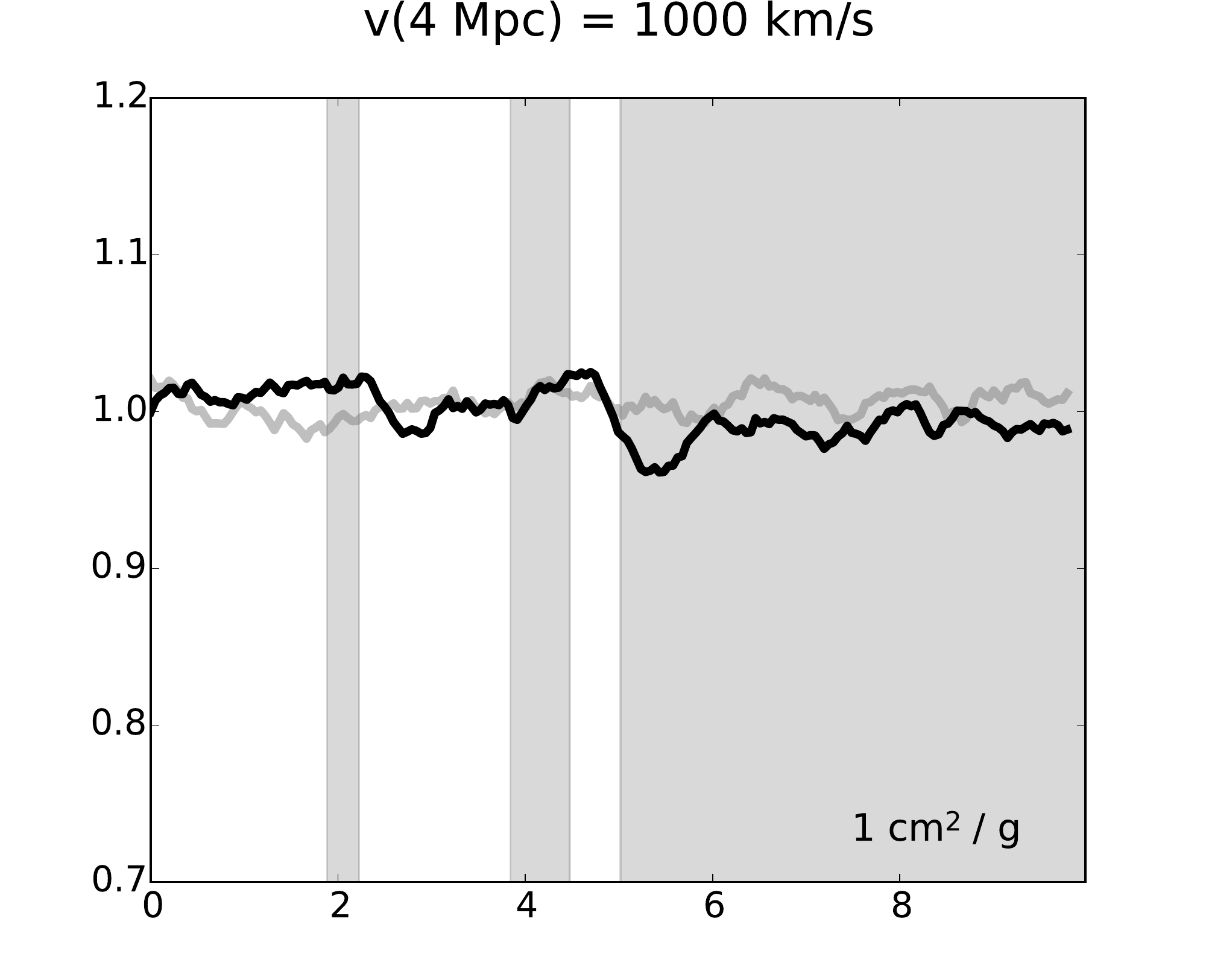}
\includegraphics[width=0.75\columnwidth,trim=1cm 1.5cm 2cm 0cm,clip=true]{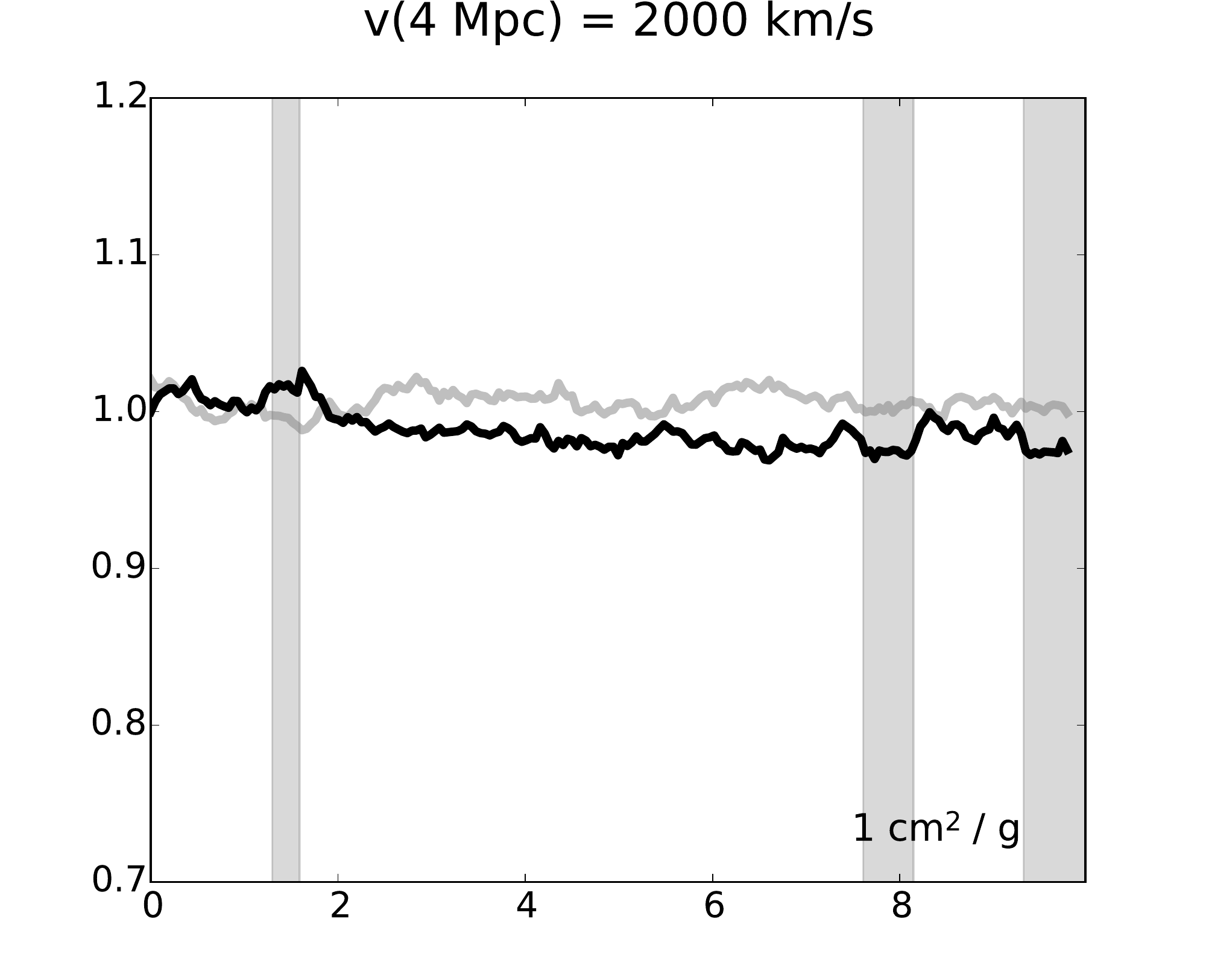}}
\centerline{
\includegraphics[width=0.75\columnwidth,trim=5mm 1.5cm 2cm 1cm,clip=true]{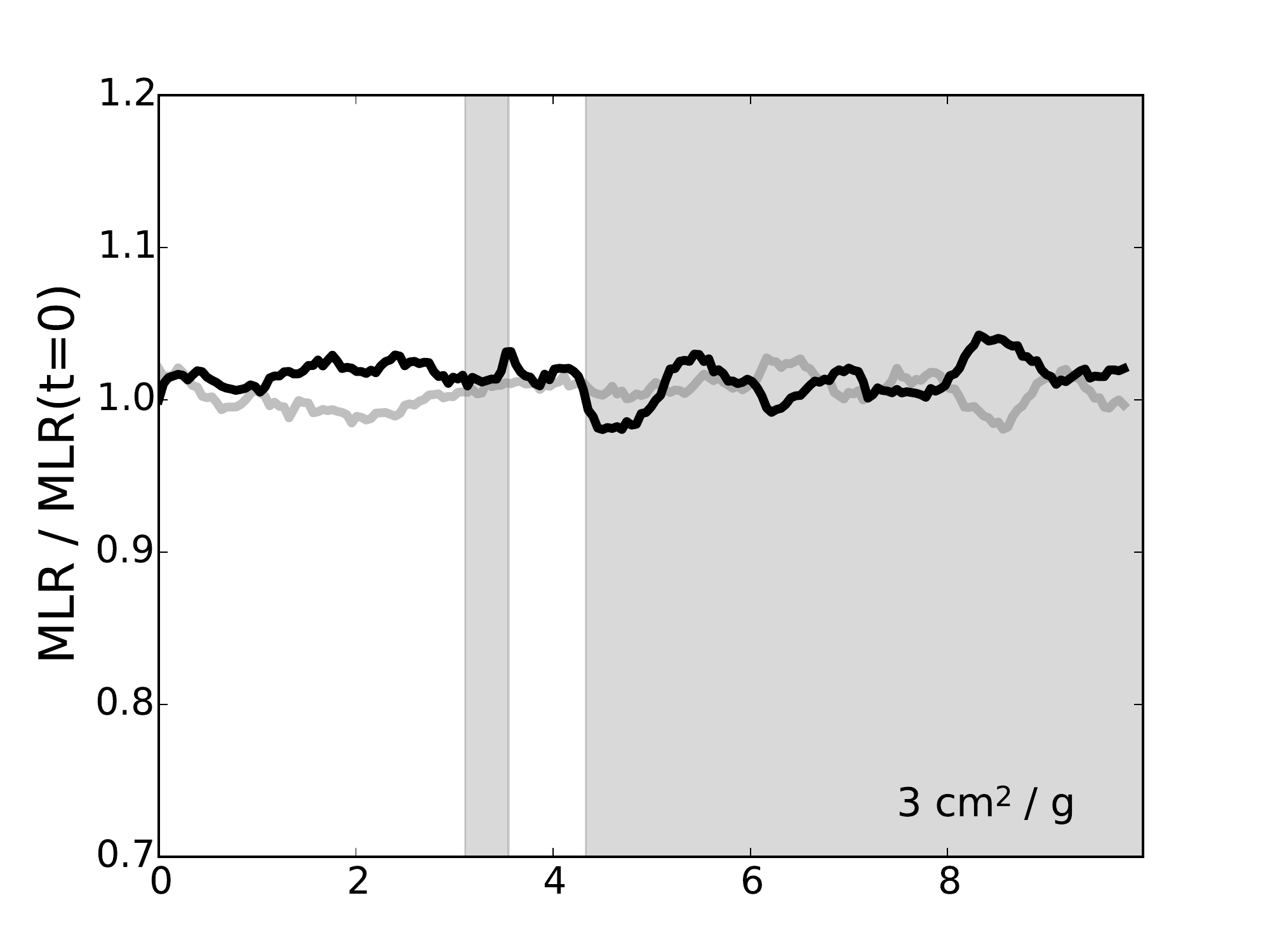}
\includegraphics[width=0.75\columnwidth,trim=1cm 1.5cm 2cm 1cm,clip=true]{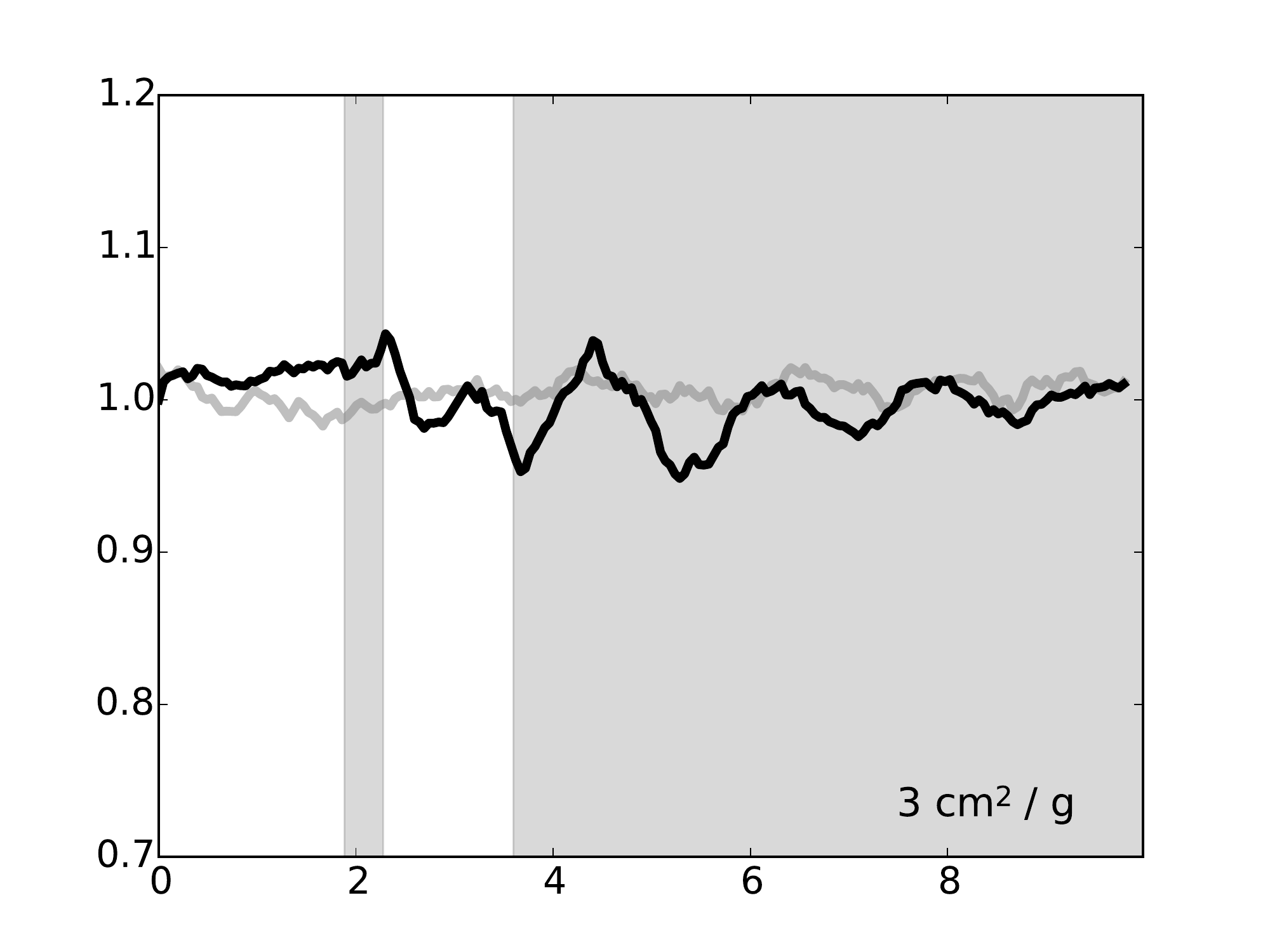}
\includegraphics[width=0.75\columnwidth,trim=1cm 1.5cm 2cm 1cm,clip=true]{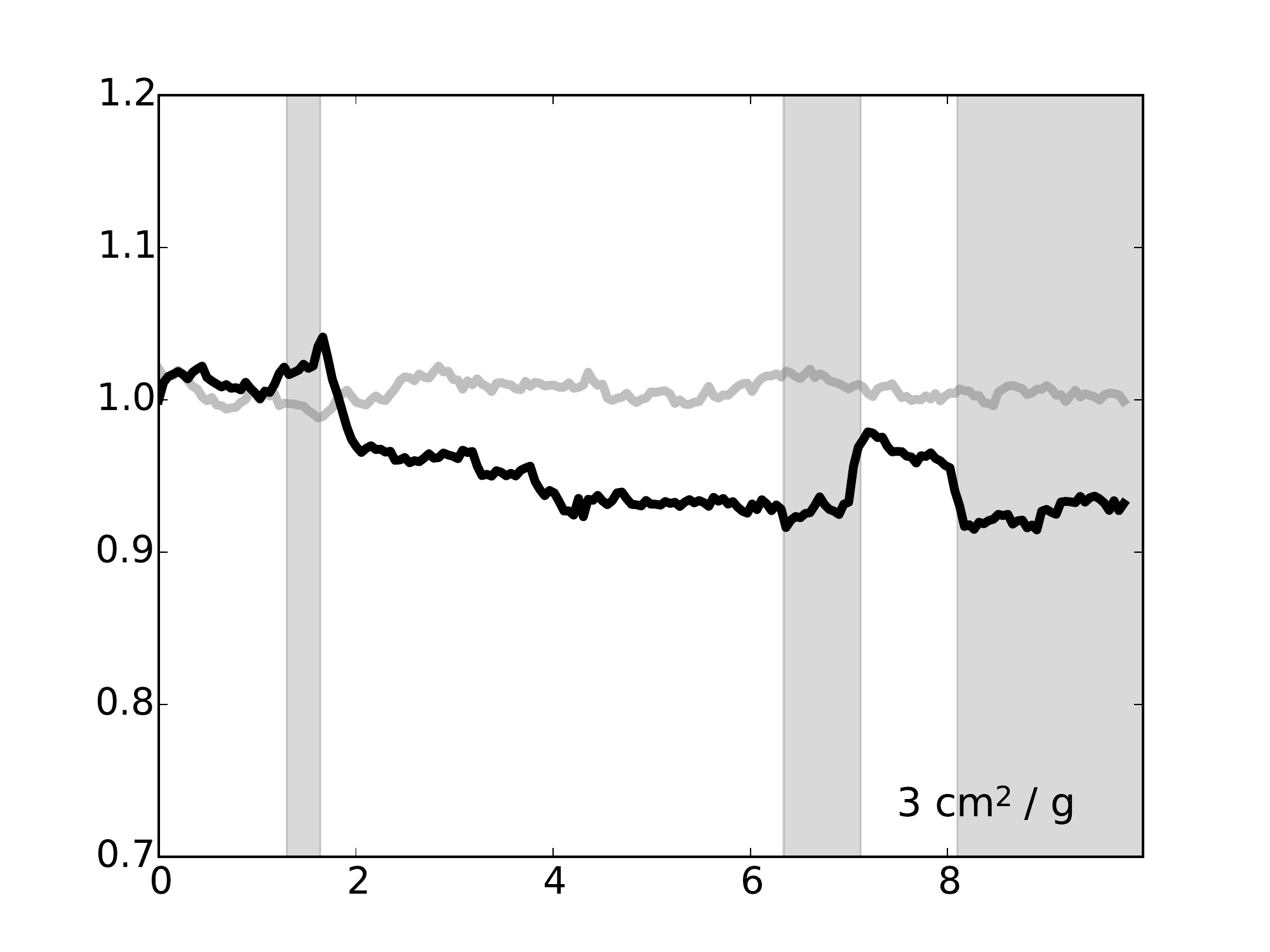}}
\centerline{
\includegraphics[width=0.75\columnwidth,trim=5mm 0mm 2cm 1cm,clip=true]{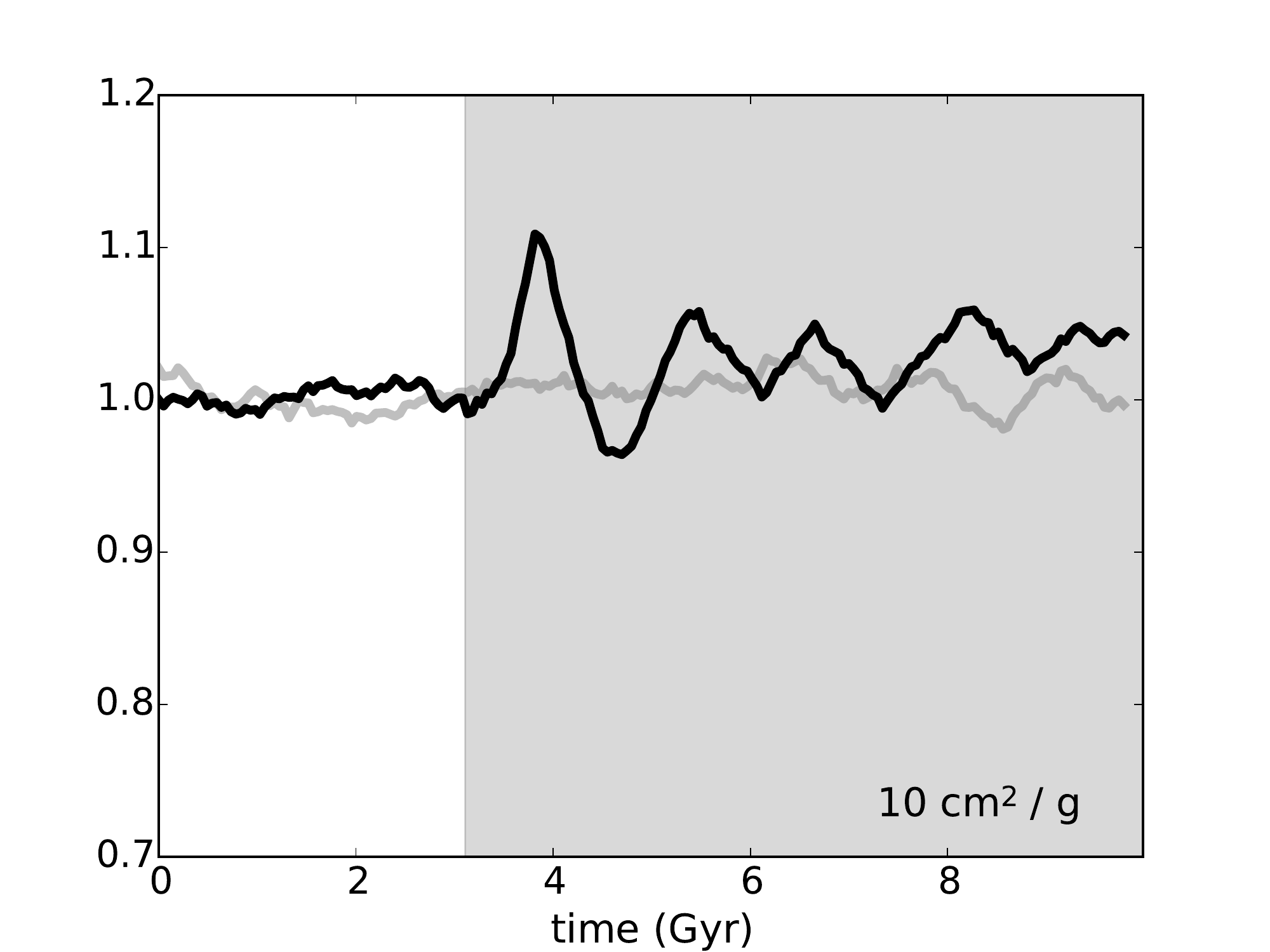}
\includegraphics[width=0.75\columnwidth,trim=1cm 0mm 2cm 1cm,clip=true]{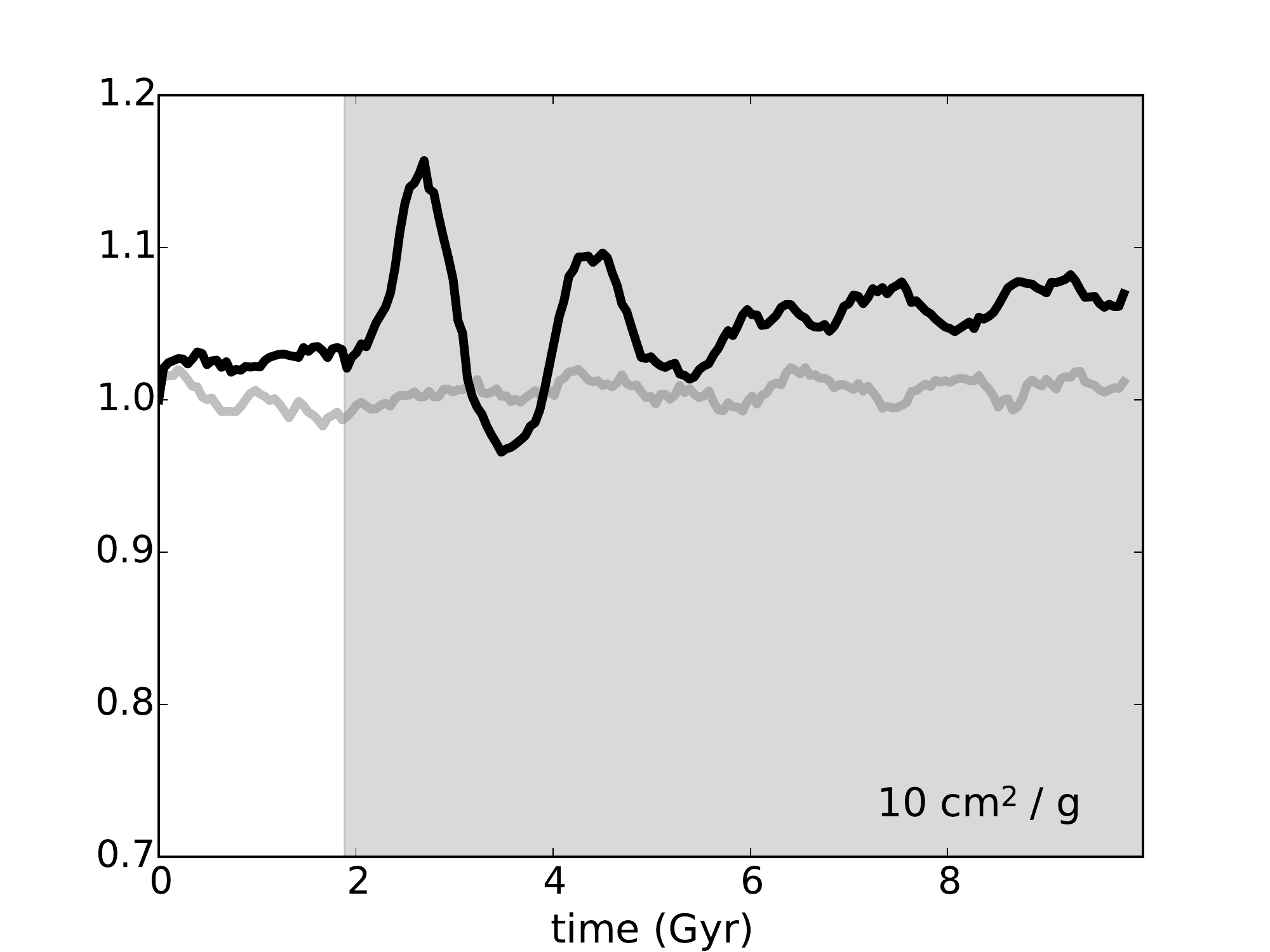}
\includegraphics[width=0.75\columnwidth,trim=1cm 0mm 2cm 1cm,clip=true]{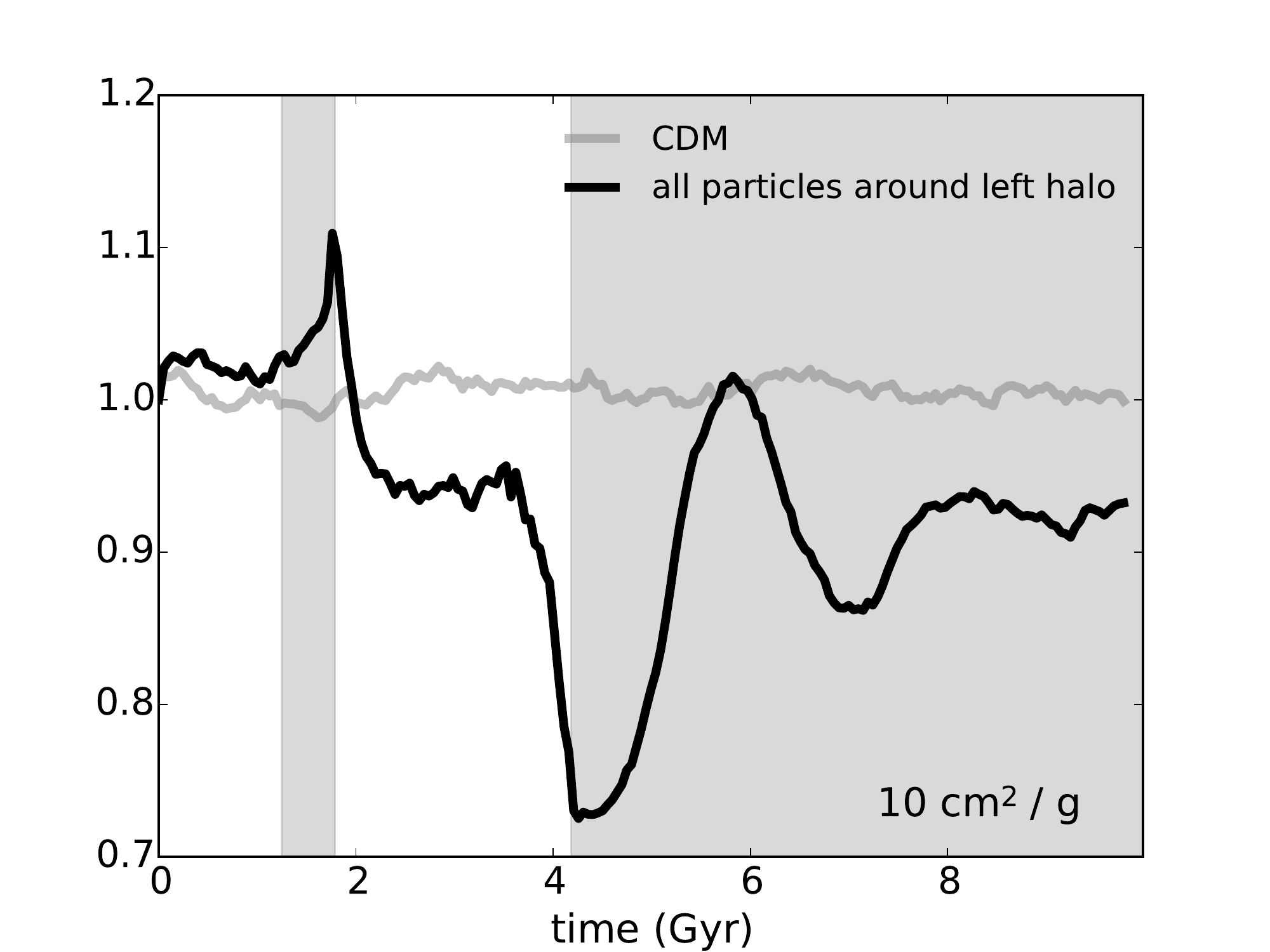}}
\caption{Mass-to-light ratios over the total ratio at $t$ = 0.  Rows represent different cross sections, from $\sigma_\text{SI}/m_\chi$ = 1, 3, and 10 (bottom) cm$^2$/g.  The evolution of the mass-to-light ratio in CDM mergers is overplotted on each panel in gray. Columns represent different initial velocities, with $v_\text{4Mpc}$ = 0 (left), 1000, and 2000 (right) km/s.  Shaded in gray are regions where the dark matter peaks were separated by less than 600 kpc.}
\label{fig:mlr}
\end{figure*}

It has been suggested that dark matter self-scatters will cause halos to preferentially lose mass over the luminous galaxies, causing a decrease in the mass to light ratio \citep[e.g.,][]{2008ApJ...679.1173R}.  Based on the arguments presented in Section \ref{sec:theory}, one expects that, like galaxy-dark matter offsets, the magnitude of this change will be dependent on merger and halo parameters.  The largest decrement in the mass-to-light ratio is expected to occur in high cross section, high velocity mergers, in which the largest fraction of dark matter should be ejected.  Assuming $\sigma_\text{SI}/m_\chi = $ 10 cm$^2$/g and a collision velocity between 4200 -- 4400 km/s ($v_\text{4Mpc}$ = 2000 km/s), about 50--60\% of the dark matter is expected to scatter out.  However, much of the dark matter that escapes is likely captured by the other halo.  In addition, the galaxies are lost with the dark matter in response to the increased shallowness of the gravitational potential.

To study the net change in the mass to light ratio due to these effects, we calculated the mass-to-light ratio in 600 kpc apertures surrounding each dark matter peak in our simulated 2D mass maps in order to mimic observations, which are shown as black solid lines as a fraction of the overall mass-to-light ratio ($M_\text{DM}/M_\text{gal} = 50$) in Figure \ref{fig:mlr}.  In CDM mergers, overplotted on each panel in gray, there are small, sub-percent level fluctuations in the mass-to-light ratio during the merger.  The introduction of self-interactions causes the mass-to-light ratio to decrease after pericenter, and only increase after the subclusters return to pericenter.  This implies that \emph{in equal mass mergers, the mass-to-light ratio is a constraint that performs best for advanced mergers}, preferrably when the clusters are approaching second pericenter.  After the dark matter halos coalesce, the mass-to-light ratio oscillates as the galaxies oscillate around barycenter, changing at the percent level for 1 and 3 cm$^2$/g, but as much as 10-15\% with 10 cm$^2$/g.  This is due to the size of the galaxy oscillations relative to the size of the aperture---for the lower cross sections, the galaxies oscillate within the 600 kpc aperture, and thus we do not observe as large an oscillation amplitude in the mass-to-light ratio, if at all.

In our simulations, the mass-to-light ratio never changes by more than $\sim$20\%, which is smaller than the intrinsic scatter in the intrinsic mass-to-light ratio of relaxed clusters \citep[$\sim$30\%; ][]{2000ApJ...530...62G, 2007ApJ...663..150M}, and much smaller than the intrinsic scatter in the mass-richness relation \citep{2009ApJ...699..768R}.  Thus, we do not expect the mass-to-light ratio to have any constraining power for equal-mass mergers.  Further compounding the problem is the uncertainty in the extent to which cluster mergers induce or quench star formation in member galaxies, which may impact cluster luminosities.

We note that in non-equal mass mergers, the mass-to-light ratio may not provide stronger constraints, despite the greater likelihood that the dark mass is stripped from the lower-mass cluster.  As in equal mass mergers, the galaxies will likely follow the unbound dark matter, and will likely also be lost to the halo.  However, one may still expect a dip, the magnitude of which will need to be investigated through more detailed numerical calculations.



\section{DISCUSSION}
\label{sec:discussion}

What cluster observations may provide the best SIDM constraints?  We have demonstrated that equal mass mergers in SIDM cosmologies cannot generically produce offsets $\gtrsim$ 50 kpc except under cosmologically disfavored high merger velocities and self-interaction cross sections.

\subsection{Future Prospects for Offsets}

\subsubsection{Offsets in Unequal Mass Mergers}
\label{sec:discussion:unEMMs}

One natural direction to turn to is unequal mass mergers, which are more common due to the sharply falling halo mass function.  The classic merging cluster for SIDM studies, the Bullet Cluster, for example, has a mass ratio of about 8:1 \citep{2008ApJ...679.1173R}, the largest among the cluster mergers for which an SIDM study has been undertaken.  How might offsets evolve in such systems?

We can again turn to our toy model in Sec. \ref{sec:theory} but consider the merger \emph{from the perspective of the less massive cluster}, which is more easily perturbed.  Two key differences must be taken into account.  Compared to a merger with an equal mass partner, a merger with a more massive partner will cause (1) the cluster to experience a higher collision velocity and (2) the dispersion in the relative particle velocities to be larger than in a merger with an equal mass partner.  Both of these cause greater mass loss in the less massive cluster: the higher collision velocity increases the likelihood that a self-interaction unbinds the interacting particles from the halo, and the higher relative dispersion increases the number of high-velocity particles that are easily unbound, lowering the energy required to unbind the speedy particles.

More quantitatively, if the mass ratio between the clusters is $q = M_1 / M_2$ (such that $q > 1$), the collision velocity roughly scales as
\begin{equation}
v_\text{col} \sim \sqrt{ M_1 ( 1 + q^{-1}) }.
\end{equation}
The relative velocity dispersion equals the velocity dispersion of the two clusters added in quadrature:
\begin{equation}
\sigma_v = \sqrt{ \sigma_{v,1}^2 + \sigma_{v,2}^2}  = \sigma_{v,1} \sqrt{1 + q^{-2/3}}
\end{equation}
assuming that $\sigma \sim v_\text{esc} \sim M^{1/3}$.  In the limit where $\sigma_{v,1} \gg \sigma_{v,2}$, $\sigma_v \approx \sigma_{v,1}$ (this is the limit considered by \citet{2004ApJ...606..819M}).

For a 10:1 merger in which the cluster masses are 10$^{15}$ and 10$^{14}$ M$_\odot$, the above estimates yield $v_\text{col} \approx 0.75 \, v_\text{col,EMM} \approx $ 3000 km/s and $\sigma_v \approx 0.75 \, \sigma_{v,\text{EMM}} \approx$ 600 km/s.  The escape probability for a two-particle collision is 56\%, which implies that 36\%, 66\%, or 93\% of the lower-mass cluster should be scattered out for $\sigma_\text{SI}/m_\chi$ = 1, 3, and 10 cm$^2$/g, respectively.  This implies that SIDM-induced momentum loss---the driver behind the largest offsets in our cosmological mergers---will be subdominant to mass loss.  We thus expect that \emph{offsets will be smaller for more unequal mass mergers}.  This prediction is in line with simulations of the Bullet Cluster by \citet{2016arXiv160504307R}, who found $\sim$10 kpc offsets assuming $\sigma_\text{SI}/m_\chi$ = 1 cm$^2$/g.

\subsubsection{Observational Challenges and Directions}

Thus it appears that offsets---in mergers across all mass scales and ratios---are expected to be small (typically $\ll 100$ kpc).  This is concerning given that there are several observational challenges that make precise and accurate offset measurement difficult.  For one, there are large uncertainties in measurements of galaxy number density peaks due to the low numbers of cluster galaxies typical in merging systems.  While we (and most other simulators) have modeled the galaxy distribution with $\sim10^5$ galaxies, a cluster merger typically has only several 10s to a few 100s of galaxies that are confirmed cluster members (see Table \ref{tab:obs_mergers}).  Additional cluster members can be confirmed through intensive spectroscopy, but the increase in precision afforded by a larger sample is likely at most a factor of 2.  A more promising alternative may lie in measuring mass-BCG offsets, as the uncertainty on a BCG's location is much smaller, and DM-BCG offsets are expected to be slightly larger.

Uncertainties on the dark matter peak or centroid, however, are typically larger.  Mass maps are initially derived via weak lensing, which is intrinsically limited by shape noise.  \citet{2016arXiv160504307R} found that the uncertainty in the dark matter peak position based on weak lensing shear maps could not be reduced below $\sim$40 kpc due to shape noise arising from the intrinsic ellipticities of galaxies.  However, while current weak lensing measurements are limited by large uncertainties, a number of innovative methods that, for instance, combine shape measurements with magnification, or for bluer galaxies, a Tully-Fisher measurement, may help decrease uncertainties by a factor of up to two or ten, respectively \citep{2013arXiv1311.1489H, 2014ApJ...780L..16H}.  More promisingly, strong lensing measurements may be subject to uncertainties smaller by as much as a factor of 10.

Another difficulty is due to the fact that weak lensing maps do not discriminate between dark and light matter.  A significant portion ($\sim$10\%) of the mass lies in gas, which lag behind the galaxies (and dark matter).  The gas can thus pull the weak lensing peaks towards barycenter, increasing the mass-galaxy offsets.  For example, in the Musket Ball cluster, the 130 kpc weak lensing-galaxy offset shrinks to an 80 kpc DM-galaxy offset after conservatively modeling out the gas mass \citep{2013PhDT.......211D}, alleviating some of the tension between our results and observed mergers.  However, the large shift in the Musket Ball is not necessarily representative: it reflects conservative modeling in the face of poor gas mass constraints.  Intensively studied systems with better X-ray and SZ data yielding tighter gas mass constraints are unlikely to show dramatic effects from gas mass modeling. Furthermore, the increased resolution obtained by strong lensing enables better separation of DM and gas mass contributions; a combined strong and weak lensing mass map of the Bullet Cluster by Bradac et al. (2009) is sensitive enough to reveal the gas mass morphology.

Despite these difficulties, should one attempt to measure offsets, the best systems to pursue are the most massive mergers (which should have larger offsets), preferably when near pericenter (either leaving or returning, when the largest offsets are expected).  Observationally, this will require highly resolved mass distributions via weak and strong lensing observations, X-ray and SZ maps for gas mass modeling, and extensive spectroscopy for galaxy member identification.

\subsection{The Future for Alternative Constraints}

Though it appears that prospects for placing more stringent constraints on $\sigma_\text{SI}/m_\chi$ by way of offsets are unfavorable, there are promising alternatives, which we have discussed in Sec. \ref{sec:bcg_miscentering} \&  \ref{sec:betterconstraints}.  In particular, statistical studies of observed mergers and post-mergers may be key, enabling the study of merger rates, the distribution of halo-halo separations, and BCG/galaxy miscentering.  However, there are two major concerns that need to be addressed before comparing our predictions to statistical observations.

Firstly, our predictions are based on dark matter-only simuations of isolated binary mergers.  True clusters have up to 20\% of their mass in gas, which plays an important role in merger evolution.  \citet{2016ApJ...820...85Z}, for example, have shown that the inclusion of baryons---a significantly more collisional cluster component---can decrease merger lifetimes by as much as a factor of 3.  Predictions for merger rates, halo-halo separations, and collision velocities may be significantly altered with the inclusion of baryons.  Furthermore, gas may help re-establish a density gradient at the centers of relaxed clusters, causing BCG orbits to damp and thereby erase SIDM-induced miscentering.  Comparisons of hydrodynamical simulations of CDM and SIDM mergers are thus essential to deriving realistic predictions.

Secondly, most cluster mergers are likely unequal mass mergers.  A study similar to the one done here that systematically explores SIDM merger evolution in asymmetric mergers will be key to determining whether our predictions remain robust for a generic sample of mergers.  This will enable us to interpret cosmological simulations that take into account clusters' full, complex merger histories, which may modify our predictions for core sloshing and merger evolution.  Such simulations can also help us quantify how substructure---the effects of which can be difficult to account for in weak lensing observations---introduces systematics that may make observed miscenterings difficult to interpret.  Studies of simulated mergers in a cosmological volume are also an essential prerequisite for making reliable comparisons to the statistics of observed mergers.

Also note that we have modeled the self-interaction cross section as velocity-independent.  However, alternative models posit that the cross section is velocity-dependent.  A velocity dependence with a fall-off at large velocities could produce small or no offsets on the cluster scale yet produce observable differences at galaxy scales.


\section{CONCLUSIONS}
\label{sec:conclusions}

We have run a suite of equal mass galaxy cluster merger simulations to systematically study how the formation of galaxy-dark matter offsets in velocity-independent SIDM cosmologies are affected by merger parameters.  We found that self-interactions produce small galaxy-dark matter offsets.  These offsets are much smaller than those found in many observations of equal or near-equal mass mergers, suggesting that the observed offsets are dominated by noise or systematic effects.  However, we have also found that self-interactions can significantly influence other aspects of merger evolution. These can be exploited to derive stronger constraints on the self-interaction cross section. Our main conclusions are as follows:

\begin{itemize} 

\item SIDM equal mass mergers exhibit complex phenomology that cannot be described solely by either a drag force on or explusion of dark matter; the overall evolution must be described as a combination of the two.  Particles that are captured by or exchanged between halos induce a loss in the halos' bulk momentum that can be described via a drag formalism.  Ejected particles induce mass loss, which give rise to the formation of tails.

\item Offsets between the collisional dark matter and collisionless galaxies are therefore dependent not only on the self-interaction cross section, which describes the number of dark matter interactions that occur, but also on the fate of the interacting particles, which is dependent on merger velocity, halo concentration, and impact parameter.

\item For cosmological merger and halo parameters, offsets tend to be small, $\lesssim$ 20 kpc for 1 cm$^2$/g and $\lesssim$ 50 kpc for 3 cm$^2$/g, for a narrow range of cross section and for halo masses $10^{15}M_\odot$.  For cross sections $\sigma_\text{SI}/m_\chi < 1\hbox{ cm}^2/\hbox{g}$, we expect offsets to essentially vanish, but for $\sigma_\text{SI}/m_\chi \gtrsim 10 \hbox{ cm}^2/\hbox{g}$, halos coalesce on impact unless the merger speed is unusually high.  Even for cross sections that admit offsets, offsets are transient phenomena.  They peak near pericenter passage and decrease thereafter, vanishing near apocenter.  We expect even smaller offsets in less massive systems, and in systems where the two subclusters are not equal in mass.  Given the uncertainties in the offset measurements, it will be difficult to generate competitive constraints on the hard-sphere SIDM cross section.

\item Our predictions \emph{cannot} explain offsets measured in observed, on-going equal and near-equal mass mergers, which are typically of order 100s of kpc.  We suggest that noise and systematic effects may explain the large measured offsets.

\item While we have identified several promising signatures of SIDM in merging clusters, BCG miscentering in relaxed systems may produce the most stringent limits on the SIDM cross section yet.  Our simulations predict that BCGs and galaxies should oscillate around barycenter on orbits with radii comparable to the SIDM cores of dark matter halos for many Gyr after the merger.  These excursions are of order 100 kpc for $\sigma/m = 1\hbox{ cm}^2/\hbox{g}$, compared to the $< 20$ kpc offsets found during the merger.  If similar behavior occurs in unequal mass mergers, we should observe that a significant fraction of BCGs are miscentered.  If this prediction is robust under the addition of gas physics and full merger histories, the lack of large BCG miscenterings suggests that $\sigma_\text{SI}/m_\chi \lesssim 0.1\hbox{ cm}^2/\hbox{g}$.  

\end{itemize}

Although we have demoted offset formation in merging clusters as a method to constrain SIDM, we have found several promising avenues to achieve competitive constraints on cluster scales.  Detailed observational studies of the mass and galaxy distributions of merging clusters are crucial to performing the analyses done here.  In addition, whether the predictions we have made for equal mass mergers carry over to unequal mass mergers, like the Bullet Cluster, remains to be seen.  While we have provided an estimate of SIDM phenomenology in unequal mass mergers, simulations---including those with hydrodynamics---are necessary to explore in detail whether our predictions for offset formation, BCG/galaxy miscentering, and other alternative constraints are robust under different mass ratios.  Such simulations are essential to enabling detailed comparisons between observation and theory.


\section*{Acknowledgements}
DMW was supported by NSF grant 1518246.  We thank Maru{\v s}a Brada{\v c}, Marcus Br{\"u}ggen, James Bullock, Will Dawson, James Jee, Manoj Kaplinghat, Reinout van Weeren, and David Sobral for discussions that shaped the direction of this project, and for their careful reading of the manuscript.  The Merging Cluster Collaboration has provided valuable input to this project.  We thank the Ohio Supercomputer Center for the use of the Ruby cluster during its commissioning phase.

\bibliographystyle{mnras}
\bibliography{ms-emm}


\appendix

\section{Derivation of post-interaction outcome probabilities}
\label{apdx:poutcomes}

\subsection{Particle loss}

Here we compute the probability that a two-body collision between dark matter particles in a halo merger will result in a net loss of particles in the halo.  Consider a collision between a particle from each of the merging halos.  Let the velocities of the particles in the rest frame of one halo be $\mathbf{v}_1$ and $\mathbf{v}_2$, and their relative velocity $\mathbf{\Delta v} = \mathbf{v}_1 - \mathbf{v}_2$.  In their center of mass frame, which moves with velocity $\mathbf{v}_\text{COM} = (\mathbf{v}_1 + \mathbf{v}_2)/2$, the particles are scattered isotropically at angle $\theta$ with velocity $\Delta \mathbf{v}/2$.  In the rest frame of the halo, the particles' post-collision velocities are
\begin{equation}
\mathbf{v}_{\pm\text{,lab}} = \mathbf{v}_\text{COM} \pm \frac{\Delta v}{2} \hat{\mathbf{n}},
\end{equation}
where $\hat{\mathbf{n}} = (\cos \theta, \sin \theta)$ denotes the direction the particles were scattered in the center of mass frame.

In order for both particles to escape from the halo at rest, both $v_{+,\text{lab}}$, $v_{-\text{,lab}} > v_\text{esc}$.  This requires
\begin{equation}
v^2_\text{COM} \pm v_\text{COM} \Delta v \cos \theta + \frac{\Delta v^2}{4} > v^2_\text{esc}.
\end{equation}
Since the particles are identical, the collision is symmetric, and we can consider scattering angles $\theta \in (0, \pi/2)$.  Both particles become unbound if the particle with the lower velocity $v_{-,\text{lab}}$ escapes if
\begin{equation}
\cos \theta < \cos \theta_\text{crit}^\text{e} = \frac{v^2_\text{COM} + \Delta v^2/4 - v^2_\text{esc}}{v_\text{COM} \Delta v}. \label{eqn:halorestframe_vcond}
\end{equation}
For those collisions in which $\cos \theta_\text{crit}^\text{e} < 0$, neither particle will escape.  Collisions in which $\cos \theta_\text{crit}^\text{e} \geqslant 1$, on the other hand, will always unbind both particles.  For intermediate cases in which $0 < \cos \theta_\text{crit}^\text{e} < 1$, both particles will escape if the scattering angle satisfies
\begin{align}
\theta_\text{crit}^\text{e} < \theta < \pi/2.
\end{align}
A pair of particles must scatter in such a way such that satisfies both \ref{eqn:halorestframe_vcond} and the angular constraints discussed above in order for both particles to become unbound.  How often does this occur in an equal mass merger?  From the angular constraint, the probability that particles escape is
\begin{equation}
P(\text{escape} \, | \, v_\text{CM}, \Delta v) = \left\{
\begin{array}{l l}
0, & \cos \theta_\text{crit}^\text{e} < 0 \\
\cos \theta_\text{crit}^\text{e}, & 0 \leqslant \cos \theta_\text{crit}^\text{e} \leqslant 1 \\
1, & \cos \theta_\text{crit}^\text{e} > 1
\end{array} \label{eqn:escape}
\right. .
\end{equation}
To derive the overall probability that a given two-particle collision in a cluster merger occuring at velocity $\mathbf{V}$ results in the unbinding of both particles, we make one simplying assumption: we assume that each cluster has a Maxwellian velocity distribution with dispersion $\sigma$.  Under this assumption, the velocity distribution of collisions between particles of opposite clusters is also Maxwellian with dispersion $\sigma_\text{col} = \sqrt{2}\sigma$ (we neglect collisions between particles in the same cluster).  The probability that a two-particle collision involves particles with velocities $\mathbf{v}_1$ and $\mathbf{v}_2$ is
\begin{equation}
P(\mathbf{v}_1,\mathbf{v}_2) = \frac{1}{(2 \pi \sigma^2)^3} \, \exp \left[ - \frac{\mathbf{v}_1^2}{2\sigma^2} \right] \exp \left[ - \frac{ (\mathbf{v}_2 - \mathbf{V})^2}{2\sigma^2} \right] d^3 \mathbf{v}_1 d^3 \mathbf{v}_2.
\end{equation}
In order to express this probability in terms of $v_\text{COM}$ and $\Delta v$, we can make the substitutions $\mathbf{v}_1 = \mathbf{v}_\text{COM} + \mathbf{\Delta v}/2$ and $\mathbf{v}_2 = \mathbf{v}_\text{COM} - \mathbf{\Delta v}/2$, which gives
\begin{multline}
P(\mathbf{v}_\text{COM},\mathbf{\Delta v}) = \frac{e^{-V^2/2\sigma^2}}{(2 \pi \sigma^2)^3} 
\exp \left( - \frac{2 v_\text{COM}^2 -2 \mathbf{v}_\text{COM} \cdot \mathbf{V}}{2 \sigma^2} \right) \nonumber \\
\times \exp \left( - \frac{\Delta v^2/2 + \mathbf{\Delta v} \cdot \mathbf{V}}{2 \sigma^2} \right)
d^3 \mathbf{v}_\text{COM} d^3 \mathbf{\Delta v}. \label{eqn:pvcom6D_vectors}
\end{multline}
The probability that two particles are lost in a given collision is thus
\begin{equation}
P(\text{escape}) = \int P(\text{escape} \, | \, v_\text{com},\Delta v ) P(\mathbf{v}_\text{com},\mathbf{\Delta v}) \, d^3 \mathbf{v}_\text{COM} d^3 \mathbf{\Delta v}. \label{eqn:pesc_vectors}
\end{equation}

\subsection{Particle capture}

The probability that a given collision will result in the capture of the incoming particle (e.g. both of the interacting particles are bound post-collision) can be calculated much like the two-particle escape probability, with a few minor modifications.  Requiring that the post-collision velocities $\mathbf{v}_{\pm,\text{lab}} < v_\text{esc}$ yields the condition
\begin{equation}
v^2_\text{COM} \pm v_\text{COM} \Delta v \cos \theta + \frac{\Delta v^2}{4} < v_\text{esc}^2.
\end{equation}
The particle with the higher velocity $v_{+,\text{lab}}$ remains bound if
\begin{equation}
\cos \theta < \cos \theta_\text{crit}^\text{c} = \frac{v_\text{esc}^2 - v_\text{COM}^2 - \Delta v^2/4}{v_\text{COM} \Delta v}.
\end{equation}
Note that $\cos \theta_\text{crit}^\text{c} = - \cos \theta_\text{crit}^\text{e}$.  The probability that both particles will be bound post-interaction is equal and opposite to the two-particle escape probability,
\begin{equation}
P(\text{capture} \, | \, v_\text{CM}, \Delta v) = \left\{
\begin{array}{l l}
0, & \cos \theta_\text{crit}^\text{c} < 0 \\
\cos \theta_\text{crit}^\text{c}, & 0 \leqslant \cos \theta_\text{crit}^\text{c} \leqslant 1 \\
1, & \cos \theta_\text{crit}^\text{c} > 1
\end{array} \label{eqn:capture}
\right. ,
\end{equation}
and the probability that a given interaction between two particles will result in capture of the incoming particle can be calculated by replacing $P(\text{escape})$ with $P(\text{capture})$ in Eq. \ref{eqn:pesc_vectors}.

\subsection{Particle exchange}

Collisions in which both particles are not bound or unbound post-encounter undergo an exchange.  This is simple to calculate given the calculations above.  Letting $\gamma = \cos \theta_\text{crit}^e = - \cos \theta_\text{crit}^c$, we have that
\begin{equation}
P(\text{exchange} \, | \, v_\text{CM}, \Delta v) = \left\{
\begin{array}{l l}
0, & \gamma < -1 \\
1 - |\gamma|, & -1 \leqslant \gamma \leqslant 1 \\
0, & \gamma > 1
\end{array} \label{eqn:pexchange}
\right. 
\end{equation}


\section{Derivation of momentum transfer}
\label{apdx:momentum}

Here we provide a derivation of the expected change in momentum of a halo undergoing a head-on collision with an identical halo due to a dark matter self-interaction.  We will work in the frame of one halo, say halo 1, and consider only interactions of a halo 1 particle with a halo 2 particle---we do not consider the effects of self-interactions of particles within each halo with each other.

We adopt the notation of the previous appendix, and let $\mathbf{v}_1$ and $\mathbf{v}_2$ denote the velocities of the halo 1 and halo 2 particles before collision, respectively, and $\mathbf{v}'_1$ and $\mathbf{v}'_2$ the particles' velocities post-collision.  The change in momentum of halo 1 $\Delta \mathbf{p} = \mathbf{p} - \mathbf{p}'$ following a self-interaction can be calculated by considering each possible outcome (capture, exchange, or escape) separately.  In the capture scenario, particle 2 becomes bound to halo 1, and halo 1 acquires the momentum of particle 2.  In the escape scenario, particle 1 is kicked out of halo 1, and thus the halo loses the momentum of particle 1.  In the exchange scenario, in which one particle remains and the other escapes, the change in momentum is the difference between the momentum of the particle that remains and the the pre-collision momentum of particle 1.  That is,
\begin{equation}
\Delta \mathbf{p} = m_\chi \left\{
\begin{array}{l l}
- \mathbf{v}_1, & \text{escape} \\
\mathbf{v}_\text{remain} - \mathbf{v}_1, & \text{exchange} \\
\mathbf{v}_2, & \text{capture}
\end{array}
\right. 
\end{equation}

We can now calculate the expected change in momentum of halo 1 following a self-interaction,
\begin{equation}
\langle \langle \Delta \mathbf{p} \rangle \rangle =  \int \Delta \langle \mathbf{p} \rangle ( \mathbf{v}_1, \mathbf{v}_2 ) \, P( \mathbf{v}_1, \mathbf{v}_2 ) \, d^3 \mathbf{v}_1 \, d^3 \mathbf{v}_2.
\end{equation}
or in the center of mass coordinates,
\begin{equation}
\langle \Delta \mathbf{p} \rangle =  \int \Delta \mathbf{p} ( \Delta \mathbf{v}, \mathbf{v}_\text{COM} ) \, P( \mathbf{v}_\text{COM} , \Delta \mathbf{v}) \, d^3 \mathbf{v}_\text{COM} \, d^3 \Delta \mathbf{v}.
\end{equation}


\section{Validation of the ``speedy SIDM" implementation for non-cosmological \textsc{GADGET2} simulations}
\label{apdx:sidm_validation}

\subsection{Modifications to the \citet{2013MNRAS.430...81R} implementation of self-interaction physics in GADGET2}

Our dark matter self-interaction algorithm is based on a GADGET2 implementation by \citet{2013MNRAS.430...81R}, in which self-interactions are modeled as hard-sphere elastic scattering events.  Although simulation particles are numerically described by a single position---an infinitesimal portion of configuration space---in actuality, they represent a finite volume of configuration space, represented by a smoothing kernel $W$ with a smoothing length $h_\text{SI}$ centered on the particle's position.  \citet{2013MNRAS.430...81R} found that in order to accurately calculate the probability that self-interactions occur between two particles, $h_\text{SI}$ should be dependent on the local dark matter density.  However, they chose to adopt a single smoothing length for all dark matter particles, which underestimates the self-interaction probability in low-density regions.  In addition, many unnecessary, costly calculations (the computational cost scales $\propto h_\text{SI}^3$) were performed in dense regions in which the constant smoothing length adopted was larger than needed to accurately estimate the self-interaction probability.

In order to address both issues, we modified \citet{2013MNRAS.430...81R}'s algorithm to have an adaptive self-interaction smoothing scale.  \citet{2013MNRAS.430...81R} found that the optimal $h_\text{SI}$ that optimizes speed and accuracy in computing the interaction probability was
\begin{equation}
h_{\text{SI}} = 0.2 (\rho/m)^{-1/3},
\end{equation}
where $\rho$ is the local dark matter density and $m$ is the mass of the dark matter simulation particles; $\rho/m$ represents the average interparticle spacing.  For each dark matter particle $i$, we compute local dark matter densities $\rho_i$ using the built-in SPH density estimator, then calculate its smoothing scale $h_{\text{SI},i}$.  During the simulation, $h_{\text{SI},i}$ is updated for all dark matter particles whenever their gravitational accelerations are updated.

\subsection{Hernquist Sphere Tests}

\begin{figure}
\centering
\includegraphics[width=\columnwidth]{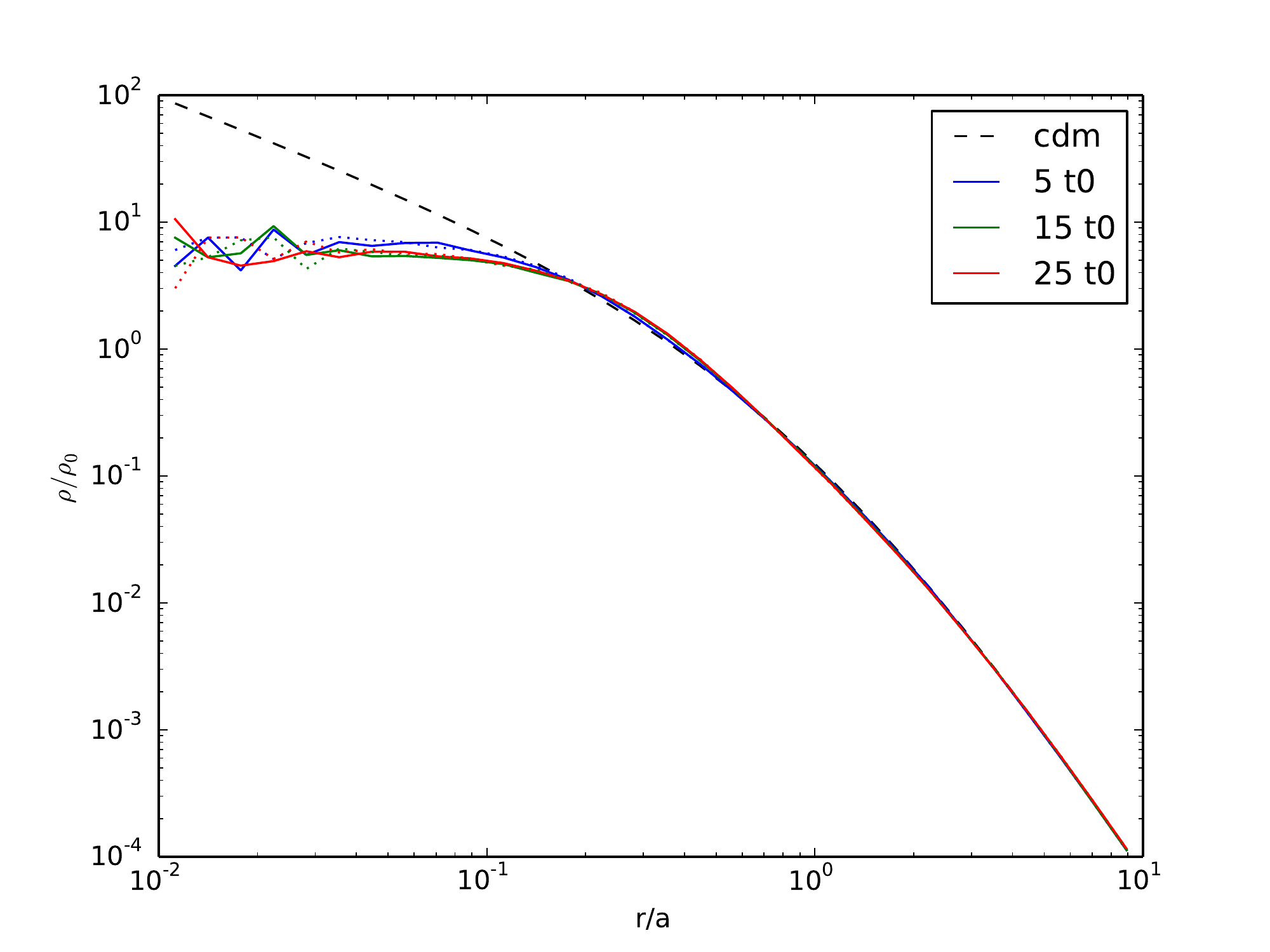}
\includegraphics[width=\columnwidth]{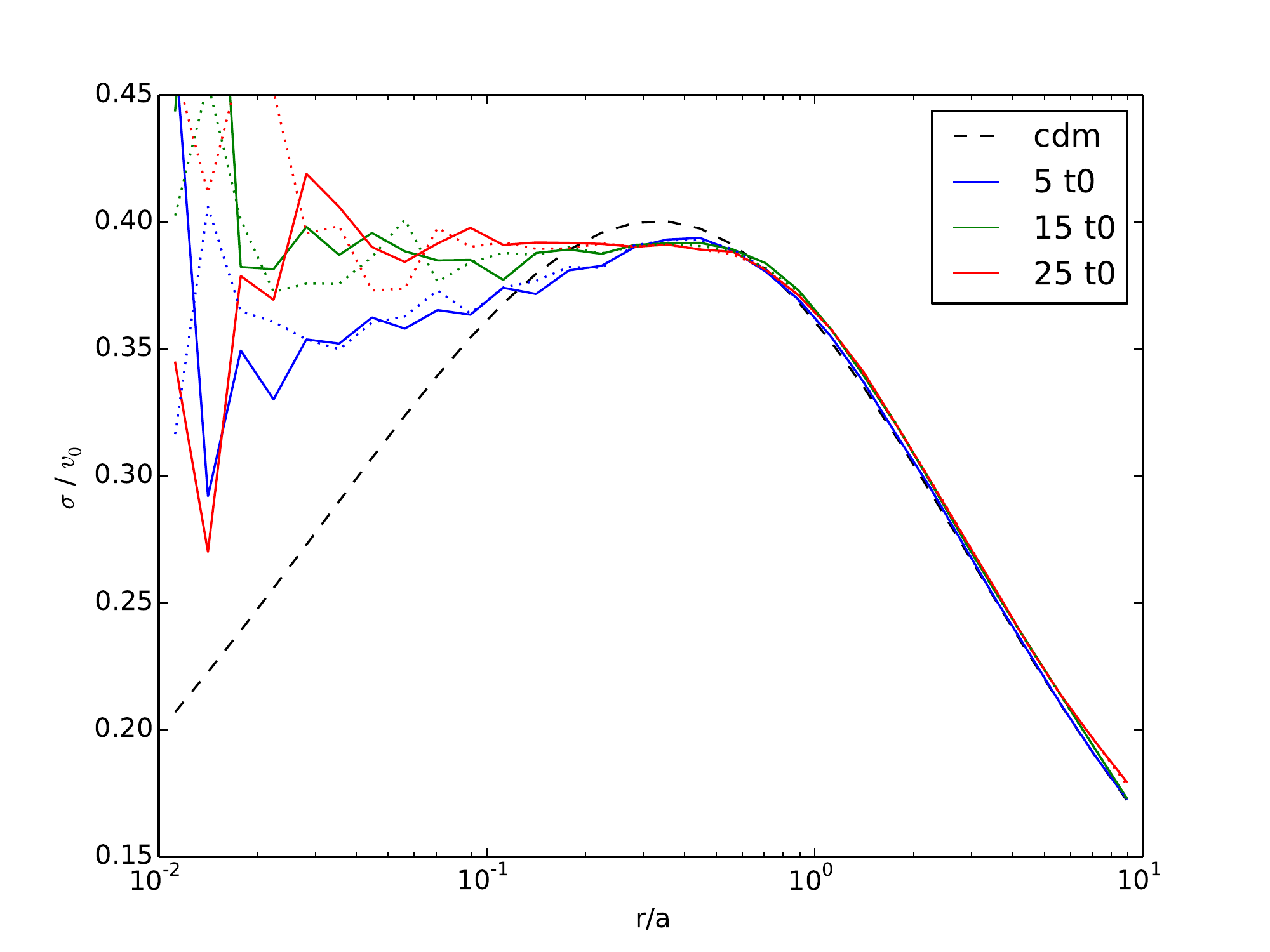}
\caption{Time evolution of density (top) and velocity dispersion (bottom) profiles of a 10$^{12}$ M$_\odot$ Hernquist SIDM halo (scale radius $a$ = 26 kpc, relaxation time $t_0$ = 39 Myr, $\sigma_\text{SI}/m_\chi$ = 10 cm$^2$ g$^{-1}$) with constant (solid lines) or variable (dotted lines) SIDM smoothing lengths.  $h_\text{SIDM}$ = 0.0063$a$ = 0.16 kpc for the former.  The black dashed line represents the theoretical profiles for a CDM Hernquist sphere.  
}
\label{fig:hernquist}
\end{figure}

We tested our smoothing modification in an astrophysically relevant context:  a Hernquist sphere \citep{1990ApJ...356..359H}.  This test allows us to test the code over a range of densities and perform speed benchmarks on a dark-matter-only simulation.  We simulated Hernquist spheres ranging from 10$^7$ to 10$^{12}$ M$_\odot$ with a self-interaction cross section of 10 cm$^2$/g for 1 Gyr as in \citet{2012MNRAS.423.3740V}.  In Figure \ref{fig:hernquist}, we compare the density and velocity dispersion profiles for the 10$^{12}$ M$_\odot$ halo modeled with constant (dotted) and adaptive (solid) SIDM smoothing lengths.  The profiles computed by the two algorithms agree.

Substantial gains are made in the runtime of the simulations with adaptive stepping, ranging between a factor of about 3 up to 12 depending on the density contrast, tree algorithm, and halo mass.

\section{Validation of Peak Finding Methods}
\label{apdx:peak_validation}

\begin{figure*}
\centering
\includegraphics[width=\columnwidth]{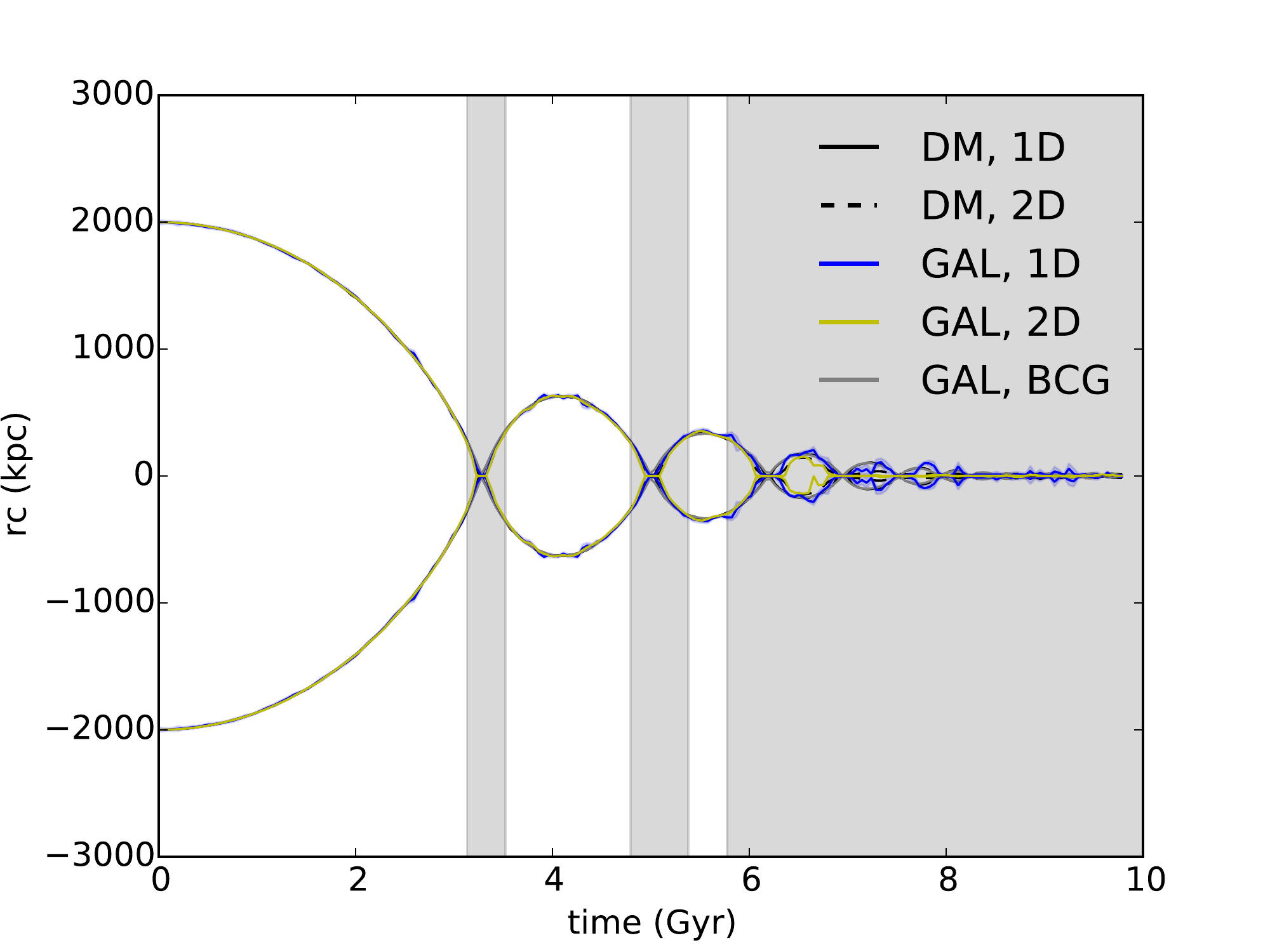}
\includegraphics[width=\columnwidth]{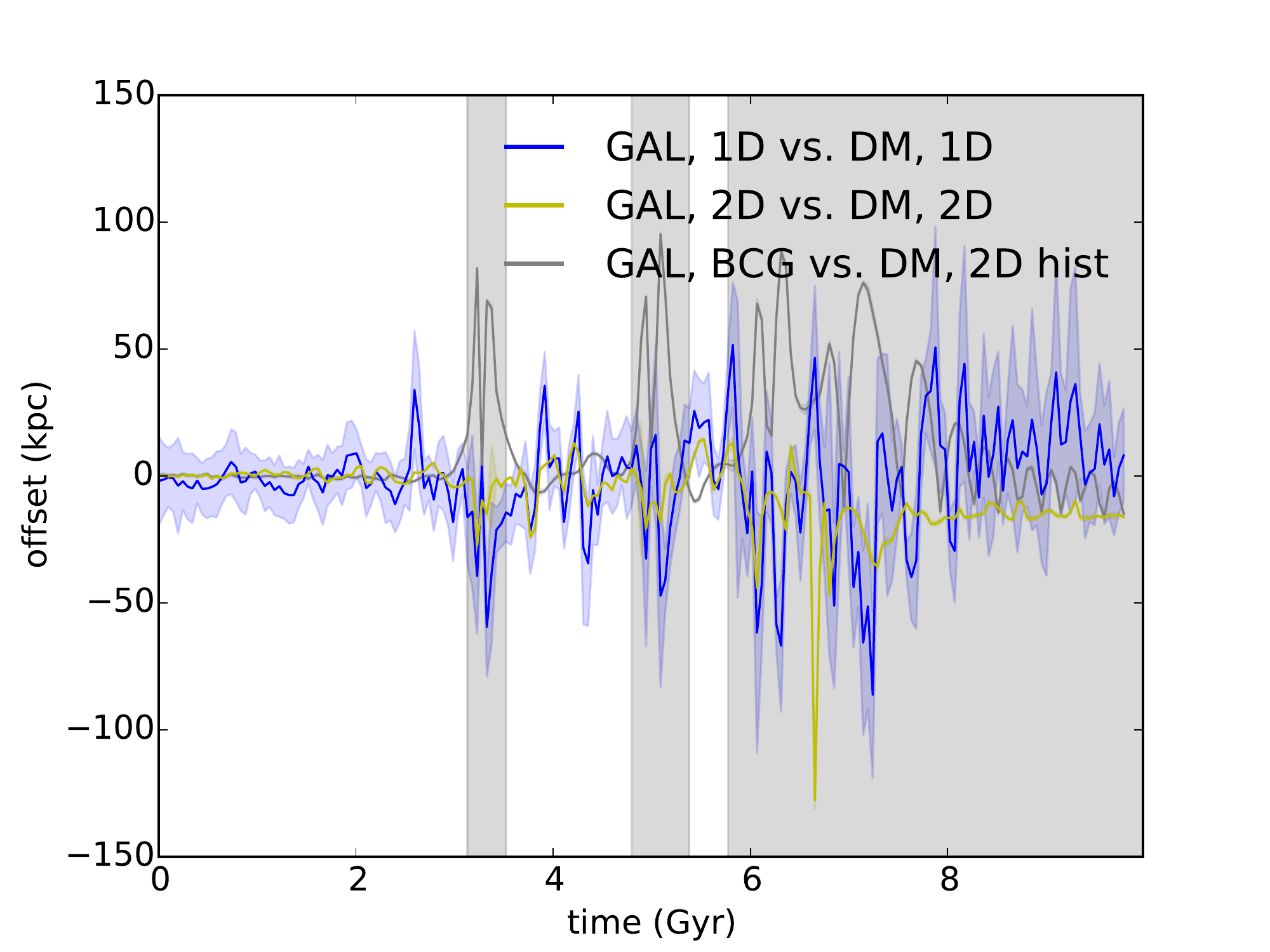}
\caption{Verification of smoothing methods.  Plotted on left are the evolution of dark matter and galaxy peaks estimated by various analysis methods and on the right are the offsets between the dark matter and galaxy peaks for our fiducial CDM $v_\text{4Mpc}$ = 0 km/s simulation. No offsets are expected.  Shaded in gray are epochs when peaks were separated by less than the dark matter scaling radius.  }
\label{fig:peak_check}
\end{figure*}

In CDM, no forces act preferentially on the dark matter, and so the dark matter and galaxy distributions and their respective peaks should evolve identically.  Thus we can check that our peak identification methods are accurate and unbiased by verifying that the dark matter and galaxy offsets do not exist in CDM mergers.  In addition, the comparison allows us to check the uncertainties on the measured offsets.

In Figure \ref{fig:peak_check}, we show such a test---plotted are the evolution of peaks (left panel) and the galaxy-dark matter offsets (right panel) calculated by all methods outlined above for our fiducial, $v_\text{4Mpc}$ = 0 km/s CDM merger.  Before pericenter passage, offsets are consistent with zero.  After pericenter passage, larger excursions from zero occur, but are typically consistent with zero.  Offsets are at most about 40 kpc for the 1D analysis and 20 kpc for 2D, which are within the range of our bootstrapped uncertainties.

Large, statistically significant deviations occur near pericenter passages.  We have shaded in gray those epochs in which the dark matter halos are separated by less than the scale radius (about 600 kpc).  When the separation between dark matter halos are small, peak identification can become increasingly biased towards barycenter as one halo ``contaminates" the peak identification of the other halo due to their large overlap.  In offset plots in the main text, we thereby do not show offsets when the halo-to-halo separation is less than the scale radius.

\subsection{Notes on Choice of Analysis Parameters}

\begin{figure}
\centering
\includegraphics[width=\columnwidth]{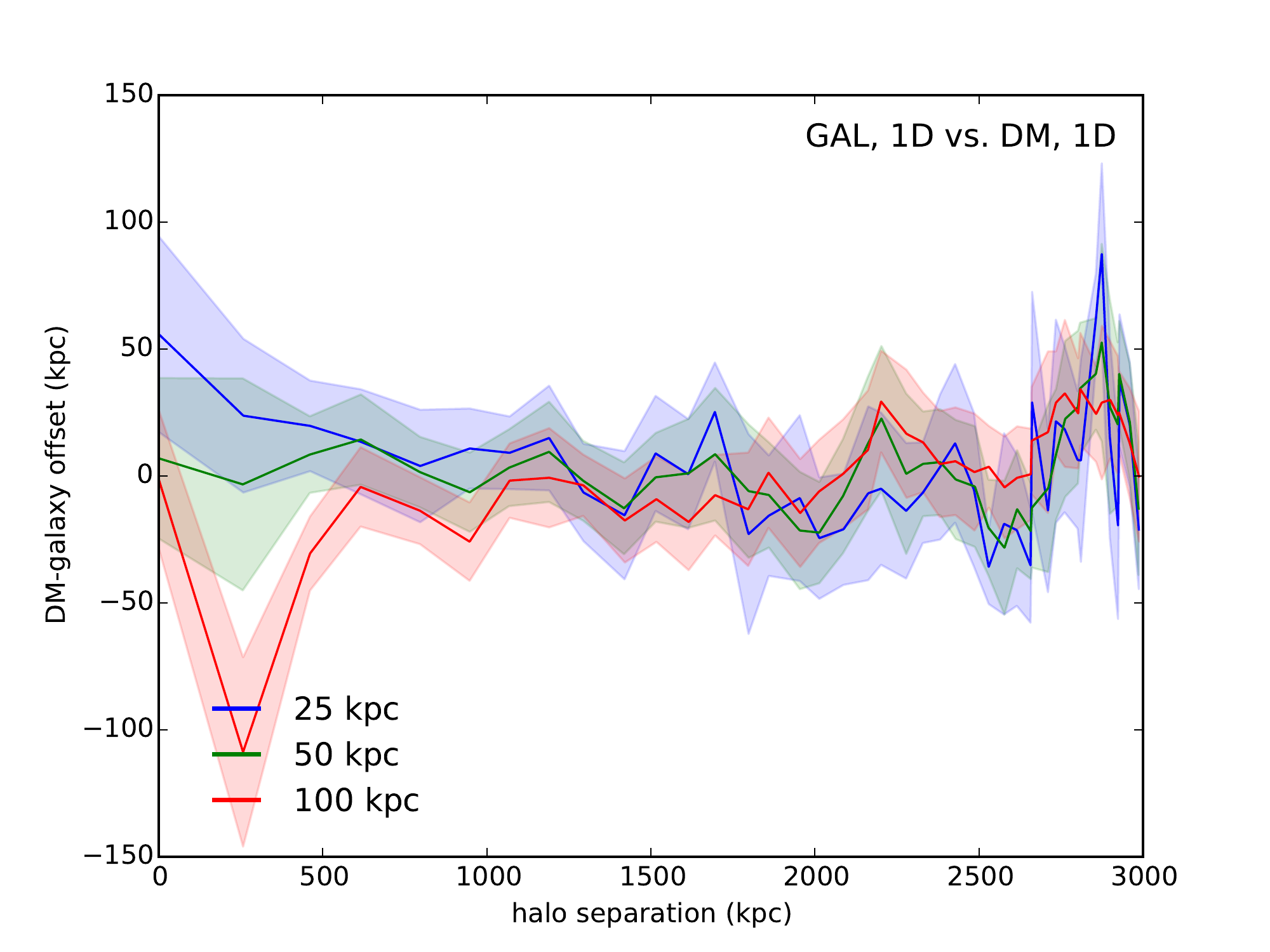}
\caption{Impact of smoothing scale on offset estimates plotted against halo-to-halo separation.  We show the evolution of the galaxy peaks in a SIDM merger with $\sigma_\text{SI}/m_\chi$ = 1 cm$^2$/g just past pericenter as a function of galaxy smoothing.  The galaxy distribution ($N$ = 5800 particles) is smoothed with a gaussian kernel with a width of 25 (blue), 50 (green), or 100 (red) kpc.  The bands show uncertainties based on bootstrap resampling.}
\label{fig:smoothing}
\end{figure}

The presence and size of offsets depends sensitively on the choice of smoothing scale.  Smoothing can modify or erase structure below the smoothing scale, which presents two difficulties: (1) two peaks separated by less than the smoothing scale can be smoothed into one, and (2) asymmetries in the peak below the smoothing scale can be smoothed out due to contamination from the second halo.  The first is largely a concern near pericenter passage, when identification of individual halos is difficult, and thus such clusters would not be identified in a merging cluster survey as such.  The second, however, can be in effect at significant halo-to-halo separations.  Figure \ref{fig:smoothing} demonstrates this effect: shown is a snapshot of the galaxy distribution for an SIDM merger with $\sigma_\text{SI}/m_\chi$ = 1 cm$^2$/g after first pericenter passage, smoothed via a gaussian kernel with smoothing widths varying from 25 to 100 kpc.  The smallest smoothing scale resolves asymmetric peak structure as the central peak of galaxies moves forward relative to the rest of the galaxies, while the largest smoothing produces symmetric peaks, causing a shift in the peak towards barycenter by about 40 kpc.  Such asymmetries can exist until peaks are separated by a few scale radii.  Our smoothing scale choices maximize smoothness of evolution yet minimize the contribution from the other halo.

\bsp	
\label{lastpage}
\end{document}